\pdfoutput=1

\documentclass[12pt,a4paper]{article}

\usepackage{ifthen} 
\newboolean{pdflatex}
\setboolean{pdflatex}{true} 

\newboolean{articletitles}
\setboolean{articletitles}{true} 

\newboolean{uprightparticles}
\setboolean{uprightparticles}{true} 

\newboolean{inbibliography}
\setboolean{inbibliography}{false} 

\newboolean{feynmanlabels}
\setboolean{feynmanlabels}{true}

\usepackage{microtype}
\usepackage{lineno}  
\usepackage{xspace} 
\usepackage{caption} 

\usepackage{graphicx}  
\usepackage{subcaption}
\usepackage{placeins}
\usepackage{rotating}
\usepackage{color}
\usepackage{colortbl}
\usepackage{booktabs}
\usepackage{multirow}
\usepackage[export]{adjustbox}
\usepackage{feynmp}
\usepackage{lipsum}
\graphicspath{{./figs/}} 

\usepackage{amsmath} 
\usepackage{amssymb}
\usepackage{amsfonts}
\usepackage{upgreek} 
\usepackage{mathtools}
\newcommand*\patchAmsMathEnvironmentForLineno[1]{%
\expandafter\let\csname old#1\expandafter\endcsname\csname #1\endcsname
\expandafter\let\csname oldend#1\expandafter\endcsname\csname
end#1\endcsname
 \renewenvironment{#1}%
   {\linenomath\csname old#1\endcsname}%
   {\csname oldend#1\endcsname\endlinenomath}%
}
\newcommand*\patchBothAmsMathEnvironmentsForLineno[1]{%
  \patchAmsMathEnvironmentForLineno{#1}%
  \patchAmsMathEnvironmentForLineno{#1*}%
}
\AtBeginDocument{%
\patchBothAmsMathEnvironmentsForLineno{equation}%
\patchBothAmsMathEnvironmentsForLineno{align}%
\patchBothAmsMathEnvironmentsForLineno{flalign}%
\patchBothAmsMathEnvironmentsForLineno{alignat}%
\patchBothAmsMathEnvironmentsForLineno{gather}%
\patchBothAmsMathEnvironmentsForLineno{multline}%
\patchBothAmsMathEnvironmentsForLineno{eqnarray}%
}

\usepackage{hyperref}    
\usepackage[all]{hypcap} 

\def\lhcb {\mbox{LHCb}\xspace}

\def\babar  {\mbox{BaBar}\xspace}

\def\cleo   {\mbox{CLEO}\xspace}

\def\lhc    {\mbox{LHC}\xspace}

\def\MagUp {\mbox{\em Mag\kern -0.05em Up}\xspace}

\ifthenelse{\boolean{uprightparticles}}%
{

 \def\Peta        {\ensuremath{\upeta}\xspace}

 \def\Pmu         {\ensuremath{\upmu}\xspace}                 
 \def\Pnu         {\ensuremath{\upnu}\xspace}                 
                  
 \def\Ppi         {\ensuremath{\uppi}\xspace}                 
                  
 \def\Prho        {\ensuremath{\uprho}\xspace}                 
                  
 \def\Ptau        {\ensuremath{\uptau}\xspace}

 \def\Ppsi        {\ensuremath{\uppsi}\xspace}

 \def\PDelta      {\ensuremath{\Delta}\xspace}                 
 \def\PXi      {\ensuremath{\Xi}\xspace}                 
 \def\PLambda      {\ensuremath{\Lambda}\xspace}                 
 \def\PSigma      {\ensuremath{\Sigma}\xspace}                 
 \def\POmega      {\ensuremath{\Omega}\xspace}                 
 \def\PUpsilon      {\ensuremath{\Upsilon}\xspace}                 
 
 \def\PA      {\ensuremath{\mathrm{A}}\xspace}                 
 \def\PB      {\ensuremath{\mathrm{B}}\xspace}                 
 \def\PC      {\ensuremath{\mathrm{C}}\xspace}                 
 \def\PD      {\ensuremath{\mathrm{D}}\xspace}

 \def\PK      {\ensuremath{\mathrm{K}}\xspace}

 \def\PR      {\ensuremath{\mathrm{R}}\xspace}

 \def\PW      {\ensuremath{\mathrm{W}}\xspace}                 
 \def\PX      {\ensuremath{\mathrm{X}}\xspace}

 \def\Pa      {\ensuremath{\mathrm{a}}\xspace}                 
 \def\Pb      {\ensuremath{\mathrm{b}}\xspace}                 
 \def\Pc      {\ensuremath{\mathrm{c}}\xspace}                 
 \def\Pd      {\ensuremath{\mathrm{d}}\xspace}                 
                  
 \def\Pf      {\ensuremath{\mathrm{f}}\xspace}                 
                  
 \def\Ph      {\ensuremath{\mathrm{h}}\xspace}                 
 \def\Pi      {\ensuremath{\mathrm{i}}\xspace}

 \def\Pp      {\ensuremath{\mathrm{p}}\xspace}

 \def\Ps      {\ensuremath{\mathrm{s}}\xspace}                 
                  
 \def\Pu      {\ensuremath{\mathrm{u}}\xspace}

}
{

 \def\Peta        {\ensuremath{\eta}\xspace}

 \def\Pmu         {\ensuremath{\mu}\xspace}                 
 \def\Pnu         {\ensuremath{\nu}\xspace}                 
                  
 \def\Ppi         {\ensuremath{\pi}\xspace}                 
                  
 \def\Prho        {\ensuremath{\rho}\xspace}                 
                  
 \def\Ptau        {\ensuremath{\tau}\xspace}

 \def\Ppsi        {\ensuremath{\psi}\xspace}                 
                  
 \mathchardef\PDelta="7101
 \mathchardef\PXi="7104
 \mathchardef\PLambda="7103
 \mathchardef\PSigma="7106
 \mathchardef\POmega="710A
 \mathchardef\PUpsilon="7107
 \def\PA      {\ensuremath{A}\xspace}                 
 \def\PB      {\ensuremath{B}\xspace}                 
 \def\PC      {\ensuremath{C}\xspace}                 
 \def\PD      {\ensuremath{D}\xspace}

 \def\PK      {\ensuremath{K}\xspace}

 \def\PR      {\ensuremath{R}\xspace}

 \def\PW      {\ensuremath{W}\xspace}                 
 \def\PX      {\ensuremath{X}\xspace}

 \def\Pa      {\ensuremath{a}\xspace}                 
 \def\Pb      {\ensuremath{b}\xspace}                 
 \def\Pc      {\ensuremath{c}\xspace}                 
 \def\Pd      {\ensuremath{d}\xspace}                 
                  
 \def\Pf      {\ensuremath{f}\xspace}                 
                  
 \def\Ph      {\ensuremath{h}\xspace}                 
 \def\Pi      {\ensuremath{i}\xspace}

 \def\Pp      {\ensuremath{p}\xspace}

 \def\Ps      {\ensuremath{s}\xspace}                 
                  
 \def\Pu      {\ensuremath{u}\xspace}

}

\makeatletter
\ifcase \@ptsize \relax
  \newcommand{\miniscule}{\@setfontsize\miniscule{4}{5}}
\or
  \newcommand{\miniscule}{\@setfontsize\miniscule{5}{6}}
\or
  \newcommand{\miniscule}{\@setfontsize\miniscule{5}{6}}
\fi
\makeatother

\DeclareRobustCommand{\optbar}[1]{\shortstack{{\miniscule (\rule[.5ex]{1.25em}{.18mm})}
  \\ [-.7ex] $#1$}}

\def\mup        {{\ensuremath{\Pmu^+}}\xspace}

\def\taum       {{\ensuremath{\Ptau^-}}\xspace}

\def\neu        {{\ensuremath{\Pnu}}\xspace}

\def\neum       {{\ensuremath{\neu_\mu}}\xspace}

\def\neut       {{\ensuremath{\neu_\tau}}\xspace}

\def\W      {{\ensuremath{\PW}}\xspace}
\def\Wp     {{\ensuremath{\PW^+}}\xspace}
\def\Wm     {{\ensuremath{\PW^-}}\xspace}

\def\uquark    {{\ensuremath{\Pu}}\xspace}
\def\uquarkbar {{\ensuremath{\overline \uquark}}\xspace}

\def\dquark    {{\ensuremath{\Pd}}\xspace}
\def\dquarkbar {{\ensuremath{\overline \dquark}}\xspace}

\def\squark    {{\ensuremath{\Ps}}\xspace}
\def\squarkbar {{\ensuremath{\overline \squark}}\xspace}

\def\cquark    {{\ensuremath{\Pc}}\xspace}

\def\bquark    {{\ensuremath{\Pb}}\xspace}

\def\pion   {{\ensuremath{\Ppi}}\xspace}
\def\piz    {{\ensuremath{\pion^0}}\xspace}

\def\pip    {{\ensuremath{\pion^+}}\xspace}
\def\pim    {{\ensuremath{\pion^-}}\xspace}
\def\pipm   {{\ensuremath{\pion^\pm}}\xspace}
\def\pimp   {{\ensuremath{\pion^\mp}}\xspace}

\def\rhomeson {{\ensuremath{\Prho}}\xspace}

\def\kaon    {{\ensuremath{\PK}}\xspace}
\def\Kbar    {{\overline{\PK}{}}\xspace}

\def\KorKbar {\optbar{\PK}{}\xspace}
\def\Kz      {{\ensuremath{\kaon^0}}\xspace}
\def\Kzb     {{\ensuremath{\Kbar{}^0}}\xspace}
\def\Kp      {{\ensuremath{\kaon^+}}\xspace}
\def\Km      {{\ensuremath{\kaon^-}}\xspace}
\def\Kpm     {{\ensuremath{\kaon^\pm}}\xspace}
\def\Kmp     {{\ensuremath{\kaon^\mp}}\xspace}
\def\KS      {{\ensuremath{\kaon^0_{\rm\scriptscriptstyle S}}}\xspace}

\def\Kstarz  {{\ensuremath{\kaon^{*0}}}\xspace}
\def\Kstarzb {{\ensuremath{\Kbar{}^{*0}}}\xspace}
\def\Kstar   {{\ensuremath{\kaon^*}}\xspace}

\def\Kstarp  {{\ensuremath{\kaon^{*+}}}\xspace}
\def\Kstarm  {{\ensuremath{\kaon^{*-}}}\xspace}
\def\Kstarpm {{\ensuremath{\kaon^{*\pm}}}\xspace}

\newcommand{\etaz}{\ensuremath{\Peta}\xspace}
\newcommand{\etapr}{\ensuremath{\Peta^{\prime}}\xspace}

\def\Dbar    {{\overline{\PD}{}}\xspace}
\def\D       {{\ensuremath{\PD}}\xspace}

\ifthenelse{\boolean{uprightparticles}}
{
\def\DorDbar {\optbar{\PD}{}\xspace}
}
{
\def\DorDbar {\kern 0.18em\optbar{\kern -0.18em \PD}{}\xspace}
}
\def\Dz      {{\ensuremath{\D^0}}\xspace}
\def\Dzb     {{\ensuremath{\Dbar{}^0}}\xspace}
\def\Dp      {{\ensuremath{\D^+}}\xspace}

\def\B       {{\ensuremath{\PB}}\xspace}
\def\Bbar    {{\ensuremath{\kern 0.18em\overline{\kern -0.18em \PB}{}}}\xspace}

\def\BorBbar    {\kern 0.18em\optbar{\kern -0.18em B}{}\xspace}

\def\Bub     {{\ensuremath{\B^-}}\xspace}

\def\Bm      {{\ensuremath{\Bub}}\xspace}

\def\psiprpr  {{\ensuremath{\Ppsi(3770)}}\xspace}

  \def\Y#1S{\ensuremath{\PUpsilon{(#1S)}}\xspace}

\def\proton      {{\ensuremath{\Pp}}\xspace}

\def\Lbar        {{\ensuremath{\kern 0.1em\overline{\kern -0.1em\PLambda}}}\xspace}
\def\LorLbar    {\kern 0.18em\optbar{\kern -0.18em \PLambda}{}\xspace}

\def\BF         {{\ensuremath{\cal B}}\xspace}

\def\BR         {\BF}
\newcommand{\decay}[2]{\ensuremath{#1\!\to #2}\xspace}         

\def\to                 {\ensuremath{\rightarrow}\xspace}

\def\grpsuthree {{\ensuremath{\mathrm{SU}(3)}}\xspace}

\def\order   {{\ensuremath{\mathcal{O}}}\xspace}

\def\eps   {{\ensuremath{\varepsilon}}\xspace}

\def\CP                {{\ensuremath{C\!P}}\xspace}

\def\Vud  {{\ensuremath{V_{\uquark\dquark}^{}}}\xspace}
\def\Vcd  {{\ensuremath{V_{\cquark\dquark}^{}}}\xspace}

\def\Vubs  {{\ensuremath{V_{\uquark\bquark}^\ast}}\xspace}
\def\Vcbs  {{\ensuremath{V_{\cquark\bquark}^\ast}}\xspace}

\newcommand{\dm}{{\ensuremath{\Delta m}}\xspace}

\def\AT#1     {\ensuremath{A_{\mathrm{T}}^{#1}}\xspace}           

\def\C#1      {\ensuremath{\mathcal{C}_{#1}}\xspace}                       
\def\Cp#1     {\ensuremath{\mathcal{C}_{#1}^{'}}\xspace}                    
\def\Ceff#1   {\ensuremath{\mathcal{C}_{#1}^{\mathrm{(eff)}}}\xspace}        
\def\Cpeff#1  {\ensuremath{\mathcal{C}_{#1}^{'\mathrm{(eff)}}}\xspace}       
\def\Ope#1    {\ensuremath{\mathcal{O}_{#1}}\xspace}                       
\def\Opep#1   {\ensuremath{\mathcal{O}_{#1}^{'}}\xspace}                    

\newcommand{\ket}[1]{\ensuremath{|#1\rangle}}              
\newcommand{\braket}[2]{\ensuremath{\langle #1|#2\rangle}} 

\newcommand{\tev}{\ifthenelse{\boolean{inbibliography}}{\ensuremath{~T\kern -0.05em eV}\xspace}{\ensuremath{\mathrm{\,Te\kern -0.1em V}}}\xspace}
\newcommand{\gev}{\ensuremath{\mathrm{\,Ge\kern -0.1em V}}\xspace}
\newcommand{\mev}{\ensuremath{\mathrm{\,Me\kern -0.1em V}}\xspace}
\newcommand{\kev}{\ensuremath{\mathrm{\,ke\kern -0.1em V}}\xspace}
\newcommand{\ev}{\ensuremath{\mathrm{\,e\kern -0.1em V}}\xspace}
\newcommand{\gevc}{\ensuremath{{\mathrm{\,Ge\kern -0.1em V\!/}c}}\xspace}
\newcommand{\mevc}{\ensuremath{{\mathrm{\,Me\kern -0.1em V\!/}c}}\xspace}
\newcommand{\gevcc}{\ensuremath{{\mathrm{\,Ge\kern -0.1em V\!/}c^2}}\xspace}
\newcommand{\gevgevcccc}{\ensuremath{{\mathrm{\,Ge\kern -0.1em V^2\!/}c^4}}\xspace}
\newcommand{\mevcc}{\ensuremath{{\mathrm{\,Me\kern -0.1em V\!/}c^2}}\xspace}

\def\m    {\ensuremath{\rm \,m}\xspace}

\def\mum  {\ensuremath{{\,\upmu\rm m}}\xspace}

\def\invfb   {\ensuremath{\mbox{\,fb}^{-1}}\xspace}

\newcommand{\stat}{\ensuremath{\mathrm{\,(stat)}}\xspace}
\newcommand{\syst}{\ensuremath{\mathrm{\,(syst)}}\xspace}

\def\order{{\ensuremath{\cal O}}\xspace}
\newcommand{\chisq}{\ensuremath{\chi^2}\xspace}
\newcommand{\chisqndf}{\ensuremath{\chi^2/\mathrm{ndf}}\xspace}

\def\gsim{{~\raise.15em\hbox{$>$}\kern-.85em
          \lower.35em\hbox{$\sim$}~}\xspace}
\def\lsim{{~\raise.15em\hbox{$<$}\kern-.85em
          \lower.35em\hbox{$\sim$}~}\xspace}

\def\sqs   {\ensuremath{\protect\sqrt{s}}\xspace}

\def\ptot       {\mbox{$p$}\xspace}
\def\pt         {\mbox{$p_{\rm T}$}\xspace}

\def\rad{\ensuremath{\rm \,rad}\xspace}

\def\evtgen     {\mbox{\textsc{EvtGen}}\xspace}

\def\geant      {\mbox{\textsc{Geant4}}\xspace}

\def\photos     {\mbox{\textsc{Photos}}\xspace}

\def\pythia     {\mbox{\textsc{Pythia}}\xspace}

\def\tell1  {TELL1\xspace}
\def\ukl1   {UKL1\xspace}

\newcommand{\eg}{\mbox{\itshape e.g.}\xspace}
\newcommand{\ie}{\mbox{\itshape i.e.}\xspace}

\def\KSKpis  {\ensuremath{\KS\kaon\pion}\xspace}
\def\KSKpi    	{\ensuremath{\KS\Kpm\pimp}\xspace}
\def\KSKppim {\ensuremath{\KS\Kp\pim}\xspace}
\def\KSKmpip {\ensuremath{\KS\Km\pip}\xspace}
\def\DzKSKpi    {\ensuremath{\decay{\Dz}{\KSKpi}}\xspace}

\def\kpi                {\ensuremath{\kaon\pion}\xspace}

\def\sKSK   	{\ensuremath{m^{2}_{\KS\kaon}}\xspace}
\def\sKSpi		{\ensuremath{m^{2}_{\KS\pion}}\xspace}
\def\sKpi		{\ensuremath{m^{2}_{\kaon\pion}}\xspace}

\def\mDz		{\ensuremath{m_{\Dz}}\xspace}

\def\mKSKpis             {\ensuremath{m(\KSKpis)}\xspace}
\def\mPi		{\ensuremath{m_{\pi}}\xspace}
\def\mAB		{\ensuremath{m_{\PA\PB}}\xspace}
\def\mABsq	{\ensuremath{m^{2}_{\PA\PB}}\xspace}
\def\mACsq	{\ensuremath{m^{2}_{\PA\PC}}\xspace}
\def\mR			{\ensuremath{m_{\PR}}\xspace}
\def\mRsq		{\ensuremath{m_{\PR}^{2}}\xspace}

\def\mcM		{\ensuremath{\mathcal{M}}\xspace}
\newcommand{\amL}[1]{\ensuremath{\Pa^{{#1}}_{0,2}}\xspace}
\newcommand{\KstL}[1]{\ensuremath{\kaon^{*{#1}}_{0,1,2}}\xspace}
\newcommand{\KbarstL}[1]{\ensuremath{\Kbar^{*{#1}}_{0,1,2}}\xspace}

\def\KstLpm     {\KstL{\pm}}
\def\KstLz      {\KstL{0}}
\def\KbstLz     {\KbarstL{0}}
\newcommand{\Kst}[3]{\ensuremath{\kaon^{*}_{{#1}}({#2})^{{#3}}}\xspace}
\newcommand{\ameson}[3]{\ensuremath{\Pa_{#1}(#2)^{{#3}}}\xspace}
\newcommand{\rmeson}[2]{\ensuremath{\rhomeson(#1)^{{#2}}}\xspace}
\newcommand{\Kbarst}[3]{\ensuremath{\Kbar^{*}_{{#1}}({#2})^{{#3}}}\xspace}
\newcommand{\Dst}[2]{\ensuremath{\D^{*}({#1})^{{#2}}}\xspace}
\newcommand{\Dstp}[1]{\Dst{{#1}}{+}}

\def\KstarzorKstarzbar  {\ensuremath{\KorKbar^{\kern -0.2em *0}}\xspace}
\def\Dsttt		{\Dstp{2010}}

\def\lass		{\mbox{LASS}\xspace}
\def\flatte		{Flatt\'{e}\xspace}
\def\gousak  {Gounaris-Sakurai\xspace}

\def\psrap		{\ensuremath{\eta}\xspace}
\def\pisoft		{\ensuremath{\Ppi^{+}_{\text{slow}}}\xspace}
\def\pislow	{\pisoft}
\def\amp		{\ensuremath{\mathcal{A}}\xspace}

\newcommand{\Msq}[1]{\ensuremath{|\mathcal{M}_{{#1}}(\sKSpi, \sKpi)|^{2}}\xspace}
\newcommand{\Mbsq}[1]{\ensuremath{|\overline{\mathcal{M}}_{{#1}}(\sKSpi, \sKpi)|^{2}}\xspace}
\newcommand{\combfn}[1]{\ensuremath{c_{{#1}}(\sKSpi, \sKpi)}\xspace}
\def\efffn              {\ensuremath{\eps(\sKSpi, \sKpi)}\xspace}

\newcommand{\chisqbin}{\ensuremath{\chi^2/\mathrm{bin}}\xspace}

\def\goofit      {\mbox{\textsc{GooFit}}\xspace}
\newcommand{\kevcc}{\ensuremath{{\mathrm{\,ke\kern -0.1em V\!/}c^2}}\xspace}
\newcommand{\tevcc}{\ensuremath{{\mathrm{\,Te\kern -0.1em V\!/}c^2}}\xspace}
\newcommand{\mevmevcccc}{\ensuremath{{\mathrm{\,Me\kern -0.1em V^2\!/}c^4}}\xspace}

\def\hp			{\ensuremath{\Ph^{+}}\xspace}
\def\hm			{\ensuremath{\Ph^{-}}\xspace}
\def\hpm		{\ensuremath{\Ph^{\pm}}\xspace}

\newcommand{\invgevc}{\ensuremath{{\,(\mathrm{Ge\kern -0.1em V\!/}c)^{-1}}}\xspace}
\def\glass		{\mbox{\texttt{GLASS}}\xspace}
\def\lassc		{\mbox{\texttt{LASS}}\xspace}

\def\neutswave {\mbox{$(\kaon\pion)^{0}_{S\text{-wave}}$}\xspace}
\def\chgswave {\mbox{$(\KS\pion)^{\pm}_{S\text{-wave}}$}\xspace}
\def\posswave {\mbox{$(\KS\pion)^{+}_{S\text{-wave}}$}\xspace}
\def\negswave {\mbox{$(\KS\pion)^{-}_{S\text{-wave}}$}\xspace}
\def\kpswave	   {\mbox{$\kaon\pion$ $S$-wave}\xspace}
\def\ntll			{\ensuremath{-2\log\mathcal{L}}\xspace}
\def\aR       {\ensuremath{a_{\PR}}\xspace}
\def\pR       {\ensuremath{\phi_{\PR}}\xspace}
\def\daR      {\ensuremath{\Delta a_{\PR}}\xspace}
\def\dpR      {\ensuremath{\Delta\phi_{\PR}}\xspace}

\usepackage{cite} 
\usepackage{mciteplus}

\usepackage{tikz}
\newcommand{\graphwithletter}[5]{
	\begin{tikzpicture}
	 	\node[anchor=south west] at (0,0) {\includegraphics[clip,trim=0 0 0 0,width=#2]{{{#1}}}};
		\node at (#3#2,#4#2) {#5};
	\end{tikzpicture}}
\newcommand{\massfitwithletter}[2]{\graphwithletter{#1}{\textwidth}{0.32}{0.75}{#2}}
\usepackage{makecell}
\renewcommand{\rothead}[2][60]{\makebox[9mm][l]{\rotatebox{#1}{\makecell[c]{#2}}}}%

\usepackage{longtable} 

\begin{document}

\renewcommand{\thefootnote}{\fnsymbol{footnote}}
\setcounter{footnote}{1}


\begin{titlepage}
\pagenumbering{roman}

\vspace*{-1.5cm}
\centerline{\large EUROPEAN ORGANIZATION FOR NUCLEAR RESEARCH (CERN)}
\vspace*{1.5cm}
\noindent
\begin{tabular*}{\linewidth}{lc@{\extracolsep{\fill}}r@{\extracolsep{0pt}}}
\ifthenelse{\boolean{pdflatex}}
{\vspace*{-2.7cm}\mbox{\!\!\!\includegraphics[width=.14\textwidth]{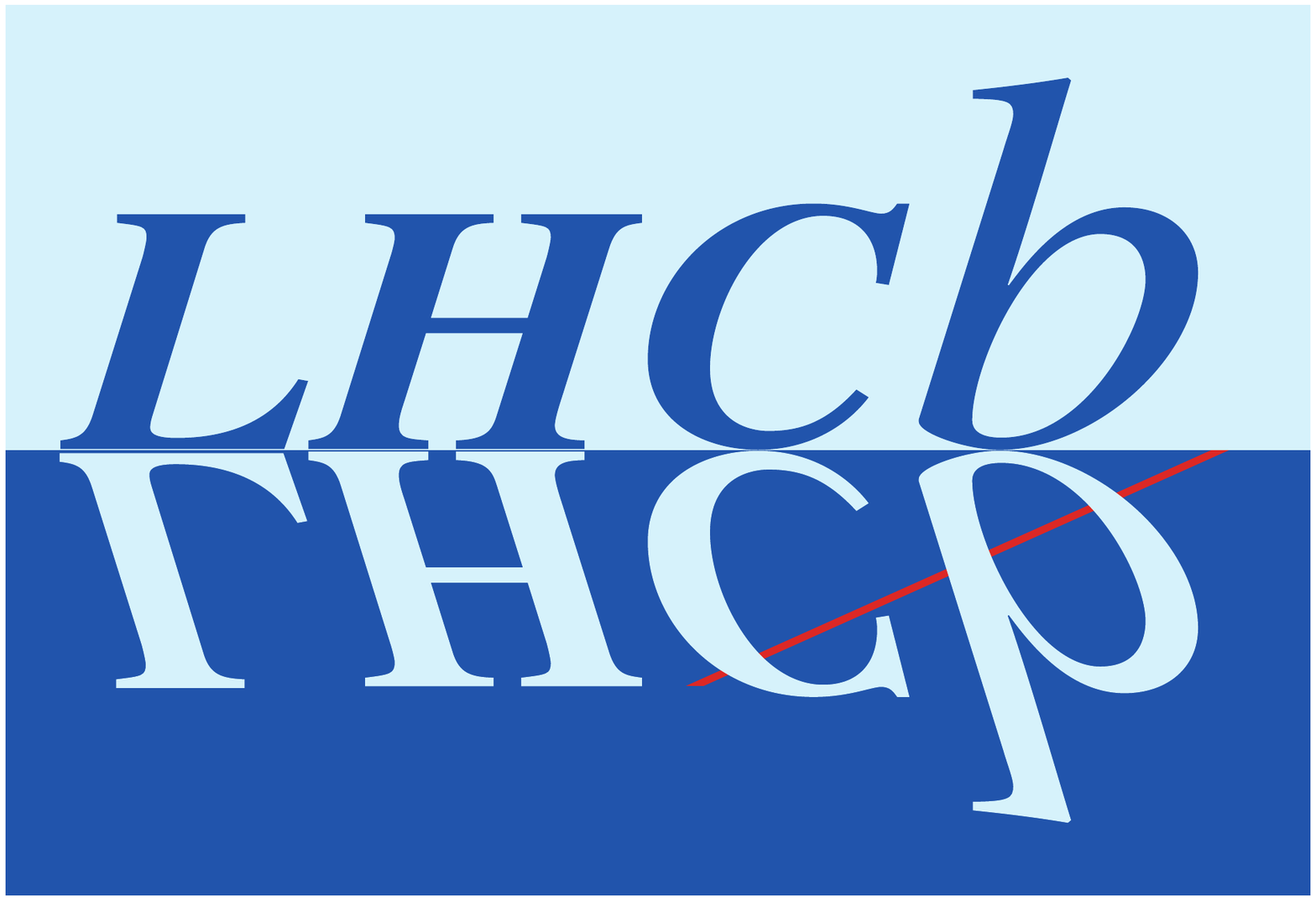}} & &}%
{\vspace*{-1.2cm}\mbox{\!\!\!\includegraphics[width=.12\textwidth]{lhcb-logo.eps}} & &}%
\\
 & & CERN-PH-EP-2015-238 \\  
 & & LHCb-PAPER-2015-026 \\  
 & & September 22, 2015 \\ 
 & & \\
\end{tabular*}

\vspace*{1.5cm}

{\bf\boldmath\huge
\begin{center}
        Studies of the resonance structure in \decay{\Dz}{\KSKpi} decays %
\end{center}
}

\vspace*{1.0cm}

\begin{center}
The LHCb collaboration\footnote{Authors are listed at the end of this paper.}
\end{center}

\vspace{\fill}

\begin{abstract}
  \noindent
  Amplitude models are applied to studies of resonance structure
  in \decay{\Dz}{\KSKmpip} and \decay{\Dz}{\KSKppim} decays using
  $\proton\proton$ collision data corresponding to an integrated
  luminosity of $3.0\invfb$ collected by the \lhcb experiment.
  Relative magnitude and phase information is determined,
  and coherence factors and related observables are computed for both the whole phase
  space and a restricted region
  of $100\mevcc$ around
  the \Kst{}{892}{\pm} resonance.
  Two formulations for the \kpswave are used,
  both of which give a good description of the data.
  The ratio of branching fractions 
  $\BR(\decay{\Dz}{\KSKppim})/\BR(\decay{\Dz}{\KSKmpip})$
  is measured to be
  $0.655\pm0.004\stat\pm0.006\syst$
  over the full phase space and
  $0.370\pm0.003\stat\pm0.012\syst$
  in the restricted region.
  A search for \CP violation is performed using the amplitude models
  and no significant effect is found.
  Predictions from \grpsuthree flavor symmetry for $\Kst{}{892}{}\kaon$ amplitudes
  of different charges are compared with the amplitude model results.
  \end{abstract}
  
\vspace*{2.0cm}

\begin{center}
  Published in Phys.~Rev.~D~93,~052018~(2016)
\end{center}

\vspace{\fill}

{\footnotesize 
\centerline{\copyright~CERN on behalf of the \lhcb collaboration, license \href{http://creativecommons.org/licenses/by/4.0/}{CC-BY-4.0}.}}
\vspace*{2mm}

\end{titlepage}


\newpage
\setcounter{page}{2}
\mbox{~}
\cleardoublepage


\renewcommand{\thefootnote}{\arabic{footnote}}
\setcounter{footnote}{0}



\pagestyle{plain} 
\setcounter{page}{1}
\pagenumbering{arabic}


\section{Introduction}
\label{sec:Introduction}
A large variety of physics can be accessed by studying
the decays\footnote{The inclusion of charge-conjugate processes is implied, except
in the definition of \CP asymmetries.} \decay{\Dz}{\KSKmpip} and \decay{\Dz}{\KSKppim}.
Analysis of the relative amplitudes of intermediate resonances contributing
to these decays can help in understanding the behavior of the
strong interaction at low energies. These modes are also of interest for improving knowledge of the
Cabibbo-Kobayashi-Maskawa (CKM)~\cite{Cabibbo:1963yz,Kobayashi:1973fv}
matrix, and \CP-violation measurements and mixing studies in the \Dz--\Dzb
system.
Both modes are singly Cabibbo-suppressed (SCS), with the \KSKmpip
final state favored by approximately $\times1.7$ with respect to its \KSKppim
counterpart~\cite{PDG2014}.
The main classes of Feynman diagrams, and the sub-decays to which they contribute,
are shown in Fig.~\ref{fig:feynman}.

Flavor symmetries are an important phenomenological tool in the study
of hadronic decays, and the presence of both charged and neutral \Kstar
resonances in each \DzKSKpi mode allows several tests of \grpsuthree flavor
symmetry to be carried out~\cite{Bhattacharya:2011gf,Bhattacharya:2012fz}.
The \KSKpi final states also provide opportunities to study
the incompletely understood \kpswave systems~\cite{PDGscalarmesons}, and
to probe several resonances in the $\KS\Kpm$ decay
channels that are poorly established.

An important goal of flavor physics is to make a precise determination
of the CKM unitarity-triangle angle $\gamma \equiv \arg({-\Vud\Vubs/\Vcd\Vcbs})$.
Information on this parameter\footnote{Another notation, $\phi_3 \equiv \gamma$,
exists in the literature.} can be obtained by studying \CP-violating
observables in the decays \decay{\Bm}{\DorDbar^{0}\Km}, where the \Dz and \Dzb are
reconstructed in a set of common final states~\cite{Atwood:1996ci,Atwood:2000ck},
such as the modes \decay{\Dz}{\KSKmpip} and \decay{\Dz}{\KSKppim}~\cite{Grossman:2002aq}.
Optimum statistical power is achieved by studying the dependence of the
\CP asymmetry on where in three-body phase space the \PD-meson decay occurs,
provided that the decay amplitude from the intermediate resonances is sufficiently
well described.
Alternatively, an inclusive analysis may be pursued, as in Ref.~\cite{LHCB-PAPER-2013-068},
with a `coherence factor'~\cite{Atwood:2003mj} parameterizing the net effect of these
resonances.
The coherence factor of these decays has been measured by the \cleo collaboration using
quantum-correlated \Dz decays at the open-charm threshold~\cite{Insler:2012pm}, but it may also
be calculated from knowledge of the contributing resonances.
In both cases, therefore, it is valuable to be able to model
the variation of the magnitude and phase of the \Dz-decay amplitudes across phase space.

The search for \CP violation in the charm system is motivated by the
fact that several theories of physics beyond the Standard Model (SM) predict enhancements above
the very small effects expected in the SM~\cite{Bianco:2003vb,Artuso:2008vf,Buccella:2013tya}.
Singly Cabibbo-suppressed decays provide a promising laboratory
in which to perform this search for direct \CP violation because
of the significant role that
loop diagrams play in these processes~\cite{Grossman:2006jg}.
Multi-body SCS decays, such as \decay{\Dz}{\KSKmpip} and \decay{\Dz}{\KSKppim},
have in addition the attractive feature that the interfering resonances may lead to
\CP violation in local regions of phase space, again motivating a good
understanding of the resonant substructure.
The same modes may also be exploited to perform a \Dz--\Dzb mixing measurement,
or to probe indirect \CP violation,
either through a time-dependent measurement of the evolution of the phase space
of the decays, or the inclusive \KSKmpip and \KSKppim final states~\cite{Malde:2011mk}.

In this paper time-integrated amplitude models of these decays are constructed and used to test
\grpsuthree flavor symmetry predictions, search for local \CP violation, and
compute coherence factors and associated parameters.
In addition, a precise measurement is performed of the ratio of branching fractions
of the two decays.
The data sample is obtained from $\proton\proton$ collisions corresponding
to an integrated luminosity of $3.0\invfb$ collected by the \lhcb detector~\cite{Alves:2008zz,LHCb-DP-2014-002}
during 2011 and 2012 at center-of-mass energies $\sqs=7\tev$ and $8\tev$, respectively.
The sample contains around one hundred times more signal decays than were analyzed in
a previous amplitude study of the same modes performed by the CLEO collaboration~\cite{Insler:2012pm}.

The paper is organized as follows. In Sect.~\ref{sec:Detector}, the detector,
data and simulation samples are described, and in Sect.~\ref{sec:selection} the
signal selection and backgrounds are discussed.
The analysis formalism, including the definition of the coherence factor,
is presented in Sect.~\ref{sec:formalism}.
The method for choosing the composition of the amplitude models,
fit results and their systematic uncertainties are described in Sect.~\ref{sec:models}.
The ratio of branching fractions, coherence factors, \grpsuthree flavor symmetry tests
and \CP violation search results are presented in Sect.~\ref{sec:otherresults}.
Finally, conclusions are drawn in Sect.~\ref{sec:conclusions}.

\begin{fmffile}{flavourless}
\begin{figure*}
  \begin{center}
    \input{feynman_impl_a}
    \hspace{2.0cm}
    \input{feynman_impl_b}
    \vspace{1.5cm}
    \input{feynman_impl_c}
    \hspace{2.0cm}
    \input{feynman_impl_d}
  \end{center}
  \caption{SCS classes of diagrams contributing to the decays \decay{\Dz}{\KSKpi}. The color-favored (tree) diagrams (a) contribute to the \decay{\KstLpm}{\KS\pipm} and \decay{(\amL{},\Prho)^{\pm}}{\KS\Kpm} channels, while the color-suppressed exchange diagrams (b) contribute to the \decay{(\amL{},\Prho)^{\pm}}{\KS\Kpm}, \decay{\KstLz}{\Kp\pim} and \decay{\KbstLz}{\Km\pip} channels. Second-order loop (penguin) diagrams (c) contribute to the \decay{(\amL{},\Prho)^{\pm}}{\KS\Kpm} and \decay{\KstLpm}{\KS\pipm} channels, and, finally, OZI-suppressed penguin annihilation diagrams (d) contribute to all decay channels.}
  \label{fig:feynman}
\end{figure*}
\end{fmffile}

\section{Detector and simulation}
\label{sec:Detector}
The \lhcb detector is a single-arm forward
spectrometer covering the \mbox{pseudorapidity} range $2<\psrap <5$,
designed for the study of particles containing \bquark or \cquark
quarks. The detector includes a high-precision tracking system
consisting of a silicon-strip vertex detector surrounding the $\proton\proton$
interaction region, a large-area silicon-strip detector located
upstream of a dipole magnet with a bending power of about
$4{\rm\,Tm}$, and three stations of silicon-strip detectors and straw
drift tubes placed downstream of the magnet.
The tracking system provides a measurement of momentum, \ptot, of charged particles with
a relative uncertainty that varies from 0.5\% at low momentum to 1.0\% at $200\gevc$.
The minimum distance of a track to a primary $\proton\proton$ interaction vertex (PV),
the impact parameter, is measured with a resolution of $(15+29/\pt)\mum$,
where \pt is the component of the momentum transverse to the beam, in\,\gevc.
Different types of charged hadrons are distinguished using information
from two ring-imaging Cherenkov (RICH) detectors. 
Photons, electrons and hadrons are identified by a calorimeter system consisting of
scintillating-pad and preshower detectors, an electromagnetic
calorimeter and a hadronic calorimeter.

The trigger~\cite{LHCb-DP-2012-004} consists of a
hardware stage, based on information from the calorimeter and muon
systems, followed by a software stage, in which all charged particles
with $\pt>500\,(300)\mevc$ are reconstructed for 2011 (2012) data.
At the hardware trigger stage, events are required to have a muon with high \pt or a
hadron, photon or electron with high transverse energy in the calorimeters. For hadrons,
the transverse energy threshold is $3.5\gev$.
Two software trigger selections are combined for this analysis.
The first reconstructs the decay chain \decay{\Dsttt}{\Dz\pislow}
with \decay{\Dz}{\hp\hm\PX}, where \hpm represents a pion or a kaon and \PX
refers to any number of additional particles.
The charged pion originating in the \Dsttt decay is referred to as `slow' due
to the small $Q$-value of the decay.
The second selection fully reconstructs the decay \decay{\Dz}{\KSKpi}, without flavor tagging.
In both cases at least one charged particle in the decay chain is required to have
a significant impact parameter with respect to any PV.

In the offline selection, trigger signals are associated with reconstructed particles.
Selection requirements can therefore be made on the trigger selection itself
and on whether the decision was due to the signal candidate, other particles produced
in the $\proton\proton$ collision, or both.
It is required that the hardware hadronic trigger decision is due
to the signal candidate, or that the hardware
trigger decision is due solely to other particles produced in the $\proton\proton$ collision.

Decays \decay{\KS}{\pip\pim} are reconstructed in two different categories:
the first involves \KS mesons that decay early enough for the
pions to be reconstructed in the vertex detector; the
second contains \KS mesons that decay later such that track segments of the
pions cannot be formed in the vertex detector. These categories are
referred to as \emph{long} and \emph{downstream}, respectively. The
long category has better mass, momentum and vertex resolution than the
downstream category, and in 2011 was the only category available in the software trigger.

In the simulation, $\proton\proton$ collisions are generated using
\pythia~\cite{Sjostrand:2006za,*Sjostrand:2007gs} 
 with a specific \lhcb
configuration~\cite{LHCb-PROC-2010-056}.  Decays of hadronic particles
are described by \evtgen~\cite{Lange:2001uf}, in which final-state
radiation is generated using \photos~\cite{Golonka:2005pn}.
The interaction of the generated particles with the detector, and its response,
are implemented using the \geant
toolkit~\cite{Allison:2006ve, *Agostinelli:2002hh} as described in
Ref.~\cite{LHCb-PROC-2011-006}.

\section{Signal selection and backgrounds}
\label{sec:selection}
The offline selection used in this analysis reconstructs the decay chain
\decay{\Dsttt}{\Dz\pislow} with \decay{\Dz}{\KSKpi}, where the charged pion \pislow
from the \Dsttt decay tags the flavor of the neutral \PD meson.
Candidates are required to pass one of the two software trigger selections
described in Sect.~\ref{sec:Detector}, as well as several offline requirements.
These use information from the RICH detectors to ensure
that the charged kaon is well-identified, which reduces the background contribution
from the decays \decay{\Dz}{\KS\pip\pim\piz} and \decay{\Dz}{\KS\pim\mup\neum}.
In addition the \KS decay vertex is required to be well-separated from the \Dz decay vertex
in order to suppress the \decay{\Dz}{\Km\pip\pip\pim} background, where a $\pip\pim$ combination
is close to the \KS mass.
\Dz candidates are required to have decay vertices well-separated from any PV,
and to be consistent with originating from a PV.
This selection suppresses the semileptonic and \decay{\Dz}{\Km\pip\pip\pim} backgrounds
to negligible levels, while a small contribution from \decay{\Dz}{\KS\pip\pim\piz}
remains in the $\dm\equiv m(\KSKpis\pion_{\text{slow}}) - \mKSKpis$ distribution.
A kinematic fit~\cite{Hulsbergen:2005pu} is applied to the reconstructed \Dsttt decay
chain to enhance the resolution in \mKSKpis, \dm and the two-body
invariant masses $m(\KS\kaon)$, $m(\KS\pion)$ and $m(\kaon\pion)$ that are used to probe the resonant
structure of these decays.
This fit constrains the \Dsttt decay vertex to coincide
with the closest PV with respect to the \Dsttt candidate,
fixes the \KS candidate mass to its nominal value,
and is required to be of good quality.

Signal yields and estimates of the various background contributions in the signal
window are determined using maximum likelihood fits to the \mKSKpis and \dm distributions.
The signal window is defined as the region less than $18\mevcc$ ($0.8\mevcc$) from the peak
value of \mKSKpis (\dm), corresponding to approximately three standard deviations
of each signal distribution.
The three categories of interest are: signal decays, mistagged background
where a correctly reconstructed \Dz meson is combined with a charged pion that incorrectly
tags the \Dz flavor, and a combinatorial background category, which also includes a small
peaking contribution in \dm from the decay \decay{\Dz}{\KS\pip\pim\piz}.
These fits use candidates in the ranges $139 < \dm < 153 \mevcc$ and $1.805 < \mKSKpis < 1.925 \gevcc$.
The sidebands of the \mKSKpis distribution are defined as those parts of the fit range where \mKSKpis
is more than $30\mevcc$ from the peak value.
The \dm (\mKSKpis) distribution in the signal region of \mKSKpis (\dm) is fitted to determine the
\Dsttt (\Dz) yield in the two-dimensional signal region~\cite{Verkerke:2003ir}.
The \Dsttt (\Dz) signal shape in the \dm (\mKSKpis) distribution is modeled using a Johnson $S_U$~\cite{Johnson:1949zj}
(Cruijff~\cite{delAmoSanchez:2010ae}) function. In the \mKSKpis distribution the combinatorial
background is modeled with an exponential function, while in \dm a power law function is used,
$f(\dm ; \mPi, p, P, b) = (\frac{\dm - \mPi}{\mPi})^p - b^{p-P}(\frac{\dm - \mPi}{\mPi})^P$,
with the parameters $p$, $P$ and $b$ determined by a fit in the \mKSKpis sidebands.
The small \decay{\Dz}{\KS\pip\pim\piz} contribution in the \dm distribution is described by a Gaussian function,
and the component corresponding
to \Dz mesons associated with a random slow pion is the sum of an exponential function
and a linear term.
These fits are shown in Fig.~\ref{fig:massfit}.
The results of the fits are used to determine the yields of interest in the two-dimensional signal region.
These yields are given in Table~\ref{tab:massfit} for both decay modes, together with the fractions
of backgrounds.

\begin{figure*}
	\begin{center}
		\begin{subfigure}{0.45\textwidth}
			\includegraphics[width=0.95\textwidth]{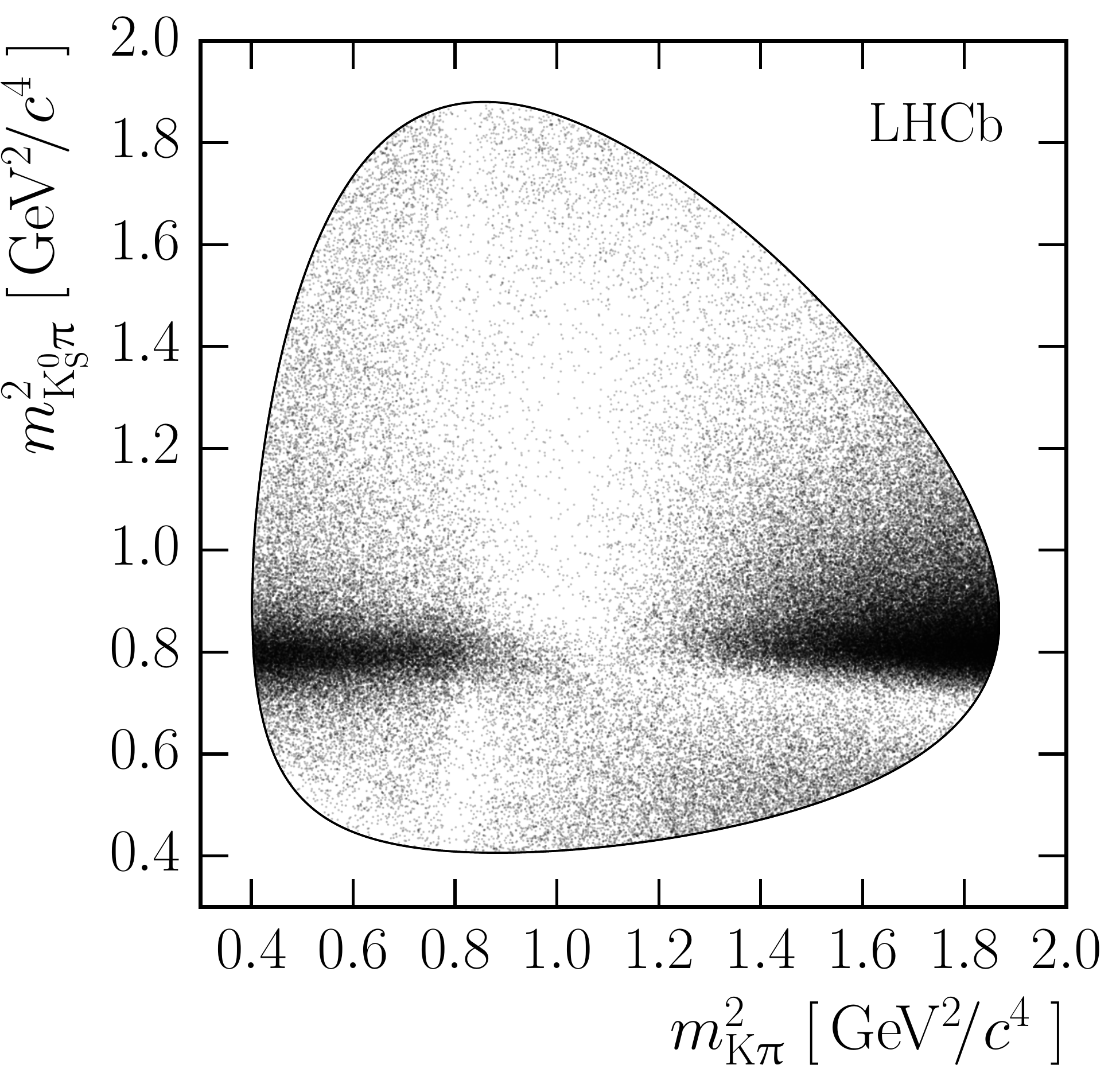}
		\end{subfigure}%
		\begin{subfigure}{0.45\textwidth}
			\includegraphics[width=0.95\textwidth]{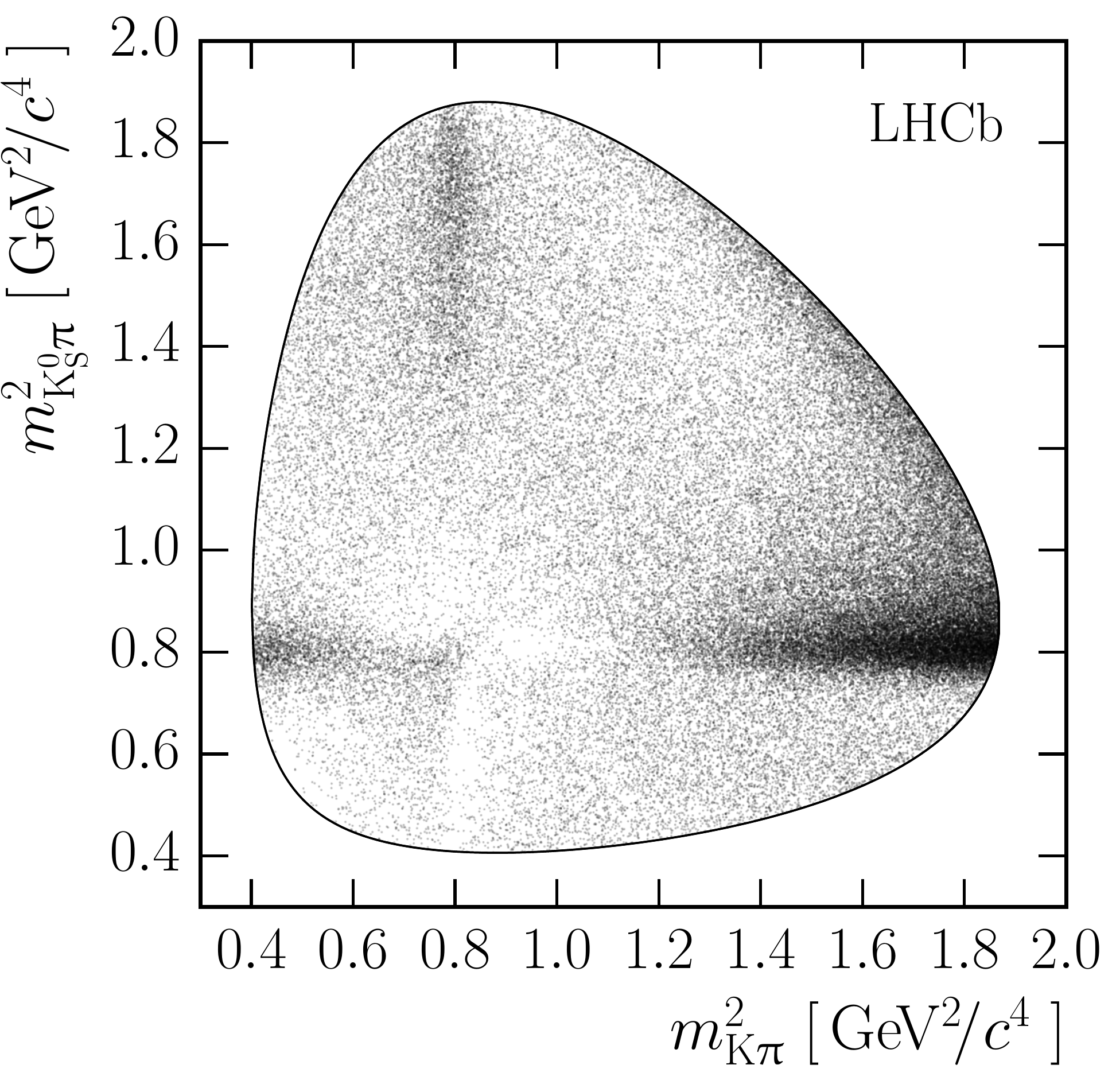}
		\end{subfigure}
	\end{center}
        \caption{Dalitz plots of the \decay{\Dz}{\KSKmpip} (left) and \decay{\Dz}{\KSKppim} (right)
                  candidates in the two-dimensional signal region.}
	\label{fig:datadalitzplots}
\end{figure*}

A second kinematic fit that also constrains the \Dz mass to
its known value is performed and used for all subsequent parts
of this analysis. This fit further improves the resolution in the two-body 
invariant mass coordinates and forces all candidates to lie within the
kinematically allowed region of the Dalitz plot.
The Dalitz plots~\cite{Dalitz:1953cp} for data in the two-dimensional signal region are shown in
Fig.~\ref{fig:datadalitzplots}. Both decays are dominated by a \Kst{}{892}{\pm}
structure.
The \Kst{}{892}{0} is also visible
as a destructively interfering contribution in the \decay{\Dz}{\KSKmpip} mode
and the low-\sKSpi region of the \decay{\Dz}{\KSKppim} mode, while a clear
excess is seen in the high-\sKSpi region.
Finally, a veto is applied to candidates close to the kinematic boundaries;
this is detailed in Sect.~\ref{sec:efficiency}.
\begin{figure*}
  \begin{center}
    \begin{subfigure}{0.45\textwidth}
      \massfitwithletter{figs/rd_tos_fav_mD0_nopull_mpl}{}
    \end{subfigure}
    \begin{subfigure}{0.45\textwidth}
	 \massfitwithletter{figs/rd_tos_fav_delM_nopull_mpl}{}
    \end{subfigure} \\
    \begin{subfigure}{0.45\textwidth}
      \massfitwithletter{figs/rd_tos_sup_mD0_nopull_mpl}{}
    \end{subfigure}
    \begin{subfigure}{0.45\textwidth}
	 \massfitwithletter{figs/rd_tos_sup_delM_nopull_mpl}{}
    \end{subfigure}
  \end{center}
  \vspace{-5mm}
  \caption{Mass (left) and \dm (right) distributions for the
            \decay{\Dz}{\KSKmpip} (top) and \decay{\Dz}{\KSKppim} (bottom)
            samples with fit results superimposed.
            The long-dashed (blue) curve represents the \Dsttt signal,
            the dash-dotted (green) curve represents the contribution
            of real \Dz mesons combined with incorrect \pislow
            and the dotted (red) curve represents
            the combined combinatorial and \decay{\Dz}{\KS\pip\pim\piz}
            background contribution. The vertical solid lines show the
            signal region boundaries, and the vertical dotted lines show
            the sideband region boundaries.}
  \label{fig:massfit}
\end{figure*}
\begin{table*}
  \caption{Signal yields and estimated background rates in the two-dimensional signal region.
           The larger mistag rate in the \decay{\Dz}{\KSKppim} mode is due to the different
           branching fractions for the two modes. Only statistical uncertainties are quoted.}
  \begin{center}
    \begin{tabular}{l r@{$\,\pm\,$}l r@{$\,\pm\,$}l r@{$\,\pm\,$}l}
      & \multicolumn{2}{l}{} & \multicolumn{2}{l}{Mistag} & \multicolumn{2}{l}{Combinatorial} \\
      Mode & \multicolumn{2}{l}{Signal yield} & \multicolumn{2}{l}{background [\%]} & \multicolumn{2}{l}{background [\%]} \\
      \midrule
      \decay{\Dz}{\KSKmpip} & 113\,290 & 130 & 0.89 & 0.09 & 3.04 & 0.14 \\
      \decay{\Dz}{\KSKppim} & 76\,380 & 120 & 1.93 & 0.16 & 2.18 & 0.15 \\
    \end{tabular}
  \end{center}
  \label{tab:massfit}
\end{table*}

\section{Analysis formalism}
\label{sec:formalism}
The dynamics of a decay \decay{\Dz}{\PA\PB\PC}, where \Dz, \PA, \PB and \PC are all pseudoscalar mesons,
can be completely described by two variables, where the conventional choice is to use a pair of squared invariant
masses.
This paper will use $\sKSpi \equiv m^{2}(\KS\pion)$ and $\sKpi \equiv m^{2}(\kpi)$ as this choice highlights the dominant resonant structure of the
\decay{\Dz}{\KSKpi} decay modes.
\subsection{Isobar models for {\boldmath\decay{\Dz}{\KSKpi}}}
\label{sec:isobarformalism}
The signal isobar models decompose the decay chain into \decay{\Dz}{(\decay{\PR}{(\PA\PB)_J})\PC} contributions,
where \PR is a resonance with spin $J$ equal to 0, 1 or 2. Resonances with spin greater than 2 should not
contribute significantly to the \decay{\Dz}{\KSKpi} decays.
The corresponding 4-momenta are denoted $p_{\Dz}$, $p_{\PA}$, $p_{\PB}$ and $p_{\PC}$.
The reconstructed invariant mass of the resonance is denoted \mAB,
and the nominal mass \mR.
The matrix element for the \decay{\Dz}{\KSKpi} decay is given by
\begin{equation}
	\mcM_{\KSKpi}(\sKSpi, \sKpi) \\
	= \sum_{\PR} a_{\PR} e^{i\phi_{\PR}} \mcM_{\PR}(\mABsq, \mACsq),
\end{equation}
where $a_{\PR}e^{i\phi_{\PR}}$ is the complex amplitude for \PR and the contributions
$\mcM_{\PR}$ from each intermediate state are given by
\begin{equation}
	\mcM_{\PR}(\mABsq, \mACsq) = B_J^{\Dz}(p, |p_0|, d_{\Dz})\Omega_J(\mABsq, \mACsq)T_{\PR}(\mABsq)B_J^{\PR}(q, q_0, d_{\PR}),
\end{equation}
where $B_J^{\Dz}(p, |p_0|, d_{\Dz})$ and $B_J^{\PR}(q, q_0, d_{\PR})$ are
the Blatt-Weisskopf centrifugal barrier factors for the production
and decay, respectively, of the resonance \PR~\cite{BlattWeisskopf}.
The parameter $p$ ($q$) is the momentum of \PC (\PA or \PB) in the \PR rest frame,
and $p_0$ ($q_0$) is the same quantity calculated using the nominal resonance mass, \mR.
The meson radius parameters are set to $d_{\Dz} = 5.0 \invgevc$
and $d_{\PR} = 1.5 \invgevc$ consistent with the literature~\cite{Insler:2012pm,Kopp:2000gv};
the systematic uncertainty due to these choices is discussed in Sect.~\ref{sec:systematics}.
Finally, $\Omega_J(\mABsq, \mACsq)$ is the spin factor for a resonance with spin $J$
and $T_{\PR}$ is the dynamical function describing the resonance \PR.
The functional forms for $B_J(q, q_0, d)$ are given in Table~\ref{tab:barrierfactors}
and those for $\Omega_J(\mABsq, \mACsq)$ in Table~\ref{tab:angularfactors} for $J = 0, 1, 2$.
As the form for $\Omega_1$ is antisymmetric in the indices \PA and \PB,
it is necessary to define the particle ordering convention used in the analysis;
this is done in Table~\ref{tab:particleordering}.
\begin{table}
	\caption{Blatt-Weisskopf centrifugal barrier penetration factors, $B_J(q, q_0, d)$~\cite{BlattWeisskopf}.}
	\begin{center}
		\begin{tabular}{c c}
			$J$ & $B_J(q, q_0, d)$ \\
			\midrule
			0 & $1$\\[6pt]
			1 & $\sqrt{\frac{1 + (q_0d)^2}{1 + (qd)^2}}$\\[6pt]
			2 & $\sqrt{\frac{9 + 3(q_0d)^2 + (q_0d)^4}{9 + 3(qd)^2 + (qd)^4}}$\\
		\end{tabular}
	\end{center}
	\label{tab:barrierfactors}
\end{table}
\begin{table}
	\caption{Angular distribution factors, $\Omega_J(p_{\Dz} + p_{\PC}, p_{\PB} - p_{\PA})$. These are expressed in terms of the tensors $T^{\mu\nu} = -g^{\mu\nu} + \frac{p^\mu_{\PA\PB}p^\nu_{\PA\PB}}{\mRsq}$ and $T^{\mu\nu\alpha\beta} = \frac{1}{2}(T^{\mu\alpha}T^{\nu\beta} + T^{\mu\beta}T^{\nu\alpha}) - \frac{1}{3}T^{\mu\nu}T^{\alpha\beta}$.}
	\begin{center}
		\begin{tabular}{c c}
			$J$ & $\Omega_J(p_{\Dz} + p_{\PC}, p_{\PB} - p_{\PA})$ \\
			\midrule
			0 & $1$\\
			1 & $(p_{\Dz}^{\mu} + p_{\PC}^{\mu})T_{\mu\nu}(p_{\PB}^{\nu} - p_{\PA}^{\nu})/(\mathrm{Ge\kern -0.1em V\!/}c)^2$\\
			2 & $(p_{\Dz}^{\mu} + p_{\PC}^{\mu})(p_{\Dz}^{\nu} + p_{\PC}^{\nu})T_{\mu\nu\alpha\beta}(p_{\PA}^{\alpha} - p_{\PB}^{\alpha})(p_{\PA}^{\beta} - p_{\PB}^{\beta})/(\mathrm{Ge\kern -0.1em V\!/}c)^4$\\
		\end{tabular}
	\end{center}
	\label{tab:angularfactors}
\end{table}
\begin{table}
	\caption{Particle ordering conventions used in this analysis.}
	\begin{center}
		\begin{tabular}{l l c c c}
			\multicolumn{2}{c}{Decay} & \PA & \PB & \PC \\
			\midrule
			\decay{\Dz}{\KS\KstarzorKstarzbar}, & \decay{\KstarzorKstarzbar}{\Kpm\pimp} & \pion & \kaon & \KS \\
			\decay{\Dz}{\Kmp\Kstarpm}, & \decay{\Kstarpm}{\KS\pipm} & \KS & \pion & \kaon \\ 
			\decay{\Dz}{\pimp(\rhomeson^{\pm},\Pa^{\pm})}, & \decay{\rhomeson^{\pm},\Pa^{\pm}}{\KS\Kpm} & \kaon & \KS & \pion \\
		\end{tabular}
	\end{center}
	\label{tab:particleordering}
\end{table}
The dynamical function $T_{\PR}$ chosen depends on the resonance \PR in question.
A relativistic Breit-Wigner form is used unless otherwise noted
\begin{equation}
T_{\PR}(\mAB) = \frac{1}{(\mRsq - \mABsq) - i\mR\Gamma_{\PR}(\mAB)},
\end{equation}
where the mass-dependent width is
\begin{equation}
\Gamma_{\PR}(\mAB) = \Gamma_{\PR}\left[B_J^{\PR}(q,q_0,d_{\PR})\right]^2\frac{\mR}{\mAB}\left(\frac{q}{q_0}\right)^{2J+1}.
\label{eq:bwrunningwidth}
\end{equation}
Several alternative forms are used for specialized cases.
The \flatte~\cite{Flatte:1976xu} form is a coupled-channel function used to describe the \ameson{0}{980}{\pm} resonance~\cite{Abele:1998qd,Aubert:2005sm,Aubert:2008bd,delAmoSanchez:2010xz,Insler:2012pm},
\begin{equation}
T_{\PR} = \frac{1}{(m_{\ameson{0}{980}{\pm}}^2 - m_{\kaon\Kbar}^2) - i[\rho_{\kaon\Kbar}g_{\kaon\Kbar}^2 + \rho_{\Peta\pion}g_{\Peta\pion}^2]},
\end{equation}
where the phase space factor is given by
\begin{equation}
	\rho_{\PA\PB} = \frac{1}{m_{\kaon\Kbar}^2}\sqrt{(m_{\kaon\Kbar}^2 - (m_{\PA} + m_{\PB})^2)(m_{\kaon\Kbar}^2 - (m_{\PA} - m_{\PB})^2)},
\end{equation}
and the coupling constants $g_{\kaon\Kbar}$ and $g_{\Peta\pion}$ are taken from Ref.~\cite{Abele:1998qd},
fixed in the isobar model fits and tabulated in Appendix~\ref{sec:appformalismdetails}.
The \gousak~\cite{Gounaris:1968mw} parameterization is used to describe the \rmeson{1450}{\pm} and \rmeson{1700}{\pm} states~\cite{Aubert:2007bs,Kusaka:2007mj,Aubert:2007jn,Aubert:2008bd,BABAR:2011ae},
\begin{equation}
T_{\PR} = \frac{1 + d(m_{\rhomeson})\frac{\Gamma_{\rhomeson}}{m_{\rhomeson}}}{(m_{\rhomeson}^2 - m_{\kaon\Kbar}^2) + f(m_{\kaon\Kbar}^2, m_{\rhomeson}^2, \Gamma_{\rhomeson}) - im_{\rhomeson}\Gamma_{\rhomeson}(m_{\kaon\Kbar})},
\end{equation}
where
\begin{equation}
d(m_{\rhomeson}) = \frac{3m_{\kaon}^{2}}{\pi q_0^2}\log\left(\frac{m_{\rhomeson} + 2q_0}{2m_{\kaon}}\right) + \frac{m_{\rhomeson}}{2\pi q_0} - \frac{m_{\kaon}^2m_{\rhomeson}}{\pi q_0^3},
\end{equation}
and
\begin{equation}
f(m_{\kaon\Kbar}^2, m_{\rhomeson}^2, \Gamma_{\rhomeson}) = \Gamma_{\rhomeson}\frac{m_{\rhomeson}^2}{q_0^3}\left\{q_0^2\left[h(m_{\kaon\Kbar}^2) - h(m_{\rhomeson}^2)\right] + q_0^2h'(m_{\rhomeson}^{2})(m_{\rhomeson}^2 - m_{\kaon\Kbar}^2)\right\}.
\end{equation}
The parameter $m_{\kaon}$ is taken as the mean of $\m_{\KS}$ and $m_{\Kpm}$,
and $h'(m_{\rhomeson}^2) \equiv \frac{dh(m_{\rhomeson}^2)}{dm_{\rhomeson}^2}$ is calculated from
\begin{equation}
h(m^2) = \frac{2q(m^2)}{\pi m}\log\left(\frac{m + 2q(m^2)}{2m_{\kaon}}\right),
\end{equation}
in the limit that $m_{\kaon} = m_{\Kpm} = m_{\KS}$.
Parameters for the \rhomeson resonances \rmeson{1450}{\pm} and \rmeson{1700}{\pm} are taken from
Ref.~\cite{Bargiotti:2003ev} and tabulated in Appendix~\ref{sec:appformalismdetails}.

This analysis uses two different parameterizations for the \kpswave contributions, dubbed \glass and \lassc, with different motivations.
These forms include both \Kst{0}{1430}{} resonance and nonresonant \kpswave contributions.
The \lassc parameterization takes the form
\begin{equation}
	T_{\PR} = f\left(\frac{m_{\kpi}}{m_{\Kst{0}{1430}{}}}\right)\frac{m_{\kpi}}{q}\sin(\delta_S + \delta_F)e^{i(\delta_S + \delta_F)},
\end{equation}
where $f(x) = A\exp\left(b_{1}x + b_{2}x^{2} + b_{3}x^{3}\right)$ is an empirical real production form factor,
and the phases are defined by
\begin{equation}
	\tan\delta_F = \frac{2aq}{2 + arq^2},\hspace{1cm}\tan\delta_S = \frac{m_{\PR} \Gamma_{\PR}(m_{\kpi})}{m_{\Kst{0}{1430}{}}^2 - m_{\kpi}^{2}}.
\end{equation}
The scattering length $a$, effective range $r$ and \Kst{0}{1430}{} mass and width are taken
from measurements~\cite{Dunwoodie} at the \lass experiment~\cite{Aston:1987ir} and are tabulated in
Appendix~\ref{sec:appformalismdetails}.
With the choice $f(x)=1$ this form has been used in previous analyses \eg Refs.~\cite{Aubert:2005ce,Aubert:2007dc,Aubert:2008yd},
and if $\delta_F$ is additionally set to zero the relativistic $S$-wave Breit-Wigner form is recovered.
The Watson theorem~\cite{Watson:1952ji} states that the phase motion, as a function of \kpi invariant mass,
is the same in elastic scattering and decay processes, in the absence of final-state interactions (\ie in the isobar model).
Studies of \kpi scattering data indicate that the $S$-wave remains elastic up to the $\kaon\Peta'$ threshold~\cite{Dunwoodie}.
The magnitude behavior is not constrained by the Watson theorem, which motivates the inclusion of
the form factor $f(x)$, but the \lassc parameterization preserves the phase behavior measured in \kpi scattering.
The real form factor parameters are allowed to take different values for the neutral and charged \Kst{0}{1430}{} resonances, as the production processes
are not the same, but the parameters taken from \lass measurements, which specify the phase behavior, are shared between both \kpi channels.
A transformed set $\mathbf{b'}=\mathbf{U}^{-1}\mathbf{b}$ of the parameters $\mathbf{b}=(b_1,b_2,b_3)$ are also defined for use in the isobar model fit,
which is described in detail in Sect.~\ref{sec:models}.
The constant matrix $\mathbf{U}$ is chosen to minimize fit correlations, and the form factor is
normalized to unity at the center of the accessible kinematic range, \eg $\frac{1}{2}(m_{\KS} + m_{\pipm} + m_{\Dz} - m_{\Kpm})$
for the charged \kpswave.

The \glass (Generalized \lass) parameterization
has been used by several recent amplitude analyses, \eg Refs.~\cite{Aubert:2008bd,delAmoSanchez:2010xz,Insler:2012pm},
\begin{equation}
	T_{\PR} = \left[F\sin(\delta_F + \phi_F)e^{i(\delta_F + \phi_F)} + \sin(\delta_S)e^{i(\delta_S + \phi_S)}e^{2i(\delta_F + \phi_F)}\right]\frac{m_{\kpi}}{q},
\end{equation}
where $\delta_F$ and $\delta_S$ are defined as before, and $F$, $\phi_F$ and $\phi_S$ are free parameters in the fit.
It should be noted that this functional form can result in phase behavior significantly different to that measured in
\lass scattering data when its parameters are allowed to vary freely.
This is illustrated in Fig.~\ref{fig:lass_glass_comparison} in Sect.~\ref{sec:modelresults}.

\subsection{Coherence factor and {\boldmath\CP}-even fraction}
\label{sec:coherence}
The coherence factor $R_{\Pf}$ and mean strong-phase difference $\delta_{\Pf}$ for the
multi-body decays \decay{\PD}{\Pf} and \decay{\PD}{\overline{\Pf}} quantify the similarity
of the two decay structures~\cite{Atwood:2003mj}.
In the limit $R_{\Pf}\to 1$ the matrix elements for the two decays are identical. 
For \decay{\Dz}{\KSKpi} the coherence factor and mean strong-phase difference are
defined by~\cite{Insler:2012pm,LHCB-PAPER-2013-068}
\begin{equation}
R_{\KSKpis}e^{-i\delta_{\KSKpis}} \equiv \frac{\int \mathcal{M}_{\KSKppim}(\sKSpi, \sKpi)\mathcal{M}^{*}_{\KSKmpip}(\sKSpi, \sKpi)d\sKSpi d\sKpi}{M_{\KSKppim}M_{\KSKmpip}},
\label{eq:coherencefactor}
\end{equation}
where
\begin{equation}
M^{2}_{\KSKpi} \equiv \int |\mathcal{M}_{\KSKpi}(\sKSpi, \sKpi)|^{2}d\sKSpi d\sKpi,
\label{eq:coherencenorm}
\end{equation}
and the integrals are over the entire available phase space. The restricted phase space
coherence factor $R_{\Kstar\kaon}e^{-i\delta_{\Kstar\kaon}}$ is defined analogously
but with all integrals restricted to an area of phase space close to the
\Kst{}{892}{\pm} resonance.
The restricted area is defined by Ref.~\cite{Insler:2012pm} as the region where the
$\KS\pipm$ invariant mass is within $100\mevcc$ of the \Kst{}{892}{\pm} mass.
The four observables $R_{\KSKpis}$, $\delta_{\KSKpis}$, $R_{\Kstar\kaon}$ and
$\delta_{\Kstar\kaon}$ were measured 
using quantum-correlated \decay{\psiprpr}{\Dz\Dzb} decays by the \cleo collaboration~\cite{Insler:2012pm},
and the coherence was found to be large for both the full and the restricted regions. 
This analysis is not sensitive to the overall phase difference between \decay{\Dz}{\KSKppim}
and \decay{\Dzb}{\KSKppim}. However, since it cancels in $\delta_{\KSKpis} - \delta_{\Kstar\kaon}$, this
combination, as well as $R_{\KSKpis}$ and $R_{\Kstar\kaon}$, can be calculated from isobar models and compared
to the respective \cleo results.

An associated parameter that it is interesting to consider is the \CP-even fraction~\cite{Gershon:2015xra},
\begin{equation}
\begin{split}
  F_+ & \equiv \frac{|\braket{\PD_+}{\KSKpi}|^2}{|\braket{\PD_+}{\KSKpi}|^2 + |\braket{\PD_-}{\KSKpi}|^2} \\
  & = \frac{1}{2}\left[1 + 2R_{\KSKpis}\cos(\delta_{\KSKpis})\sqrt{\BR_{\KSKpis}}(1 + \BR_{\KSKpis})^{-1}\right],
  \label{eq:cpeven}
\end{split}
\end{equation}
where the \CP eigenstates
$\ket{\PD_\pm}$ are given by $\frac{1}{\sqrt{2}}\left[\ket{\Dz} \pm \ket{\Dzb}\right]$
and $\BR_{\KSKpis}$ is the ratio of branching fractions of the two \decay{\Dz}{\KSKpi}
modes.
As stated above, the relative strong phase $\delta_{\KSKpis}$ is not predicted
by the amplitude models and requires external input.
\subsection{Efficiency modeling}
\label{sec:efficiency}
\begin{figure}
	\begin{center}
		\includegraphics[width=0.5\textwidth]{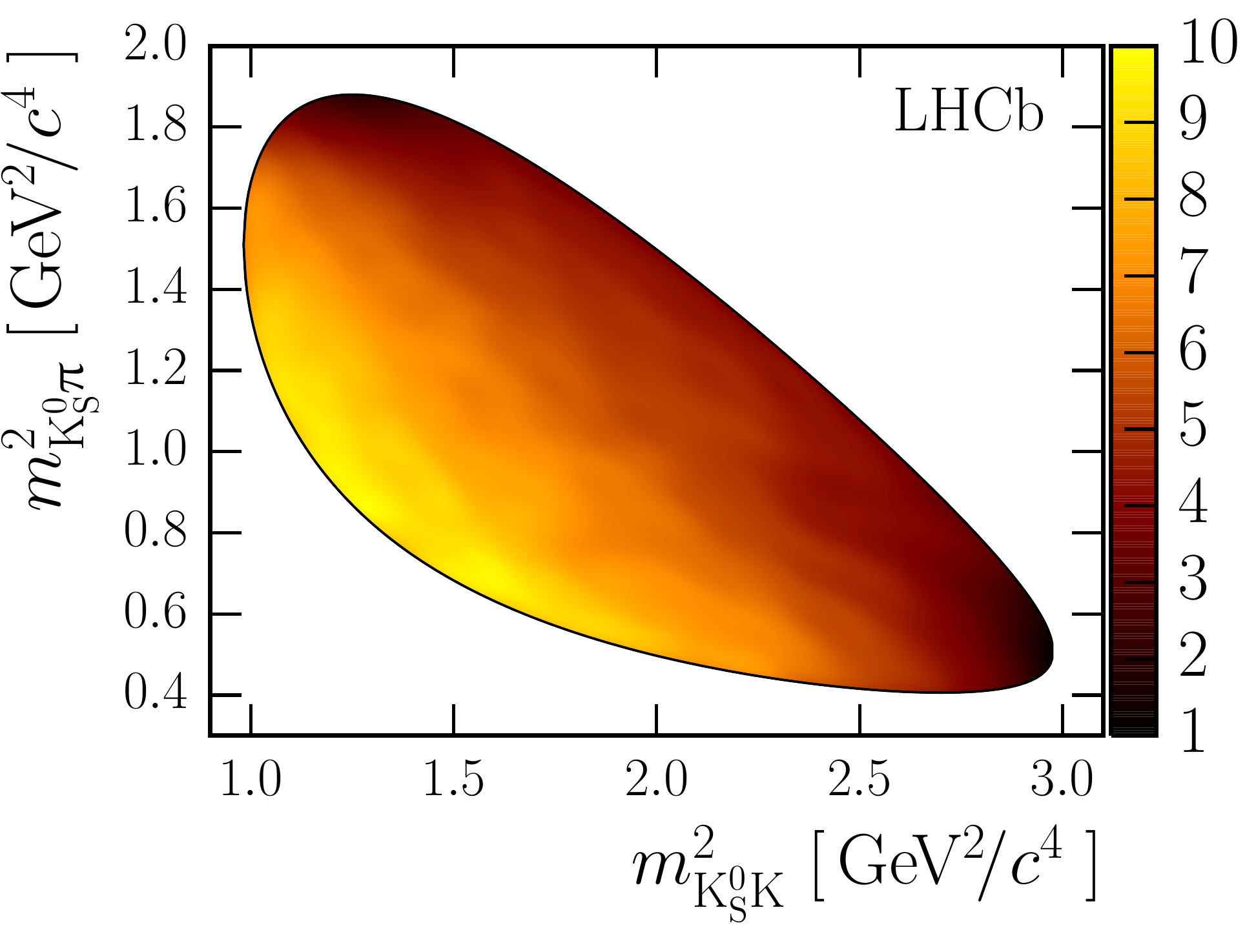}
	\end{center}
	\caption{Efficiency function used in the isobar model fits, corresponding
                 to the average efficiency over the full dataset. The coordinates \sKSpi and
  		 \sKSK are used to highlight the approximate symmetry of the efficiency function.
                 The $z$ units are arbitrary.}
	\label{fig:efficiency}
\end{figure}

The trigger strategy described in Sect.~\ref{sec:Detector}, and to a lesser
extent the offline selection, includes requirements on variables
such as the impact parameter
and \pt of the various charged particles correlated with the 2-body
invariant masses \sKSpi and \sKpi.
There is, therefore, a significant variation in reconstruction efficiency
as a function of \sKSpi and \sKpi.
This efficiency variation is modeled using simulated events
generated with a uniform distribution in these variables and propagated
through the full \lhcb detector simulation, trigger emulation and offline selection.
Weights are applied to the simulated events to ensure that various 
subsamples are present in the correct proportions.
These weights correct for known discrepancies between the simulation and real data
in the relative reconstruction efficiency for long and downstream
tracks, and take into account
the ratios of $\sqs=7\tev$ to $\sqs=8\tev$ and
\decay{\Dz}{\KSKmpip} to \decay{\Dz}{\KSKppim} simulated events to improve the description of the data.
The efficiencies of offline selection requirements based on information from the RICH
detectors are calculated using a data-driven method
based on calibration samples~\cite{LHCb-DP-2012-003} of \decay{\Dsttt}{\Dz\pislow} decays,
where \decay{\Dz}{\Km\pip}.
These efficiencies are included as additional weights.
A non-parametric kernel estimator~\cite{Cranmer:2000du} is used to produce a smooth
function \efffn describing the efficiency variation
in the isobar model fits.
The average model corresponding to the full dataset recorded in 2011 and 2012,
which is used unless otherwise noted, is shown in Fig.~\ref{fig:efficiency}.
Candidates very near to the boundary of the allowed
kinematic region of the Dalitz plot are excluded, as the kinematics in this region lead to
variations in efficiency that are difficult to model.
It is required that $\max(|\cos(\theta_{\KS\pion})|, |\cos(\theta_{\pion\kaon})|, |\cos(\theta_{\kaon\KS})|)<0.98$,
where $\theta_{\PA\PB}$ is the angle between the \PA and \PB momenta
in the $\PA\PC$ rest frame.
This criterion removes $5\%$ of the candidates.
The simulated events are also used to verify that the resolution in \sKSpi and \sKpi
is around $0.004\gevgevcccc$, corresponding to $\order(2\mevcc)$ resolution in $m(\KS\pipm)$.
Although this is not explicitly accounted for in the isobar model fits,
it has a small effect which is measurable only on the parameters of the \Kst{}{892}{\pm}
resonance and is accounted for in the systematic uncertainties.
\subsection{Fit components}
There are three event categories described in Sect.~\ref{sec:selection} that
must be treated separately in the isobar model fits.
The signal and mistagged components are described by terms proportional to $\efffn\Msq{\KSKpi}$, while the
combinatorial component is described by a smooth function, \combfn{\KSKpi}, obtained by
applying to data in the \mKSKpis sidebands the same non-parametric kernel estimator used to
model the efficiency variation.
The same combinatorial background model is used for both \Dz
flavors, and the same efficiency function is used for both modes and \Dz flavors.
The overall function used in the fit to \decay{\Dz}{\KSKpi} decays is therefore
\begin{align}
  P_{\KSKpi}(\sKSpi, \sKpi) & = (1 - f_{\text{m}} - f_{\text{c}})\efffn\Msq{\KSKpi} \nonumber\\
                              & + f_{\text{m}}\efffn\Mbsq{\KSKpi} \\\nonumber
                              & + f_{\text{c}}\combfn{\KSKpi}, 
  \label{eq:fullpdf}
\end{align}
where the mistagged contribution consists of
\decay{\Dzb}{\KSKpi} decays and $f_{\text{m}}$ ($f_{\text{c}}$)
denotes the mistagged (combinatorial) fraction tabulated in Table~\ref{tab:massfit}.

All parameters except the complex amplitudes $\aR e^{i\pR}$ are shared between the PDFs
for both modes and both \Dz flavors.
For the other parameters, Gaussian constraints are included unless stated otherwise.
The nominal values used in the constraints are tabulated
in Appendix~\ref{sec:appformalismdetails}.
No constraints are applied for the \kpswave parameters $b_{1..3}$, $F$, $\phi_S$
and $\phi_F$, as these have no suitable nominal values.
The \kpswave parameters $a$ and $r$ are treated differently in the \glass and \lassc models.
In the \lassc case these parameters are shared between the neutral and charged \kpi
channels and a Gaussian constraint to the \lass	 measurements~\cite{Dunwoodie} is included.
In the \glass case these are allowed to vary freely and take different values for the two channels.

\section{Isobar model fits}
\label{sec:models}
This section summarizes the procedure by which the amplitude models
are constructed, describes the various systematic uncertainties considered
for the models and finally discusses the models and the coherence
information that can be calculated from them.

Amplitude models are fitted using the isobar formalism and an unbinned
maximum-likelihood method, using the \goofit~\cite{Andreassen:2014xfa}
package to exploit massively-parallel Graphics Processing Unit (GPU) architectures.
Where \chisqbin values are quoted these are simply to indicate the fit quality.
Statistical uncertainties on derived quantities, such as the resonance fit fractions, are
calculated using a pseudoexperiment method based on the fit covariance matrix.
\subsection{Model composition}
Initially, 15 resonances are considered for inclusion in each of the isobar models:
\Kst{}{892,1410,1680}{0,\pm}, \Kst{0,2}{1430}{0,\pm},
\ameson{0}{980,1450}{\pm}, \ameson{2}{1320}{\pm} and
\rmeson{1450,1700}{\pm}.
Preliminary studies showed that models containing the \Kst{}{1680}{}
resonances tend to include large interference terms, which are cancelled
by other large components.
Such fine-tuned interference effects are in general unphysical, and
are therefore disfavored in the model building~\cite{Aitala:2002kr,Aubert:2005sm}.
The \Kst{}{1680}{} resonances are not considered
further, and additionally the absolute value of the sum of
interference fractions~\cite{PDGdalitzformalism} is required to be less than $30\%$ in all models.
In the absence of the \Kst{}{1680}{} resonances, large interference terms
are typically generated by the \kpswave contributions.
The requirement on the sum of interference fractions, while arbitrary, allows an iterative
procedure to be used to search for the best amplitude models.
This procedure explores a large number of possible starting configurations
and sets of resonances; it begins with the most general models containing
all 13 resonances and considers progressively simpler configurations,
trying a large number of initial fit configurations for each set of resonances,
until no further improvement in fit quality is found among models simple
enough to satisfy the interference fraction limit.
Higher values of this limit lead to a large number of candidate models
with similar fit quality.

A second procedure iteratively removes resonances from the models if
they do not significantly improve the fit quality.
In this step a resonance must improve the value of \ntll, where $\mathcal{L}$
is the likelihood of the full dataset, by
at least 16 units in order to be retained.
Up to this point, the \Kst{}{892}{} mass and width parameters
and \kpswave parameters have been allowed to vary in the fit,
but mass and width parameters for other resonances have been fixed.
To improve the quality of fit further, in a third step,
$S$ and $P$-wave resonance parameters are allowed to vary.
The tensor resonance parameters are known
precisely~\cite{PDG2014}, so remain fixed.
At this stage, resonances that no longer significantly improve the fit
quality are removed, with the threshold tightened so that
each resonance must increase \ntll by 25 units in order to be retained.

Finally, parameters that are consistent with their nominal values to
within $1\sigma$ are fixed to the nominal value. The nominal values used
are tabulated in Appendix~\ref{sec:appformalismdetails}.
The entire procedure is performed in parallel using the \glass and \lassc
parameterizations of the \kpswave.
The data are found to prefer a solution where the \glass parameterization
of the charged \kpswave has a poorly constrained degree of freedom.
The final change to the \glass models is, therefore, to fix the charged \kpswave
$F$ parameter in order to stabilize the uncertainty calculation for the two corresponding
\aR parameters by reducing the correlations among the free parameters.
\subsection{Systematic uncertainties}
\label{sec:systematics}
Several sources of systematic uncertainty are considered.
Those due to experimental issues are described first,
followed by uncertainties related to the amplitude model formalism.
Unless otherwise stated, the uncertainty assigned to each parameter
using an alternative fit is the absolute difference in its value
between the nominal and alternative fit.

As mentioned in Sect.~\ref{sec:efficiency}, candidates
extremely close to the edges of the allowed kinematic region of
the Dalitz plot are excluded. The requirement made
is that the largest of the three $|\cos(\theta_{\PA\PB})|$
values is less than 0.98.
A systematic uncertainty due to this process is estimated by changing the
threshold to $0.96$, as this excludes a similar additional area
of the Dalitz plot as the original requirement.

The systematic uncertainty related to the efficiency model \efffn
is evaluated in four ways. The first probes the process by which
a smooth curve is produced from simulated events; this uncertainty
is evaluated using an alternative fit that substitutes the
non-parametric estimator with a polynomial parameterization.
The second uncertainty is due to the limited sample size
of simulated events. This is evaluated by generating several
alternative polynomial efficiency models according to the covariance
matrix of the polynomial model parameters; the spread in parameter
values from this ensemble is assigned as the uncertainty due to the
limited sample size.
The third contribution is due to possible imperfections in the
description of the data by the simulation.
This uncertainty is assigned using an alternative simultaneous fit
that separates the sample into three categories according to
the year in which the data were collected and the type of \KS
candidate used. As noted in Sect.~\ref{sec:Detector}, the sample
recorded during 2011 does not include downstream \KS candidates.
These sub-samples have different kinematic distributions
and \efffn behavior, so this procedure tests the
ability of the simulation process to reproduce the variation seen in the data.
The final contribution is due to the re-weighting procedures used
to include the effect of offline selection requirements based on
information from the RICH detectors, and to
correct for discrepancies between data and simulation in the
reconstruction efficiencies of long and downstream \KS candidates.
This is evaluated using alternative efficiency models where the
relative proportion of the track types is altered, and the weights
describing the efficiency of selection requirements using information
from the RICH detectors are modified to account for the limited
calibration sample size.
Additional robustness checks have been performed to probe the
description of the efficiency function by the simulated events.
In these checks the data are divided into two equally populated
bins of the \Dz meson \ptot, \pt or $\eta$ and
the amplitude models are re-fitted using each bin separately.
The fit results in each pair of bins are found to be compatible
within the assigned uncertainties, indicating that
the simulated \Dz kinematics adequately match the data.

An uncertainty is assigned due to the description of the hardware
trigger efficiency in simulated events. Because the hardware trigger is not
only required to fire on the signal decay, it is important that the underlying
$\proton\proton$ interaction is well described, and a systematic
uncertainty is assigned due to possible imperfections.
This uncertainty is obtained using an alternative efficiency
model generated from simulated events that have been weighted
to adjust the fraction where the hardware trigger was fired by the signal
candidate.

The uncertainty due to the description of the combinatorial background
is evaluated by recomputing the $c_{\KSKpi}(\sKSpi, \sKpi)$ function using
\mDz sideband events to which an alternative kinematic fit has been applied,
without a constraint on the \Dz mass.
The alternative model is expected to describe the edges of the phase
space less accurately, while providing an improved description
of peaking features.

An alternative set of models is produced using a threshold of 9 units
in the value of \ntll instead of the thresholds of 16 and 25 used for
the model building procedure.
These models contain more resonances, as fewer
are removed during the model building process. A systematic uncertainty
is assigned using these alternative models for those parameters which
are common between the two sets of models.

Two parameters of the \flatte dynamical function, which is used to
describe the \ameson{0}{980}{\pm} resonance, are fixed to nominal
values in the isobar model fits. Alternative fits are performed, where
these parameters are fixed to different values according to their quoted
uncertainties, and the largest changes to the fit parameters are
assigned as systematic uncertainties.

The effect of resolution in the \sKSpi and \sKpi coordinates is neglected
in the isobar model fits, and this is expected to have an effect
on the measured \Kst{}{892}{\pm} decay width. An uncertainty is calculated using
a pseudoexperiment method, and is found to be small.

The uncertainty due to the yield determination process
described in Sect.~\ref{sec:selection} is measured by changing
the fractions $f_{\text{m}}$ and $f_{\text{c}}$ in the isobar model
fit according to their statistical uncertainties, and taking the
largest changes with respect to the nominal result as the systematic
uncertainty.

There are two sources of systematic uncertainty due to the amplitude
model formalism considered.
The first is that due to varying the meson radius
parameters $d_{\Dz}$ and $d_{\PR}$, defined in
Sect.~\ref{sec:isobarformalism}.
These are changed from $d_{\Dz} = 5.0\invgevc$ and
$d_{\PR} = 1.5\invgevc$ to $2.5\invgevc$ and $1.0\invgevc$, respectively.
The second is due to the
dynamical function $T_{\PR}$ used to describe the \rmeson{1450,1700}{\pm}
resonances. These resonances are described by the \gousak
functional form in the nominal models, which is replaced with a relativistic
$P$-wave Breit-Wigner function to calculate a systematic uncertainty due to
this choice.

The uncertainties described above are added in quadrature to produce the
total systematic uncertainty quoted for the various results.
For most quantities the dominant systematic uncertainty is due to the
meson radius parameters $d_{\Dz}$ and $d_{\PR}$. The largest sources of 
experimental uncertainty relate to the description of the efficiency
variation across the Dalitz plot.
The fit procedure and statistical uncertainty calculation have been
validated using pseudoexperiments and no bias was found.

Tables summarizing the various sources of systematic uncertainty and their
relative contributions are included in Appendix~\ref{sec:app_systematics}.

\subsection{Isobar model results}
\label{sec:modelresults}
\begin{figure}
  \begin{center}
    \resizebox{\textwidth}{!}{
    \begin{tabular}{l r}
      \includegraphics[width=0.5\textwidth]{{{figs/glass_fav_x_proj}}} & \includegraphics[width=0.5\textwidth]{{{figs/glass_fav_y_proj}}} \\
      \includegraphics[trim=0 0 0 8.6mm,valign=b,width=0.5\textwidth]{{{figs/glass_fav_z_proj}}} & \includegraphics[valign=b,height=0.4\textwidth,margin*=0 1.4cm 3.5mm 0]{{{figs/glass_fav_legend}}} \\
    \end{tabular}}
  \end{center}
  \vspace{-8mm}
  \caption{Distributions of \sKpi (upper left), \sKSpi (upper right) and \sKSK (lower left) in the \decay{\Dz}{\KSKmpip} mode
           with fit curves from the best \glass model. The solid (blue) curve shows the full PDF $P_{\KSKmpip}(\sKSpi,\sKpi)$,
           while the other curves show the components with the largest integrated fractions.}
  \label{fig:favglass}
\end{figure}
\begin{figure}
  \begin{center}
    \resizebox{\textwidth}{!}{
    \begin{tabular}{l r}
      \includegraphics[width=0.5\textwidth]{{{figs/lass_fav_x_proj}}} & \includegraphics[width=0.5\textwidth]{{{figs/lass_fav_y_proj}}} \\
      \includegraphics[trim=0 0 0 8.6mm,valign=b,width=0.5\textwidth]{{{figs/lass_fav_z_proj}}} & \includegraphics[valign=b,height=0.4\textwidth,margin*=0 1.4cm 3.5mm 0]{{{figs/lass_fav_legend}}} \\
    \end{tabular}}
  \end{center}
  \vspace{-8mm}
  \caption{Distributions of \sKpi (upper left), \sKSpi (upper right) and \sKSK (lower left) in the \decay{\Dz}{\KSKmpip} mode
           with fit curves from the best \lassc model. The solid (blue) curve shows the full PDF $P_{\KSKmpip}(\sKSpi,\sKpi)$,
           while the other curves show the components with the largest integrated fractions.}
  \label{fig:favlass}
\end{figure}
\begin{figure}
  \begin{center}
    \resizebox{\textwidth}{!}{
    \begin{tabular}{l r}
      \includegraphics[width=0.5\textwidth]{{{figs/glass_sup_x_proj}}} & \includegraphics[width=0.5\textwidth]{{{figs/glass_sup_y_proj}}} \\
      \includegraphics[trim=0 0 0 6.5mm,valign=b,width=0.5\textwidth]{{{figs/glass_sup_z_proj}}} & \includegraphics[valign=b,height=0.4\textwidth,margin*=0 1.4cm 3.5mm 0]{{{figs/glass_sup_legend}}} \\
    \end{tabular}}
  \end{center}
  \vspace{-8mm}
  \caption{Distributions of \sKpi (upper left), \sKSpi (upper right) and \sKSK (lower left) in the \decay{\Dz}{\KSKppim} mode
           with fit curves from the best \glass model. The solid (blue) curve shows the full PDF $P_{\KSKppim}(\sKSpi,\sKpi)$,
           while the other curves show the components with the largest integrated fractions.}
  \label{fig:supglass}
\end{figure}
\begin{figure}
  \begin{center}
    \resizebox{\textwidth}{!}{      
    \begin{tabular}{l r}
      \includegraphics[width=0.5\textwidth]{{{figs/lass_sup_x_proj}}} & \includegraphics[width=0.5\textwidth]{{{figs/lass_sup_y_proj}}} \\
      \includegraphics[trim=0 0 0 6.5mm,valign=b,width=0.5\textwidth]{{{figs/lass_sup_z_proj}}} & \includegraphics[valign=b,height=0.4\textwidth,margin*=0 1.4cm 3.5mm 0]{{{figs/lass_sup_legend}}} \\
    \end{tabular}}
  \end{center}
  \vspace{-8mm}
  \caption{Distributions of \sKpi (upper left), \sKSpi (upper right) and \sKSK (lower left) in the \decay{\Dz}{\KSKppim} mode
           with fit curves from the best \lassc model. The solid (blue) curve shows the full PDF $P_{\KSKppim}(\sKSpi,\sKpi)$,
           while the other curves show the components with the largest integrated fractions.}
  \label{fig:suplass}
\end{figure}
\begin{figure}
  \begin{center}
    \begin{subfigure}{0.45\textwidth}
      \includegraphics[width=\textwidth]{{{figs/glass_fav_amp2}}}
    \end{subfigure}%
    \hspace{5mm}
    \begin{subfigure}{0.45\textwidth}
      \includegraphics[width=\textwidth]{{{figs/glass_sup_amp2}}}
    \end{subfigure}
    \begin{subfigure}{0.45\textwidth}
      \includegraphics[width=\textwidth]{{{figs/glass_fav_phase}}}
    \end{subfigure}%
    \hspace{5mm}
    \begin{subfigure}{0.45\textwidth}
      \includegraphics[width=\textwidth]{{{figs/glass_sup_phase}}}
    \end{subfigure}
    \begin{subfigure}{0.45\textwidth}
      \includegraphics[width=\textwidth]{{{figs/fav_lassminusglass_phase}}}
    \end{subfigure}%
    \hspace{5mm}
    \begin{subfigure}{0.45\textwidth}
      \includegraphics[width=\textwidth]{{{figs/sup_lassminusglass_phase}}}
    \end{subfigure}
  \end{center}
  \vspace{-8mm}
  \caption{Decay rate and phase variation across the Dalitz plot. The top row shows
           $|\mcM_{\KSKpi}(\sKSpi, \sKpi)|^2$ in the best \glass isobar models,
           the center row shows the phase behavior of the same models and 
           the bottom row shows the same function subtracted from the phase behavior in the best
           \lassc isobar models. The left column shows the \decay{\Dz}{\KSKmpip} mode with
           \decay{\Dz}{\KSKppim} on the right.
           The small inhomogeneities that are visible in the bottom row
           relate to the \glass and \lassc models preferring
           slightly different values of the \Kst{}{892}{\pm} mass and width.}
  \label{fig:2dplots}
\end{figure}
\begin{sidewaystable}
  \caption{Isobar model fit results for the \decay{\Dz}{\KSKmpip} mode. The first uncertainties are statistical and the second systematic.}
  \begin{center}
    \resizebox{\textheight}{!}{
    \begin{tabular}{l r@{$\,\pm\,$}l r@{$\,\pm\,$}l r@{$\,\pm\,$}l r@{$\,\pm\,$}l r@{$\,\pm\,$}l r@{$\,\pm\,$}l}
      & \multicolumn{4}{c}{\aR} & \multicolumn{4}{c}{\pR $(^{\circ})$} & \multicolumn{4}{c}{Fit fraction [\%]} \\
      Resonance & \multicolumn{2}{c}{\glass} & \multicolumn{2}{c}{\lassc} & \multicolumn{2}{c}{\glass} & \multicolumn{2}{c}{\lassc} & \multicolumn{2}{c}{\glass} & \multicolumn{2}{c}{\lassc} \\
      \midrule
      \Kst{}{892}{+} & \multicolumn{2}{c}{1.0 (fixed)} & \multicolumn{2}{c}{1.0 (fixed)} & \multicolumn{2}{c}{0.0 (fixed)} & \multicolumn{2}{c}{0.0 (fixed)} & $57.0$ & $0.8\pm2.6$ & $56.9$ & $0.6\pm1.1$ \\
      \Kst{}{1410}{+} & $4.3$ & $0.3\pm0.7$ & $5.83$ & $0.29\pm0.29$ & $-160$ & $6\pm24$ & $-143$ & $3\pm6$ & $5$ & $1\pm4$ & $9.6$ & $1.1\pm2.9$ \\
      \posswave & $0.62$ & $0.05\pm0.18$ & $1.13$ & $0.09\pm0.21$ & $-67$ & $5\pm15$ & $-59$ & $4\pm13$ & $12$ & $2\pm9$ & $11.7$ & $1.0\pm2.3$ \\
      \Kbarst{}{892}{0} & $0.213$ & $0.007\pm0.018$ & $0.210$ & $0.006\pm0.010$ & $-108$ & $2\pm4$ & $-101.5$ & $2.0\pm2.8$ & $2.5$ & $0.2\pm0.4$ & $2.47$ & $0.15\pm0.23$ \\
      \Kbarst{}{1410}{0} & $6.0$ & $0.3\pm0.5$ & $3.9$ & $0.2\pm0.4$ & $-179$ & $4\pm17$ & $-174$ & $4\pm7$ & $9$ & $1\pm4$ & $3.8$ & $0.5\pm2.0$ \\
      \Kbarst{2}{1430}{0} & $3.2$ & $0.3\pm1.0$ & \multicolumn{2}{c}{---} & $-172$ & $5\pm23$ & \multicolumn{2}{c}{---} & $3.4$ & $0.6\pm2.7$ & \multicolumn{2}{c}{---} \\
      \neutswave & $2.5$ & $0.2\pm1.3$ & $1.28$ & $0.12\pm0.23$ & $50$ & $10\pm80$ & $75$ & $3\pm8$ & $11$ & $2\pm10$ & $18$ & $2\pm4$ \\
      \ameson{0}{980}{-} & \multicolumn{2}{c}{---} & $1.07$ & $0.09\pm0.14$ & \multicolumn{2}{c}{---} & $82$ & $5\pm7$ & \multicolumn{2}{c}{---} & $4.0$ & $0.7\pm1.1$ \\
      \ameson{2}{1320}{-} & $0.19$ & $0.03\pm0.09$ & $0.17$ & $0.03\pm0.05$ & $-129$ & $8\pm17$ & $-128$ & $10\pm8$ & $0.20$ & $0.06\pm0.21$ & $0.15$ & $0.06\pm0.13$ \\
      \ameson{0}{1450}{-} & $0.52$ & $0.04\pm0.15$ & $0.43$ & $0.05\pm0.10$ & $-82$ & $7\pm31$ & $-49$ & $11\pm19$ & $1.2$ & $0.2\pm0.6$ & $0.74$ & $0.15\pm0.34$ \\
      \rmeson{1450}{-} & $1.6$ & $0.2\pm0.5$ & $1.3$ & $0.1\pm0.4$ & $-177$ & $7\pm32$ & $-144$ & $7\pm9$ & $1.3$ & $0.3\pm0.7$ & $1.4$ & $0.2\pm0.7$ \\
      \rmeson{1700}{-} & $0.38$ & $0.08\pm0.15$ & \multicolumn{2}{c}{---} & $-70$ & $10\pm60$ & \multicolumn{2}{c}{---} & $0.12$ & $0.05\pm0.14$ & \multicolumn{2}{c}{---} \\
    \end{tabular}}
  \end{center}
  \label{tab:favresults}
\end{sidewaystable}
\begin{sidewaystable}
  \caption{Isobar model fit results for the \decay{\Dz}{\KSKppim} mode. The first uncertainties are statistical and the second systematic.}
  \begin{center}
    \resizebox{\textheight}{!}{
    \begin{tabular}{l r@{$\,\pm\,$}l r@{$\,\pm\,$}l r@{$\,\pm\,$}l r@{$\,\pm\,$}l r@{$\,\pm\,$}l r@{$\,\pm\,$}l}
      & \multicolumn{4}{c}{\aR} & \multicolumn{4}{c}{\pR $(^{\circ})$} & \multicolumn{4}{c}{Fit fraction [\%]} \\
      Resonance & \multicolumn{2}{c}{\glass} & \multicolumn{2}{c}{\lassc} & \multicolumn{2}{c}{\glass} & \multicolumn{2}{c}{\lassc} & \multicolumn{2}{c}{\glass} & \multicolumn{2}{c}{\lassc} \\
      \midrule
      \Kst{}{892}{-} & \multicolumn{2}{c}{1.0 (fixed)} & \multicolumn{2}{c}{1.0 (fixed)} & \multicolumn{2}{c}{0.0 (fixed)} & \multicolumn{2}{c}{0.0 (fixed)} & $29.5$ & $0.6\pm1.6$ & $28.8$ & $0.4\pm1.3$ \\
      \Kst{}{1410}{-} & $4.7$ & $0.5\pm1.1$ & $9.1$ & $0.6\pm1.5$ & $-106$ & $6\pm25$ & $-79$ & $3\pm7$ & $3.1$ & $0.6\pm1.6$ & $11.9$ & $1.5\pm2.2$ \\
      \negswave & $0.58$ & $0.05\pm0.11$ & $1.16$ & $0.11\pm0.32$ & $-164$ & $6\pm31$ & $-101$ & $6\pm21$ & $5.4$ & $0.9\pm1.7$ & $6.3$ & $0.9\pm2.1$ \\
      \Kst{}{892}{0} & $0.410$ & $0.010\pm0.021$ & $0.427$ & $0.010\pm0.013$ & $176$ & $2\pm9$ & $-175.0$ & $1.7\pm1.4$ & $4.82$ & $0.23\pm0.35$ & $5.17$ & $0.21\pm0.32$ \\
      \Kst{}{1410}{0} & $6.2$ & $0.5\pm1.4$ & $4.2$ & $0.5\pm0.9$ & $175$ & $4\pm14$ & $165$ & $5\pm10$ & $5.2$ & $0.7\pm1.6$ & $2.2$ & $0.6\pm2.1$ \\
      \Kst{2}{1430}{0} & $6.3$ & $0.5\pm1.7$ & \multicolumn{2}{c}{---} & $-139$ & $5\pm21$ & \multicolumn{2}{c}{---} & $7$ & $1\pm4$ & \multicolumn{2}{c}{---} \\
      \neutswave & $3.7$ & $0.3\pm1.8$ & $1.7$ & $0.2\pm0.4$ & $100$ & $10\pm70$ & $144$ & $3\pm6$ & $12$ & $1\pm8$ & $17$ & $2\pm6$ \\
      \ameson{0}{980}{+} & $1.8$ & $0.1\pm0.6$ & $3.8$ & $0.2\pm0.7$ & $64$ & $5\pm24$ & $126$ & $3\pm6$ & $11$ & $1\pm6$ & $26$ & $2\pm10$ \\
      \ameson{0}{1450}{+} & $0.44$ & $0.05\pm0.13$ & $0.86$ & $0.10\pm0.12$ & $-140$ & $9\pm35$ & $-110$ & $8\pm7$ & $0.45$ & $0.09\pm0.34$ & $1.5$ & $0.3\pm0.4$ \\
      \rmeson{1450}{+} & $2.3$ & $0.4\pm0.8$ & \multicolumn{2}{c}{---} & $-60$ & $6\pm18$ & \multicolumn{2}{c}{---} & $1.5$ & $0.5\pm0.9$ & \multicolumn{2}{c}{---} \\
      \rmeson{1700}{+} & $1.04$ & $0.12\pm0.32$ & $1.25$ & $0.15\pm0.33$ & $4$ & $11\pm20$ & $39$ & $9\pm15$ & $0.5$ & $0.1\pm0.5$ & $0.53$ & $0.11\pm0.23$ \\
    \end{tabular}}
  \end{center}
  \label{tab:supresults}
\end{sidewaystable}
\begin{table}
  \caption{Modulus and phase of the relative amplitudes between
           resonances that appear in both the \decay{\Dz}{\KSKmpip}
           and \decay{\Dz}{\KSKppim} modes. Relative phases are calculated
           using the value of $\delta_{\KSKpis}$ measured in \psiprpr decays~\cite{Insler:2012pm},
           and the uncertainty on this value is included in the statistical uncertainty.
           The first uncertainties are statistical and the second systematic.}
  \begin{center}
    \begin{tabular}{l c r@{$\,\pm\,$}l r@{$\,\pm\,$}l}
      Relative \\
      amplitude & & \multicolumn{2}{c}{\glass} & \multicolumn{2}{c}{\lassc} \\
      \midrule
      \multirow{2}{*}{$\frac{\amp(\Kst{}{892}{-})}{\amp(\Kst{}{892}{+})}$} & mod & $0.582$ & $0.007\pm0.008$ & $0.576$ & $0.005\pm0.010$ \\
       & arg $(^{\circ})$ & $-2$ & $15\pm2$ & $-2$ & $15\pm1$ \\
      \cmidrule{2-6}
      \multirow{2}{*}{$\frac{\amp(\Kst{}{1410}{-})}{\amp(\Kst{}{1410}{+})}$} & mod & $0.64$ & $0.08\pm0.22$ & $0.90$ & $0.08\pm0.15$ \\
       & arg $(^{\circ})$ & $52$ & $17\pm20$ & $62$ & $16\pm6$ \\
      \cmidrule{2-6}
      \multirow{2}{*}{$\frac{\amp((\KS\pion)^{-}_{S-\text{wave}})}{\amp((\KS\pion)^{+}_{S-\text{wave}})}$} & mod & $0.54$ & $0.06\pm0.26$ & $0.59$ & $0.05\pm0.08$ \\
       & arg $(^{\circ})$ & $-100$ & $20\pm40$ & $-44$ & $17\pm10$ \\
      \cmidrule{2-6}
      \multirow{2}{*}{$\frac{\amp(\Kst{}{892}{0})}{\amp(\Kbarst{}{892}{0})}$} & mod & $1.12$ & $0.05\pm0.11$ & $1.17$ & $0.04\pm0.05$ \\
       & arg $(^{\circ})$ & $-78$ & $16\pm10$ & $-75$ & $15\pm2$ \\
      \cmidrule{2-6}
      \multirow{2}{*}{$\frac{\amp(\Kst{}{1410}{0})}{\amp(\Kbarst{}{1410}{0})}$} & mod & $0.60$ & $0.05\pm0.12$ & $0.62$ & $0.09\pm0.12$ \\
       & arg $(^{\circ})$ & $-9$ & $16\pm14$ & $-23$ & $17\pm11$ \\
      \cmidrule{2-6}
      \multirow{2}{*}{$\frac{\amp(\Kst{2}{1430}{0})}{\amp(\Kbarst{2}{1430}{0})}$} & mod & $1.1$ & $0.1\pm0.5$ & \multicolumn{2}{c}{---} \\
       & arg $(^{\circ})$ & $31$ & $17\pm12$ & \multicolumn{2}{c}{---} \\
      \cmidrule{2-6}
      \multirow{2}{*}{$\frac{\amp((\Kp\pim)_{S-\text{wave}})}{\amp((\Km\pip)_{S-\text{wave}})}$} & mod & $0.87$ & $0.08\pm0.14$ & $0.78$ & $0.06\pm0.18$ \\
       & arg $(^{\circ})$ & $49$ & $25\pm16$ & $68$ & $16\pm6$ \\
      \cmidrule{2-6}
      \multirow{2}{*}{$\frac{\amp(\ameson{0}{980}{+})}{\amp(\ameson{0}{980}{-})}$} & mod & \multicolumn{2}{c}{---} & $2.1$ & $0.2\pm0.6$ \\
       & arg $(^{\circ})$ & \multicolumn{2}{c}{---} & $42$ & $16\pm5$ \\
      \cmidrule{2-6}
      \multirow{2}{*}{$\frac{\amp(\ameson{0}{1450}{+})}{\amp(\ameson{0}{1450}{-})}$} & mod & $0.49$ & $0.06\pm0.28$ & $1.14$ & $0.16\pm0.30$ \\
       & arg $(^{\circ})$ & $-60$ & $19\pm34$ & $-63$ & $20\pm19$ \\
      \cmidrule{2-6}
      \multirow{2}{*}{$\frac{\amp(\rmeson{1450}{+})}{\amp(\rmeson{1450}{-})}$} & mod & $0.86$ & $0.16\pm0.26$ & \multicolumn{2}{c}{---} \\
       & arg $(^{\circ})$ & $110$ & $20\pm50$ & \multicolumn{2}{c}{---} \\
      \cmidrule{2-6}
      \multirow{2}{*}{$\frac{\amp(\rmeson{1700}{+})}{\amp(\rmeson{1700}{-})}$} & mod & $1.6$ & $0.4\pm0.4$ & \multicolumn{2}{c}{---} \\
       & arg $(^{\circ})$ & $70$ & $20\pm70$ & \multicolumn{2}{c}{---} \\
    \end{tabular}
  \end{center}
  \label{tab:relativeamplitudes}
\end{table}
The fit results for the best isobar models using the \glass and \lassc parameterizations
of the \kpswave are given in Tables~\ref{tab:favresults} and \ref{tab:supresults}.
Distributions of \sKpi, \sKSpi and \sKSK are shown alongside the best model of the
\decay{\Dz}{\KSKmpip} mode using the \glass parameterization in Fig.~\ref{fig:favglass}.
In Fig.~\ref{fig:favglass} and elsewhere the nomenclature $\PR_1 \times \PR_2$ denotes
interference terms.
The corresponding distributions showing the best model using the \lassc parameterization
are shown in Fig.~\ref{fig:favlass}. Distributions for the \decay{\Dz}{\KSKppim} mode are
shown in Figs.~\ref{fig:supglass} and \ref{fig:suplass}.
Figure~\ref{fig:2dplots} shows the \glass isobar models in two dimensions, and demonstrates that
the \glass and \lassc choices of \kpswave parameterization both lead to similar descriptions
of the overall phase variation.
Figures~\ref{fig:favglass}--\ref{fig:suplass} show distributions distorted by
efficiency effects, while Fig.~\ref{fig:2dplots} shows the decay rate without distortion.
Lookup tables for the complex amplitude variation across the Dalitz plot in
all four isobar models are available in the supplemental material.

The data are found to favor solutions that have a significant
neutral \kpswave contribution,
even though the exchange (Fig.~\ref{fig:feynman}b) and penguin annihilation
(Fig.~\ref{fig:feynman}d) processes that contribute to the neutral channel are expected to
be suppressed. The expected suppression is observed for the $P$-wave \Kst{}{892}{} resonances, with the
neutral mode fit fractions substantially lower. The models using the \lassc parameterization
additionally show this pattern for the \Kst{}{1410}{} states.
The sums of the fit fractions~\cite{PDGdalitzformalism}, excluding interference terms,
in the \decay{\Dz}{\KSKmpip} and \decay{\Dz}{\KSKppim}
models are, respectively, 103\% (109\%) and 81\% (99\%) using the \glass (\lassc)
\kpswave parameterization.

Using measurements of the mean strong-phase difference between the
\decay{\Dz}{\KSKpi} modes available from \psiprpr decays~\cite{Insler:2012pm},
the relative complex amplitudes between each resonance in one \Dz decay mode and its conjugate contribution
to the other \Dz decay mode are computed. These values are summarized in Table~\ref{tab:relativeamplitudes}.

Additional information about the models is listed in Appendices~\ref{sec:extraisobarinfo} and \ref{sec:app_systematics},
including the interference fractions and decomposition of the systematic uncertainties.
The best models also contain contributions from the \rmeson{1450}{\pm} and \rmeson{1700}{\pm} resonances
in the $\KS\Kpm$ channels, supporting evidence in Ref.~\cite{Bargiotti:2003ev} of the $\kaon\Kbar$
decay modes for these states. Alternative models are fitted where one $\rhomeson^{\pm}$ contribution is removed
from the best models; in these the value of \ntll is found to degrade by at least 162 units.
Detailed results are tabulated in Appendix~\ref{sec:extraisobarinfo}.

The \kpswave systems are poorly understood~\cite{PDGscalarmesons}, and there is no clear
theoretical guidance regarding the correct description of these systems in an isobar model.
As introduced in Sect.~\ref{sec:isobarformalism}, the \lassc parameterization is motivated by
the Watson theorem, but this assumes that three-body interactions are negligible and is not,
therefore, expected to be precisely obeyed in nature.
\begin{figure}
  \begin{center}
    \includegraphics[width=0.5\textwidth]{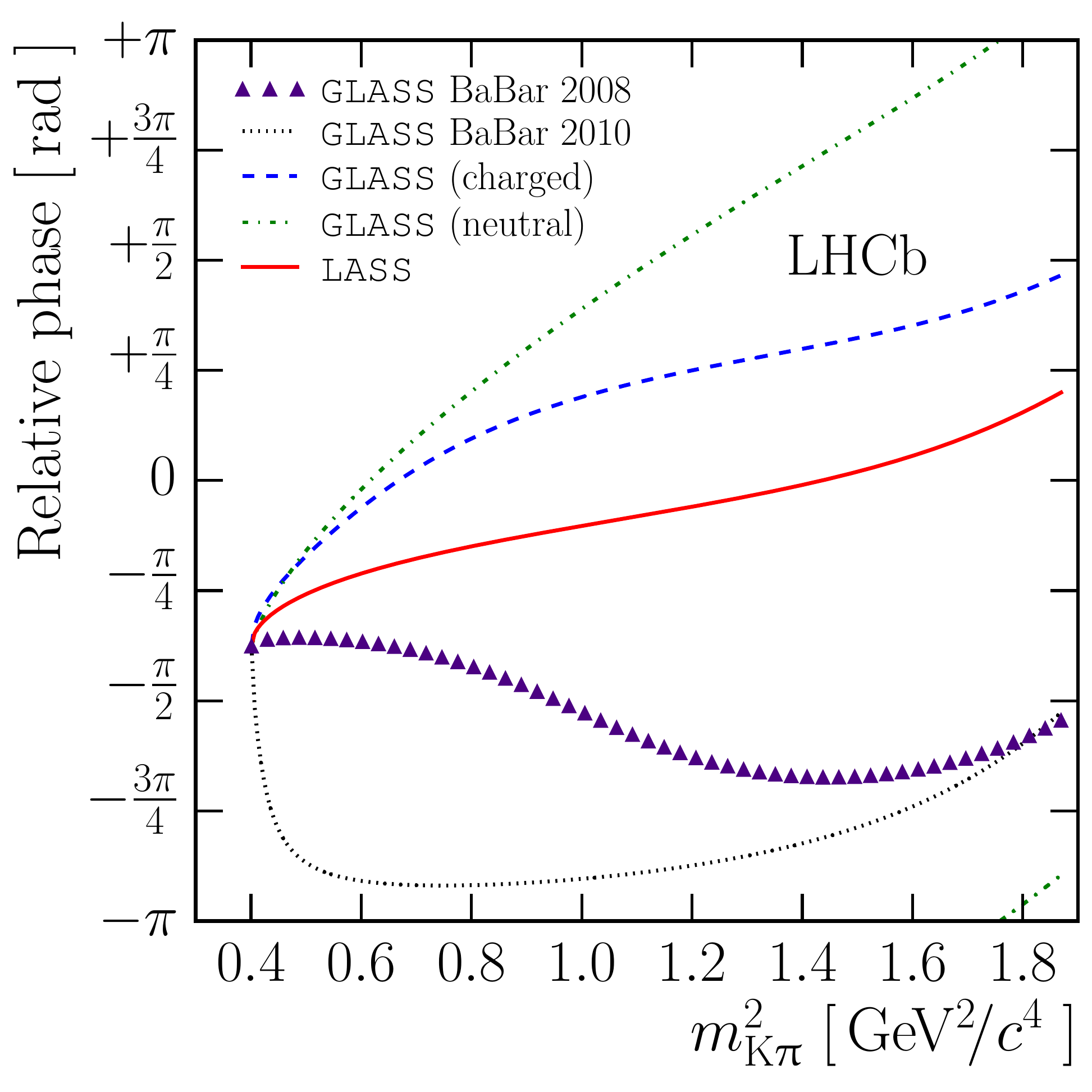}
  \end{center}
  \caption{Comparison of the phase behavior of the various \kpswave
           parameterizations used. The solid (red) curve shows the \lassc parameterization,
           while the dashed (blue) and dash-dotted (green) curves show, respectively,
           the \glass functional form fitted to the charged and neutral $S$-wave channels.
           The final two curves show the \glass forms fitted to the charged \kpswave in
           \decay{\Dz}{\KS\pip\pim} decays in Ref.~\cite{Aubert:2008bd} (triangular markers, purple)
           and Ref.~\cite{delAmoSanchez:2010xz} (dotted curve, black). The latter of these was
           used in the analysis of \decay{\Dz}{\KSKpi} decays by the \cleo collaboration~\cite{Insler:2012pm}.}
  \label{fig:lass_glass_comparison}
\end{figure}
The isobar models using the \glass parameterization
favor solutions with qualitatively similar phase behavior to those
using the \lassc parameterization. This is illustrated in
Fig.~\ref{fig:lass_glass_comparison}, which also shows the \glass
forms obtained in fits to \decay{\Dz}{\KS\pip\pim} decays by the \babar
collaboration~\cite{Aubert:2008bd,delAmoSanchez:2010xz}
and previously used in fits to the \decay{\Dz}{\KSKpi} decay modes~\cite{Insler:2012pm}.
This figure shows that the \glass functional form has substantial freedom
to produce different phase behavior to the \lassc form, but that this
is not strongly favored in the \decay{\Dz}{\KSKpi} decays.
The good quality of fit obtained using the \lassc parameterization
indicates that large differences in phase behavior with respect to
\kpi scattering data~\cite{Dunwoodie,Aston:1987ir} are not required
in order to describe the \decay{\Dz}{\KSKpi} decays.
A similar conclusion was drawn in Ref.~\cite{Pennington:2007se} for the decay
\decay{\Dp}{\Km\pip\pip}, while Ref.~\cite{Aitala:2005yh} found behavior
inconsistent with scattering data using the same \Dp decay mode but a
slightly different technique.
Ref.~\cite{Epifanov:2007rf} studied
the \kpswave in \decay{\taum}{\KS\pim\neut} decays
and found that a parameterization based on the \lass \kpi scattering
data, but without a real production form factor, gave a poor description
of the \taum decay data.

The quality of fit for each model is quantified by calculating \chisq using a dynamic binning scheme.
The values are summarized
in Table~\ref{tab:fitquality}, while the binning scheme and two-dimensional quality of fit are shown in
Appendix~\ref{sec:extraisobarinfo}.
This binning scheme is generated by iteratively sub-dividing the Dalitz plot to produce
new bins of approximately equal population until further sub-division would result in a
bin population of fewer than 15 candidates, or a bin dimension smaller than $0.02\gevgevcccc$
in \sKSpi or \sKpi. This minimum size corresponds to five times the average resolution in these variables.
\begin{table}
	\caption{Values of \chisqbin indicating the fit quality obtained using both \kpswave parameterizations
	              in the two decay modes. The binning scheme for the \decay{\Dz}{\KSKmpip} (\decay{\Dz}{\KSKppim}) mode contains
	              2191 (2573) bins.}
	\begin{center}
		\begin{tabular}{l c c}
			& \multicolumn{2}{c}{Isobar model} \\
			& \glass & \lassc \\
			\midrule
			\decay{\Dz}{\KSKmpip} & 1.12 & 1.10 \\
			\decay{\Dz}{\KSKppim} & 1.07 & 1.09 \\
		\end{tabular}
	\end{center}
	\label{tab:fitquality}
\end{table}

The overall fit quality is slightly better in the isobar models using the \glass \kpswave parameterization,
but this is not a significant effect and it should be noted that these models contain more degrees of freedom,
with 57 parameters fitted in the final \glass model compared to 50 when using the \lassc parameterization.

\section{Additional measurements}
In this section, several additional results, including those
derived from the amplitude models, are presented.
\label{sec:otherresults}
\subsection{Ratio of branching fractions measurement}
\label{sec:relativebr}
The ratio of branching fractions
\begin{equation}
  \BF_{\KSKpis} \equiv \frac{\BF(\decay{\Dz}{\KSKppim})}{\BF(\decay{\Dz}{\KSKmpip})},
\end{equation}
and the restricted region ratio $\BF_{\Kstar\kaon}$, defined in the same region
near the \Kst{}{892}{\pm} resonance as the coherence factor $R_{\Kstar\kaon}$ (Sect.~\ref{sec:coherence}), are also measured.
The efficiency correction due to the reconstruction efficiency \efffn
is evaluated using the best isobar models, and the difference between the results
obtained with the two \kpswave parameterizations is taken as a systematic uncertainty
in addition to those effects described in Sect.~\ref{sec:systematics}.
This efficiency correction modifies the ratio of yields quoted in Table~\ref{tab:massfit}
by approximately 3\%.

The two ratios are measured to be
\begin{align*}
  \BF_{\KSKpis} &= 0.655\pm 0.004\stat\pm0.006\syst, \\
  \BF_{\Kstar\kaon} &= 0.370\pm0.003\stat\pm0.012\syst.
\end{align*}
These are the most precise measurements to date.

\subsection{Coherence factor and {\boldmath\CP}-even fraction results}
\begin{table}
  \caption{Coherence factor observables to which the isobar models are sensitive.
           The third column summarizes the \cleo results measured in quantum-correlated
           decays~\cite{Insler:2012pm}, where the uncertainty on $\delta_{\KSKpis} - \delta_{\Kstar\kaon}$
           is calculated assuming maximal correlation between $\delta_{\KSKpis}$ and
           $\delta_{\Kstar\kaon}$.}
  \begin{center}
    \begin{tabular}{l r@{$\,\pm\,$}l r@{$\,\pm\,$}l | r@{$\pm$}l}
      Variable & \multicolumn{2}{c}{\glass} & \multicolumn{2}{c}{\lassc} & \multicolumn{2}{c}{\cleo} \\
      \midrule
      $R_{\KSKpis}$ & 0.573 & $0.007\pm0.019$ & 0.571 & $0.005\pm0.019$ & 0.73 & 0.08 \\
      $R_{\Kstar\kaon}$ & 0.831 & $0.004\pm0.010$ & 0.835 & $0.003\pm0.011$ & 1.00 & 0.16 \\
      $\delta_{\KSKpis} - \delta_{\Kstar\kaon}$ & $(0.2$ & $0.6\pm1.1)^{\circ}$ & $(-0.0$ & $0.5\pm0.7)^{\circ}$ & $(-18$ & $31)^{\circ}$ \\
    \end{tabular}
  \end{center}
  \label{tab:coherence}
\end{table}
The amplitude models are used to calculate the coherence factors
$R_{\KSKpis}$ and $R_{\Kstar\kaon}$, and the strong-phase difference
$\delta_{\KSKpis} - \delta_{\Kstar\kaon}$, as described in Sect.~\ref{sec:coherence}.
The results are summarized in Table~\ref{tab:coherence}, alongside
the corresponding values measured in \psiprpr decays by the \cleo collaboration.
Lower, but compatible, coherence is calculated using the isobar models than was measured
at \cleo, with the discrepancy larger for the coherence factor
calculated over the full phase space.
The results from the \glass and \lassc isobar models are very similar, showing that
the coherence variables are not sensitive to the \kpswave parameterization.

The coherence factor $R_{\KSKpis}$, and the ratio of branching fractions
$\BR_{\KSKpis}$, are combined with the mean phase difference between the
two final states measured in \psiprpr decays~\cite{Insler:2012pm} to
calculate the \CP-even fraction $F_+$, defined in Eq.~\ref{eq:cpeven},
which is determined to be
\begin{equation*}
  F_+ = 0.777 \pm 0.003\stat \pm 0.009\syst,
\end{equation*}
using the \glass amplitude models. A consistent result is obtained using
the alternative (\lassc) amplitude models.
This model-dependent value is compatible with the direct measurement using
only \psiprpr decay data~\cite{Insler:2012pm,Gershon:2015xra}.

\subsection{{\boldmath\grpsuthree} flavor symmetry tests}
\label{sec:flavoursym}
\grpsuthree flavor symmetry can be used to relate decay amplitudes in
several \PD meson decays, such that a global fit to many such amplitudes
can provide predictions for the neutral and charged \Kst{}{892}{} complex amplitudes in
\decay{\Dz}{\KSKpi} decays~\cite{Bhattacharya:2011gf,Bhattacharya:2012fz}.
Predictions are available for the $\Kstarp\Km$, $\Kstarm\Kp$, $\Kstarz\Kzb$
and $\Kstarzb\Kz$ complex amplitudes, where \Kstar refers to the \Kst{}{892}{}
resonances. There are therefore three relative
amplitudes and two relative phases that can be determined from the isobar
models, with an additional relative phase accessible if the isobar results
are combined with the \cleo measurement of the mean strong phase
difference~\cite{Insler:2012pm}.
The results are summarized in Table~\ref{tab:suthree}.

The isobar model results are found to follow broadly the patterns predicted
by \grpsuthree flavor symmetry.
The amplitude ratio between the \Kst{}{892}{+} and \Kbarst{}{892}{0} resonances,
which is derived from the \decay{\Dz}{\KSKmpip} isobar model alone, shows
good agreement. The two other amplitude ratios additionally depend on the
ratio $\BF_{\KSKpis}$, and these are more discrepant with the \grpsuthree
predictions.
The relative phase between the charged and neutral \Kst{}{892}{} resonances
shows better agreement with the flavor symmetry prediction in the \decay{\Dz}{\KSKppim}
mode, where both resonances have clear peaks in the data.
The \glass and \lassc isobar models are found to agree well,
suggesting the problems are not related to the \kpswave.
\begin{table}
	\caption{\grpsuthree flavor symmetry predictions~\cite{Bhattacharya:2012fz} and
				 results. The uncertainties on phase difference predictions are calculated
				 from the quoted magnitude and phase uncertainties. Note that some
				 theoretical predictions depend on the \etaz--\etapr mixing angle $\theta_{\etaz-\etapr}$
				 and are quoted for two different values.
                                 The bottom entry in the table relies on the \cleo measurement~\cite{Insler:2012pm}
				 of the coherence factor phase $\delta_{\KSKpis}$, and the uncertainty on this
				 phase is included in the statistical uncertainty, while the other entries are calculated
				 directly from the isobar models and relative branching ratio.
                                 Where two uncertainties are quoted the first is statistical and the second systematic.}
	\begin{center}
                \hspace{-8mm}
                \resizebox{\textwidth}{!}{
		\begin{tabular}{l r@{$\,\pm\,$}l r@{$\,\pm\,$}l r@{$\,\pm\,$}l r@{$\,\pm\,$}l}
			 & \multicolumn{4}{c}{Theory} & \multicolumn{4}{c}{Experiment} \\
			Ratio & \multicolumn{2}{c}{$\theta_{\Peta-\Peta'}=19.5^{\circ}$} & \multicolumn{2}{c}{$\theta_{\Peta-\Peta'}=11.7^{\circ}$} & \multicolumn{2}{c}{\glass} & \multicolumn{2}{c}{\lassc} \\
			\midrule
			$\frac{|\amp(\Kst{}{892}{-}\Kp)|}{|\amp(\Kst{}{892}{+}\Km)|}$ & 0.685 & 0.032 & 0.685 & 0.032 & $0.582$ & $0.007\pm0.007$ & $0.576$ & $0.005\pm0.010$ \\
			\addlinespace[2pt]
			$\frac{|\amp(\Kbarst{}{892}{0}\Kz)|}{|\amp(\Kst{}{892}{+}\Km)|}$ & 0.138 & 0.033 & 0.307 & 0.035 & $0.297$ & $0.010\pm0.024$ & $0.295$ & $0.009\pm0.014$ \\
			\addlinespace[2pt]
			$\frac{|\amp(\Kst{}{892}{0}\Kzb)|}{|\amp(\Kst{}{892}{+}\Km)|}$ & 0.138 & 0.033 & 0.307 & 0.035 & $0.333$ & $0.008\pm0.016$ & $0.345$ & $0.007\pm0.010$ \\
			\midrule
			Argument & \multicolumn{4}{c}{Theory ($^{\circ}$)} & \multicolumn{4}{c}{Experiment ($^{\circ}$)}\\
			\midrule
			$\frac{\amp(\Kbarst{}{892}{0}\Kz)}{\amp(\Kst{}{892}{+}\Km)}$ & 151 & 14 & 112 & 8 & $72$ & $2\pm4$ & $78.5$ & $2.0\pm2.8$ \\
			\addlinespace[2pt]
			$\frac{\amp(\Kst{}{892}{0}\Kzb)}{\amp(\Kst{}{892}{-}\Kp)}$ & $-9$ & 13 & $-37$ & 6 & $-4$ & $2\pm9$ & $5.0$ & $1.7\pm1.4$ \\
			\addlinespace[2pt]
			$\frac{\amp(\Kst{}{892}{0}\Kzb)}{\amp(\Kbarst{}{892}{0}\Kz)}$ & \multicolumn{2}{c}{180} & \multicolumn{2}{c}{180} & $-78$ & $16\pm10$ & $-75$ & $15\pm2$ \\
		\end{tabular}}
		\label{tab:suthree}
	\end{center}
\end{table}

\subsection{{\boldmath\CP} violation tests}
\label{sec:cpv}
Searches for time-integrated \CP-violating effects in the resonant structure of
these decays are performed using the best
isobar models.
The resonance amplitude and phase parameters $a_{\PR}$ and $\phi_{\PR}$
are substituted with $a_{\PR}(1 \pm \Delta a_{\PR})$ and
$\phi_{\PR} \pm \Delta \phi_{\PR}$, respectively, where the
signs are set by the flavor tag.
The convention adopted is that
a positive sign produces the \Dzb complex amplitude.
The full fit results are tabulated in Appendix~\ref{sec:cpvappendix}.

A subset of the $\Delta$ parameters is used to perform a \chisq
test against the no-\CP violation hypothesis: only those parameters
corresponding to resonances that are present in the best isobar
models using both the \glass and \lassc \kpswave parameterizations
are included. The absolute difference $|\Delta_{{\footnotesize\glass}} - \Delta_{{\footnotesize\lassc}}|$
is assigned as the systematic uncertainty due to dependence on the
choice of isobar model.
This subset of parameters is shown in Table~\ref{tab:cpvresults}, where
the change in fit fraction between the \Dz and \Dzb solutions is included
for illustrative purposes.
In the \chisq test the statistical and systematic uncertainties are added
in quadrature.

Using the best \glass (\lassc) isobar models the test result is
$\chisqndf = 30.5/32 = 0.95$ ($32.3/32 = 1.01$),
corresponding to a $p$-value of 0.54 (0.45). 
Therefore, the data are compatible with the hypothesis of \CP-conservation.
\begin{sidewaystable}
  \caption{\CP violation fit results. Results are only shown for those
           resonances that appear in both the \glass and \lassc models.
           The first uncertainties are statistical and the second systematic;
           the only systematic uncertainty is that due to the choice of
           isobar model.}
    \begin{subtable}{\textwidth}
      \centering		
      \caption{Results for the \decay{\Dz}{\KSKmpip} mode.}
      \resizebox{\textwidth}{!}{
      \begin{tabular}{l r@{$\,\pm\,$}l r@{$\,\pm\,$}l r@{$\,\pm\,$}l r@{$\,\pm\,$}l r@{$\,\pm\,$}l r@{$\,\pm\,$}l r@{$\,\pm\,$}l r@{$\,\pm\,$}l r@{$\,\pm\,$}l r@{$\,\pm\,$}l r@{$\,\pm\,$}l r@{$\,\pm\,$}l}
        & \multicolumn{4}{c}{\daR} & \multicolumn{4}{c}{\dpR$ (^{\circ})$} & \multicolumn{4}{c}{$\Delta(\text{Fit fraction})$ [\%]} \\
        Resonance & \multicolumn{2}{c}{\glass} & \multicolumn{2}{c}{\lassc} & \multicolumn{2}{c}{\glass} & \multicolumn{2}{c}{\lassc} & \multicolumn{2}{c}{\glass} & \multicolumn{2}{c}{\lassc} \\
        \midrule
        \Kst{}{892}{+} & \multicolumn{2}{c}{0.0 (fixed)} & \multicolumn{2}{c}{0.0 (fixed)} & \multicolumn{2}{c}{0.0 (fixed)} & \multicolumn{2}{c}{0.0 (fixed)} & $0.6$ & $1.0\pm0.3$ & $0.9$ & $1.0\pm0.3$\\
        \Kst{}{1410}{+} & $0.07$ & $0.06\pm0.04$ & $0.03$ & $0.06\pm0.04$ & $3.9$ & $3.5\pm1.9$ & $2.0$ & $2.9\pm1.9$ & $1.4$ & $0.8\pm0.2$ & $1.2$ & $1.6\pm0.2$\\
        \posswave & $0.02$ & $0.08\pm0.07$ & $-0.05$ & $0.08\pm0.07$ & $2.0$ & $1.7\pm0.0$ & $2.0$ & $1.7\pm0.0$ & $1$ & $4\pm3$ & $-2.3$ & $3.5\pm3.3$\\
        \Kbarst{}{892}{0} & $-0.046$ & $0.031\pm0.005$ & $-0.051$ & $0.030\pm0.005$ & $1.2$ & $1.6\pm0.3$ & $1.5$ & $1.7\pm0.3$ & $-0.43$ & $0.30\pm0.03$ & $-0.47$ & $0.29\pm0.03$\\
        \Kbarst{}{1410}{0} & $0.006$ & $0.034\pm0.017$ & $0.02$ & $0.04\pm0.02$ & $2$ & $5\pm5$ & $-3$ & $6\pm5$ & $0.3$ & $1.0\pm0.1$ & $0.4$ & $0.7\pm0.1$\\
        \neutswave & $0.05$ & $0.04\pm0.02$ & $0.03$ & $0.04\pm0.02$ & $0.4$ & $1.6\pm0.6$ & $1.0$ & $1.4\pm0.6$ & $2.2$ & $1.3\pm0.4$ & $2.6$ & $2.2\pm0.4$\\
        \ameson{2}{1320}{-} & $-0.25$ & $0.14\pm0.01$ & $-0.24$ & $0.13\pm0.01$ & $2$ & $9\pm3$ & $-1$ & $9\pm3$ & $-0.20$ & $0.13\pm0.05$ & $-0.15$ & $0.10\pm0.05$\\
        \ameson{0}{1450}{-} & $-0.01$ & $0.14\pm0.12$ & $-0.13$ & $0.14\pm0.12$ & $0$ & $5\pm4$ & $-4$ & $6\pm4$ & $-0.0$ & $0.4\pm0.4$ & $-0.4$ & $0.4\pm0.4$\\
        \rmeson{1450}{-} & $0.06$ & $0.13\pm0.11$ & $-0.05$ & $0.12\pm0.11$ & $-13$ & $10\pm9$ & $-5$ & $9\pm9$ & $0.3$ & $0.7\pm0.6$ & $-0.3$ & $0.7\pm0.6$\\
      \end{tabular}}
    \end{subtable}
    \\[12pt]
    \begin{subtable}{\textwidth}
      \centering
      \caption{Results for the \decay{\Dz}{\KSKppim} mode.}
      \resizebox{\textwidth}{!}{
      \begin{tabular}{l r@{$\,\pm\,$}l r@{$\,\pm\,$}l r@{$\,\pm\,$}l r@{$\,\pm\,$}l r@{$\,\pm\,$}l r@{$\,\pm\,$}l r@{$\,\pm\,$}l r@{$\,\pm\,$}l r@{$\,\pm\,$}l r@{$\,\pm\,$}l r@{$\,\pm\,$}l r@{$\,\pm\,$}l}
        & \multicolumn{4}{c}{\daR} & \multicolumn{4}{c}{\dpR$ (^{\circ})$} & \multicolumn{4}{c}{$\Delta(\text{Fit fraction})$ [\%]} \\
        Resonance & \multicolumn{2}{c}{\glass} & \multicolumn{2}{c}{\lassc} & \multicolumn{2}{c}{\glass} & \multicolumn{2}{c}{\lassc} & \multicolumn{2}{c}{\glass} & \multicolumn{2}{c}{\lassc} \\
        \midrule
        \Kst{}{892}{-} & \multicolumn{2}{c}{0.0 (fixed)} & \multicolumn{2}{c}{0.0 (fixed)} & \multicolumn{2}{c}{0.0 (fixed)} & \multicolumn{2}{c}{0.0 (fixed)} & $-1.1$ & $0.7\pm0.2$ & $-0.9$ & $0.7\pm0.2$\\
        \Kst{}{1410}{-} & $0.05$ & $0.12\pm0.08$ & $-0.03$ & $0.10\pm0.08$ & $-6$ & $4\pm3$ & $-3.0$ & $3.6\pm2.8$ & $0.6$ & $2.7\pm2.4$ & $-2$ & $4\pm2$\\
        \negswave & $0.10$ & $0.25\pm0.24$ & $-0.14$ & $0.25\pm0.24$ & $-7.7$ & $3.4\pm0.0$ & $-8$ & $4\pm0$ & $2$ & $6\pm6$ & $-4$ & $6\pm6$\\
        \Kst{}{892}{0} & $-0.010$ & $0.024\pm0.001$ & $-0.012$ & $0.022\pm0.001$ & $-1.4$ & $2.9\pm2.2$ & $0.8$ & $2.8\pm2.2$ & $-0.4$ & $0.4\pm0.0$ & $-0.4$ & $0.4\pm0.0$\\
        \Kst{}{1410}{0} & $0.10$ & $0.10\pm0.09$ & $0.19$ & $0.13\pm0.09$ & $-1$ & $9\pm8$ & $-9$ & $9\pm8$ & $1.9$ & $1.1\pm0.2$ & $1.6$ & $0.8\pm0.2$\\
        \neutswave & $-0.07$ & $0.06\pm0.05$ & $-0.12$ & $0.06\pm0.05$ & $-2$ & $4\pm4$ & $2$ & $4\pm4$ & $-4$ & $5\pm5$ & $-9$ & $6\pm5$\\
        \ameson{0}{980}{+} & $0.06$ & $0.04\pm0.01$ & $0.052$ & $0.025\pm0.008$ & $-3$ & $5\pm2$ & $-0.9$ & $3.1\pm2.2$ & $2.2$ & $2.8\pm2.4$ & $4.6$ & $3.3\pm2.4$\\
        \ameson{0}{1450}{+} & $-0.11$ & $0.10\pm0.04$ & $-0.07$ & $0.07\pm0.04$ & $10$ & $8\pm5$ & $5$ & $6\pm5$ & $-0.21$ & $0.30\pm0.23$ & $-0.4$ & $0.4\pm0.2$\\
        \rmeson{1700}{+} & $-0.03$ & $0.13\pm0.09$ & $-0.12$ & $0.13\pm0.09$ & $4$ & $6\pm2$ & $2$ & $5\pm2$ & $-0.07$ & $0.25\pm0.19$ & $-0.27$ & $0.27\pm0.19$\\
      \end{tabular}}
    \end{subtable}
  \label{tab:cpvresults}
\end{sidewaystable}

\section{Conclusions}
\label{sec:conclusions}
The decay modes \decay{\Dz}{\KSKpi} have been studied using unbinned,
time-integrated, fits to a high purity sample of 189\,670 candidates,
and two amplitude models have been constructed for each decay mode.
These models are compared to data in a large number of bins
in the relevant Dalitz plots and a \chisq test indicates a good description
of the data.

Models are presented using two different parameterizations of the \kpswave
systems, which have been found to be an important component of these decays.
These systems are poorly understood, and comparisons have been made
to previous results and alternative parameterizations, but the treatment
of the \kpswave is found to have little impact on the other results
presented in this paper. The large fractions attributed to the neutral
\kpswave channels could indicate larger than expected
contributions from the penguin annihilation diagrams shown in Fig.~\ref{fig:feynman}d.

The models are seen to favor small, but significant, contributions
from the \decay{\rmeson{1450,1700}{\pm}}{\KS\Kpm} resonances,
modes which were seen by the OBELIX experiment~\cite{Bargiotti:2003ev}
but are not well established. 
All models contain clear contributions from both the \Kst{}{892}{\pm} and
\Kst{}{892}{0} resonances, with the \Kst{}{892}{0} contribution found to be
suppressed as expected from the diagrams shown in Fig.~\ref{fig:feynman}.
This allows the full set of amplitudes in these decays that are predicted
by \grpsuthree flavor symmetry to be tested, in contrast to the previous
analysis by the \cleo collaboration~\cite{Insler:2012pm}.
Partial agreement is found with these predictions.

The ratio of branching fractions between the two \decay{\Dz}{\KSKpi} modes
is also measured, both across the full Dalitz plot area and in a restricted
region near the \Kst{}{892}{\pm} resonance, with much improved precision
compared to previous results.

Values for the \decay{\Dz}{\KSKpis} coherence factor are computed using the amplitude
models, again both for the whole Dalitz plot area and in the restricted
region, and are found to be in reasonable agreement with direct measurements
by \cleo~\cite{Insler:2012pm} using quantum-correlated \decay{\psiprpr}{\Dz\Dzb}
decays. The \CP-even fraction of the \decay{\Dz}{\KSKpis} decays is also computed, using
input from the quantum-correlated decays, and is found to be in agreement with
the direct measurement~\cite{Insler:2012pm,Gershon:2015xra}.
A search for time-integrated \CP violation is carried out using the amplitude models,
but no evidence is found with either choice of parameterization for the \kpswave.

The models presented here will be useful for future \Dz--\Dzb mixing,
indirect \CP violation and CKM angle $\gamma$ studies, where knowledge
of the strong-phase variation across the Dalitz plot can improve the
attainable precision.
These improvements will be particularly valuable for studies of the
large dataset that is expected to be accumulated in Run 2 of the \lhc.

\section*{Acknowledgements}
\noindent We express our gratitude to our colleagues in the CERN
accelerator departments for the excellent performance of the LHC. We
thank the technical and administrative staff at the LHCb
institutes. We acknowledge support from CERN and from the national
agencies: CAPES, CNPq, FAPERJ and FINEP (Brazil); NSFC (China);
CNRS/IN2P3 (France); BMBF, DFG and MPG (Germany); INFN (Italy); 
FOM and NWO (The Netherlands); MNiSW and NCN (Poland); MEN/IFA (Romania); 
MinES and FANO (Russia); MinECo (Spain); SNSF and SER (Switzerland); 
NASU (Ukraine); STFC (United Kingdom); NSF (USA).
We acknowledge the computing resources that are provided by CERN,
IN2P3 (France), KIT and DESY (Germany), INFN (Italy), SURF (The
Netherlands), PIC (Spain), GridPP (United Kingdom), RRCKI (Russia),
CSCS (Switzerland), IFIN-HH (Romania), CBPF (Brazil), PL-GRID (Poland)
and OSC (USA). We are indebted to the communities behind the multiple open 
source software packages on which we depend. We are also thankful for the 
computing resources and the access to software R\&D tools provided by Yandex LLC (Russia).
Individual groups or members have received support from AvH Foundation (Germany),
EPLANET, Marie Sk\l{}odowska-Curie Actions and ERC (European Union), 
Conseil G\'{e}n\'{e}ral de Haute-Savoie, Labex ENIGMASS and OCEVU, 
R\'{e}gion Auvergne (France), RFBR (Russia), XuntaGal and GENCAT (Spain), The Royal Society 
and Royal Commission for the Exhibition of 1851 (United Kingdom).
We acknowledge the use of the Advanced Research
Computing (ARC) facilities (Oxford) in carrying out this work.

\clearpage

{\noindent\bf\Large Appendices}
\appendix
\section{Additional isobar formalism information}
\label{sec:appformalismdetails}
This appendix contains Table~\ref{tab:nominalparameters}, which
summarizes the nominal values used for the various resonance and
form factor parameters.
\begin{table}
  \caption{Nominal values for isobar model parameters that are fixed in the model fits, 
  			   or used in constraint terms. These values are taken from Refs.~\cite{PDG2014,Abele:1998qd,Bargiotti:2003ev,Dunwoodie} as
			   described in Sect.~\ref{sec:formalism}.}
  \begin{center}
    \begin{tabular}{l l r@{$\pm$}l l}
      \multicolumn{2}{c}{Parameter} & \multicolumn{2}{c}{Value} \\
      \midrule
      \multirow{2}{*}{\Kst{}{892}{\pm}} & $m_{\PR}$ & 891.66 & 0.26 & \mevcc \\
       & $\Gamma_{\PR}$ & 50.8 & 0.9 & \mevcc \\
      \cmidrule{2-5}
      \multirow{2}{*}{\Kst{}{1410}{\pm}} & $m_{\PR}$ & 1.414 & 0.015 & \gevcc \\
       & $\Gamma_{\PR}$ & 0.232 & 0.021 & \gevcc \\
      \cmidrule{2-5}
      \multirow{2}{*}{\chgswave} & $m_{\PR}$ & 1.435 & 0.005 & \gevcc \\
       & $\Gamma_{\PR}$ & 0.279 & 0.006 & \gevcc \\
      \cmidrule{2-5}
      \multirow{2}{*}{\Kst{}{892}{0}} & $m_{\PR}$ & 895.94 & 0.22 & \mevcc \\
       & $\Gamma_{\PR}$ & 48.7 & 0.8 & \mevcc \\
      \cmidrule{2-5}
      \multirow{2}{*}{\Kst{}{1410}{0}} & $m_{\PR}$ & 1.414 & 0.015 & \gevcc \\
       & $\Gamma_{\PR}$ & 0.232 & 0.021 & \gevcc \\
      \cmidrule{2-5}
      \multirow{2}{*}{\Kst{2}{1430}{0}} & $m_{\PR}$ & 1.4324 & 0.0013 & \gevcc \\
       & $\Gamma_{\PR}$ & 0.109 & 0.005 & \gevcc \\
      \cmidrule{2-5}
      \multirow{2}{*}{\neutswave} & $m_{\PR}$ & 1.435 & 0.005 & \gevcc \\
       & $\Gamma_{\PR}$ & 0.279 & 0.006 & \gevcc \\
      \cmidrule{2-5}
      \multirow{2}{*}{\kpswave} & $r$ & 1.8 & 0.4 & \invgevc \\
       & $a$ & 1.95 & 0.09 & \invgevc \\
      \cmidrule{2-5}
      \multirow{3}{*}{\ameson{0}{980}{\pm}} & $m_{\PR}$ & 0.980 & 0.020 & \gevcc \\
       & $g_{\Peta\pion}$ & 324 & 15 & \mev \\
       & $\frac{g^2_{\kaon\Kbar}}{g^2_{\Peta\pion}}$ & 1.03 & 0.14 &  \\
      \cmidrule{2-5}
      \multirow{2}{*}{\ameson{2}{1320}{\pm}} & $m_{\PR}$ & 1.3181 & 0.0007 & \gevcc \\
       & $\Gamma_{\PR}$ & 0.1098 & 0.0024 & \gevcc \\
      \cmidrule{2-5}
      \multirow{2}{*}{\ameson{0}{1450}{\pm}} & $m_{\PR}$ & 1.474 & 0.019 & \gevcc \\
       & $\Gamma_{\PR}$ & 0.265 & 0.013 & \gevcc \\
      \cmidrule{2-5}
      \multirow{2}{*}{\rmeson{1450}{\pm}} & $m_{\PR}$ & 1.182 & 0.030 & \gevcc \\
       & $\Gamma_{\PR}$ & 0.389 & 0.020 & \gevcc \\
      \cmidrule{2-5}
      \multirow{2}{*}{\rmeson{1700}{\pm}} & $m_{\PR}$ & 1.594 & 0.020 & \gevcc \\
       & $\Gamma_{\PR}$ & 0.259 & 0.020 & \gevcc \\
    \end{tabular}
  \end{center}
  \label{tab:nominalparameters}
\end{table}

\section{Additional isobar model information}
\label{sec:extraisobarinfo}
This appendix contains additional information about the various isobar model
parameters that are used and allowed to vary freely in the model fits, \eg resonance mass
and width parameter values, and parameters of the \glass and \lass \kpswave
functional forms.

Tables~\ref{tab:favinterference} and \ref{tab:supinterference} summarize the most
significant interference terms in the \decay{\Dz}{\KSKmpip} and \decay{\Dz}{\KSKppim}
models, respectively. Table~\ref{tab:lassrotations} defines the matrices $\mathbf{U}$
used to define the \lassc \kpswave form factor. Tables~\ref{tab:extraglassparameters} (\glass)
and~\ref{tab:extralassparameters} (\lassc) summarize the various resonance and
form factor parameters. The nominal values that are used in Gaussian constraint terms
are given in Appendix~\ref{sec:appformalismdetails}.

\begin{figure}
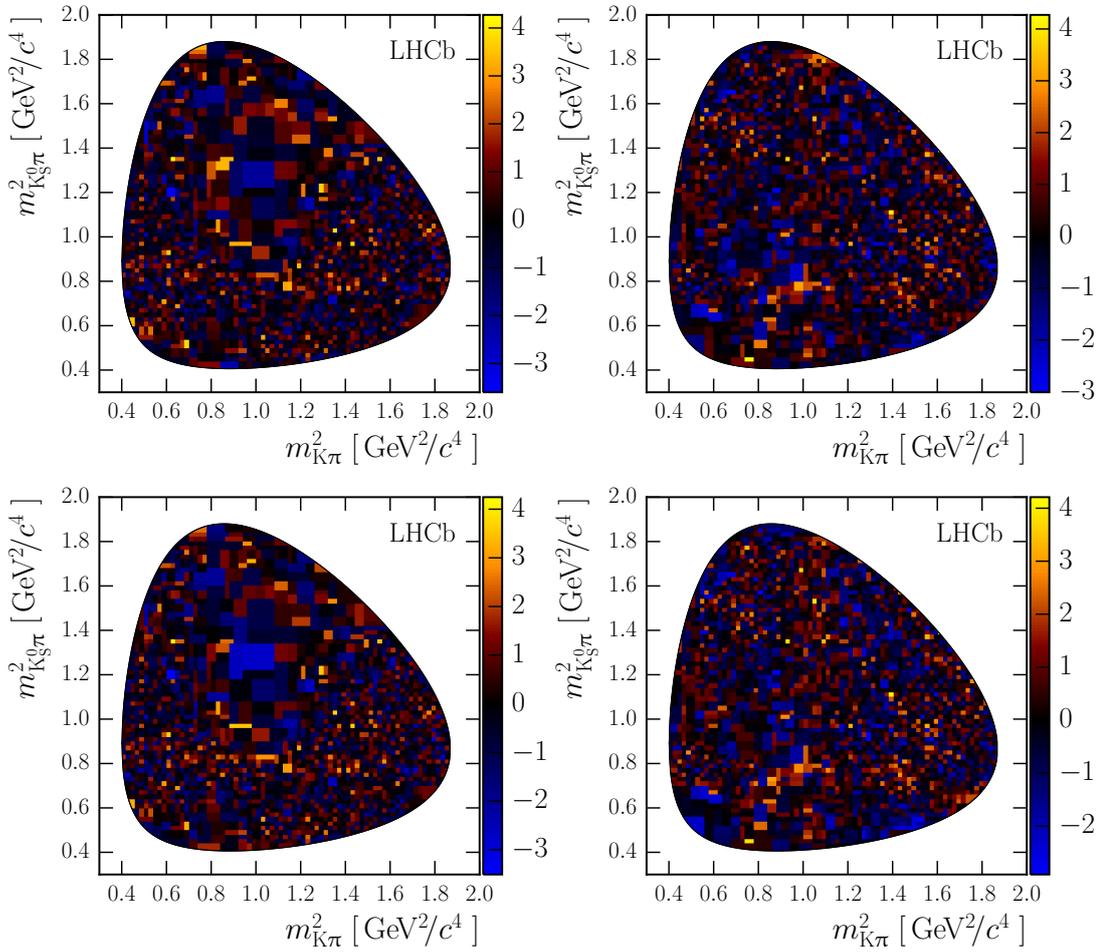

  \begin{center}
    \begin{subfigure}{0.45\textwidth}
      \centering
      \includegraphics[width=\textwidth]{{{figs/glass_fav_chi}}}
    \end{subfigure}%
    \begin{subfigure}{0.45\textwidth}
      \centering
      \includegraphics[width=\textwidth]{{{figs/glass_sup_chi}}}
    \end{subfigure}
    \begin{subfigure}{0.45\textwidth}
      \centering
      \includegraphics[width=\textwidth]{{{figs/lass_fav_chi}}}
    \end{subfigure}%
    \begin{subfigure}{0.45\textwidth}
      \centering
      \includegraphics[width=\textwidth]{{{figs/lass_sup_chi}}}
    \end{subfigure}
  \end{center}
  \caption{Two-dimensional quality-of-fit distributions illustrating the dynamic
           binning scheme used to evaluate \chisq. The variable shown
           is $\frac{d_i - p_i}{\sqrt{p_i}}$ where $d_i$ and $p_i$ are the number of
           events and the fitted value, respectively, in bin $i$. The
           \decay{\Dz}{\KSKmpip} (\decay{\Dz}{\KSKppim}) mode is shown in the
           left (right) column, and the \glass (\lassc) isobar models are shown
           in the top (bottom) row.}
  \label{fig:chiplots}
\end{figure}

Figure~\ref{fig:2dcomb} shows the smooth functions that describe the combinatorial
background in the isobar model fits.
Figure~\ref{fig:chiplots} illustrates the two-dimensional quality of fit achieved in the
four isobar models and shows the binning scheme used to derive \chisqbin values.
\begin{sidewaystable}
  \caption{Interference fractions for the \decay{\Dz}{\KSKmpip} mode. The first uncertainties are statistical and the second systematic. Only the 25 largest terms are shown.}
  \begin{subtable}{0.5\textwidth}
    \centering
    \caption{Best \glass isobar model.}
    \begin{tabular}{l@{ $\times$ }l r@{$\,\pm\,$}l}
      \multicolumn{2}{c}{Resonances} & \multicolumn{2}{c}{Fit fraction [\%]} \\
      \midrule
      \neutswave & \Kst{}{892}{+} & $12$ & $1\pm4$ \\
      \posswave & \Kbarst{}{1410}{0} & $10.4$ & $0.7\pm2.0$ \\
      \Kbarst{}{1410}{0} & \Kst{}{892}{+} & $-9.0$ & $0.4\pm1.0$ \\
      \posswave & \neutswave & $-8$ & $2\pm11$ \\
      \posswave & \Kbarst{2}{1430}{0} & $-7.8$ & $1.1\pm1.6$ \\
      \Kst{}{1410}{+} & \Kbarst{}{1410}{0} & $-4.6$ & $0.5\pm3.4$ \\
      \Kbarst{2}{1430}{0} & \Kbarst{}{1410}{0} & $-4$ & $1\pm4$ \\
      \posswave & \Kst{}{892}{+} & $4.0$ & $0.4\pm2.4$ \\
      \posswave & \Kst{}{1410}{+} & $-4.0$ & $0.5\pm2.1$ \\
      \Kst{}{892}{+} & \rmeson{1450}{-} & $3.7$ & $0.4\pm1.0$ \\
      \neutswave & \Kbarst{}{1410}{0} & $-3.4$ & $0.5\pm3.5$ \\
      \Kst{}{892}{+} & \ameson{0}{1450}{-} & $3.2$ & $0.3\pm1.3$ \\
      \Kbarst{}{1410}{0} & \Kbarst{}{892}{0} & $-2.7$ & $0.2\pm0.5$ \\
      \Kbarst{}{1410}{0} & \ameson{0}{1450}{-} & $2.6$ & $0.2\pm0.8$ \\
      \posswave & \ameson{0}{1450}{-} & $2.3$ & $0.4\pm1.6$ \\
      \neutswave & \rmeson{1450}{-} & $2.1$ & $0.3\pm1.2$ \\
      \neutswave & \Kst{}{1410}{+} & $1.9$ & $0.5\pm2.9$ \\
      \Kst{}{1410}{+} & \ameson{0}{1450}{-} & $-1.8$ & $0.2\pm1.1$ \\
      \Kst{}{1410}{+} & \Kst{}{892}{+} & $1.7$ & $0.7\pm3.5$ \\
      \posswave & \ameson{2}{1320}{-} & $-1.6$ & $0.3\pm1.5$ \\
      \Kst{}{1410}{+} & \Kbarst{}{892}{0} & $1.36$ & $0.11\pm0.32$ \\
      \Kbarst{2}{1430}{0} & \ameson{0}{1450}{-} & $-1.3$ & $0.2\pm1.0$ \\
      \Kbarst{2}{1430}{0} & \Kst{}{1410}{+} & $-1.3$ & $0.2\pm0.5$ \\
      \Kst{}{892}{+} & \Kbarst{}{892}{0} & $1.2$ & $0.2\pm0.4$ \\
      \posswave & \rmeson{1450}{-} & $1.1$ & $0.3\pm0.9$ \\
    \end{tabular}
  \end{subtable}%
  \begin{subtable}{0.5\textwidth}
    \centering
    \caption{Best \lassc isobar model.}
    \begin{tabular}{l@{ $\times$ }l r@{$\,\pm\,$}l}
      \multicolumn{2}{c}{Resonances} & \multicolumn{2}{c}{Fit fraction [\%]} \\
      \midrule
      \neutswave & \Kst{}{892}{+} & $11.7$ & $0.5\pm1.5$ \\
      \neutswave & \Kst{}{1410}{+} & $-11.5$ & $1.7\pm3.4$ \\
      \posswave & \neutswave & $-10.8$ & $1.6\pm3.4$ \\
      \Kbarst{}{1410}{0} & \Kst{}{892}{+} & $-5.7$ & $0.4\pm0.8$ \\
      \posswave & \Kst{}{892}{+} & $5.6$ & $0.4\pm1.2$ \\
      \posswave & \Kbarst{}{1410}{0} & $5.2$ & $0.5\pm0.9$ \\
      \neutswave & \ameson{0}{980}{-} & $-5.1$ & $0.8\pm0.8$ \\
      \Kst{}{892}{+} & \rmeson{1450}{-} & $4.0$ & $0.3\pm1.2$ \\
      \Kst{}{1410}{+} & \Kbarst{}{1410}{0} & $-3.7$ & $0.4\pm2.0$ \\
      \neutswave & \rmeson{1450}{-} & $3.1$ & $0.4\pm0.9$ \\
      \Kst{}{892}{+} & \ameson{0}{1450}{-} & $2.7$ & $0.3\pm0.5$ \\
      \posswave & \rmeson{1450}{-} & $2.5$ & $0.2\pm0.8$ \\
      \posswave & \Kst{}{1410}{+} & $-2.5$ & $0.4\pm0.9$ \\
      \Kst{}{1410}{+} & \ameson{0}{1450}{-} & $-2.1$ & $0.2\pm0.9$ \\
      \Kst{}{1410}{+} & \Kbarst{}{892}{0} & $1.91$ & $0.12\pm0.14$ \\
      \Kst{}{1410}{+} & \ameson{0}{980}{-} & $-1.76$ & $0.27\pm0.28$ \\
      \posswave & \Kbarst{}{892}{0} & $-1.75$ & $0.10\pm0.19$ \\
      \posswave & \ameson{0}{1450}{-} & $1.7$ & $0.5\pm1.5$ \\
      \Kbarst{}{892}{0} & \ameson{0}{980}{-} & $-1.68$ & $0.14\pm0.29$ \\
      \Kbarst{}{1410}{0} & \Kbarst{}{892}{0} & $-1.7$ & $0.1\pm0.5$ \\
      \Kbarst{}{1410}{0} & \ameson{0}{980}{-} & $1.49$ & $0.16\pm0.21$ \\
      \Kst{}{1410}{+} & \Kst{}{892}{+} & $-1.5$ & $0.9\pm1.2$ \\
      \neutswave & \ameson{0}{1450}{-} & $1.3$ & $0.8\pm1.8$ \\
      \posswave & \ameson{2}{1320}{-} & $-1.2$ & $0.3\pm0.8$ \\
      \neutswave & \Kbarst{}{892}{0} & $1.21$ & $0.07\pm0.19$ \\
    \end{tabular}
  \end{subtable}%
  \label{tab:favinterference}
\end{sidewaystable}
\begin{sidewaystable}
  \caption{Interference fractions for the \decay{\Dz}{\KSKppim} mode. The first uncertainties are statistical and the second systematic. Only the 25 largest terms are shown.}
  \begin{subtable}{0.5\textwidth}
    \centering
    \caption{Best \glass isobar model.}
    \begin{tabular}{l@{ $\times$ }l r@{$\,\pm\,$}l}
      \multicolumn{2}{c}{Resonances} & \multicolumn{2}{c}{Fit fraction [\%]} \\
      \midrule
      \Kst{2}{1430}{0} & \ameson{0}{980}{+} & $10.3$ & $0.7\pm3.5$ \\
      \negswave & \neutswave & $6$ & $1\pm5$ \\
      \Kst{}{892}{-} & \ameson{0}{980}{+} & $-5$ & $1\pm4$ \\
      \Kst{}{1410}{0} & \Kst{}{892}{-} & $-5.0$ & $0.3\pm1.0$ \\
      \neutswave & \Kst{}{892}{-} & $5$ & $1\pm4$ \\
      \Kst{2}{1430}{0} & \Kst{}{1410}{0} & $-4.1$ & $0.7\pm2.2$ \\
      \Kst{}{1410}{0} & \ameson{0}{980}{+} & $3.8$ & $0.2\pm1.0$ \\
      \negswave & \Kst{2}{1430}{0} & $4$ & $1\pm7$ \\
      \Kst{}{892}{-} & \Kst{}{892}{0} & $3.61$ & $0.10\pm0.32$ \\
      \neutswave & \rmeson{1450}{+} & $3.4$ & $0.6\pm1.4$ \\
      \Kst{}{1410}{-} & \Kst{}{892}{-} & $-3.4$ & $0.4\pm0.6$ \\
      \Kst{2}{1430}{0} & \Kst{}{892}{-} & $3.2$ & $0.4\pm1.3$ \\
      \neutswave & \Kst{2}{1430}{0} & $-3.1$ & $1.2\pm1.7$ \\
      \Kst{2}{1430}{0} & \rmeson{1450}{+} & $-2.6$ & $0.5\pm1.6$ \\
      \Kst{}{1410}{0} & \rmeson{1450}{+} & $2.3$ & $0.4\pm0.8$ \\
      \negswave & \Kst{}{892}{-} & $1.9$ & $0.2\pm1.3$ \\
      \negswave & \Kst{}{1410}{0} & $-1.9$ & $0.6\pm2.6$ \\
      \Kst{}{892}{0} & \ameson{0}{980}{+} & $-1.8$ & $0.3\pm1.8$ \\
      \ameson{0}{1450}{+} & \ameson{0}{980}{+} & $1.7$ & $0.4\pm0.8$ \\
      \negswave & \Kst{}{1410}{-} & $1.7$ & $0.3\pm1.1$ \\
      \neutswave & \ameson{0}{980}{+} & $-1$ & $1\pm4$ \\
      \negswave & \rmeson{1450}{+} & $-1.4$ & $0.2\pm0.4$ \\
      \Kst{}{1410}{0} & \Kst{}{892}{0} & $-1.3$ & $0.2\pm0.8$ \\
      \negswave & \ameson{0}{1450}{+} & $1.33$ & $0.17\pm0.29$ \\
      \Kst{2}{1430}{0} & \rmeson{1700}{+} & $-1.3$ & $0.2\pm1.2$ \\
    \end{tabular}
  \end{subtable}%
  \begin{subtable}{0.5\textwidth}
    \centering
    \caption{Best \lassc isobar model.}
    \begin{tabular}{l@{ $\times$ }l r@{$\,\pm\,$}l}
      \multicolumn{2}{c}{Resonances} & \multicolumn{2}{c}{Fit fraction [\%]} \\
      \midrule
      \negswave & \ameson{0}{980}{+} & $11.5$ & $0.8\pm1.8$ \\
      \neutswave & \Kst{}{1410}{-} & $-11$ & $2\pm5$ \\
      \Kst{}{892}{-} & \ameson{0}{980}{+} & $9.6$ & $0.7\pm2.2$ \\
      \Kst{}{1410}{-} & \Kst{}{892}{-} & $-9.4$ & $0.7\pm1.0$ \\
      \ameson{0}{1450}{+} & \ameson{0}{980}{+} & $7.0$ & $0.6\pm1.1$ \\
      \neutswave & \ameson{0}{980}{+} & $-5.9$ & $1.2\pm2.6$ \\
      \negswave & \neutswave & $-5$ & $2\pm5$ \\
      \neutswave & \ameson{0}{1450}{+} & $-5.0$ & $0.6\pm1.3$ \\
      \negswave & \Kst{}{1410}{-} & $4.4$ & $0.5\pm0.9$ \\
      \Kst{}{892}{-} & \Kst{}{892}{0} & $3.73$ & $0.08\pm0.30$ \\
      \Kst{}{892}{0} & \ameson{0}{980}{+} & $3.7$ & $0.3\pm1.0$ \\
      \Kst{}{1410}{-} & \ameson{0}{980}{+} & $-3.6$ & $0.4\pm1.0$ \\
      \Kst{}{1410}{0} & \Kst{}{892}{-} & $-3.5$ & $0.5\pm0.9$ \\
      \neutswave & \Kst{}{1410}{0} & $-2.8$ & $0.3\pm1.1$ \\
      \negswave & \ameson{0}{1450}{+} & $2.6$ & $0.5\pm1.6$ \\
      \neutswave & \Kst{}{892}{0} & $-2.0$ & $0.1\pm0.4$ \\
      \Kst{}{1410}{-} & \ameson{0}{1450}{+} & $1.9$ & $0.4\pm0.6$ \\
      \negswave & \Kst{}{892}{0} & $-1.86$ & $0.13\pm0.18$ \\
      \Kst{}{892}{-} & \ameson{0}{1450}{+} & $1.6$ & $0.3\pm0.4$ \\
      \negswave & \Kst{}{892}{-} & $1.5$ & $0.3\pm1.3$ \\
      \Kst{}{1410}{-} & \rmeson{1700}{+} & $1.5$ & $0.2\pm0.5$ \\
      \Kst{}{892}{-} & \rmeson{1700}{+} & $-1.45$ & $0.16\pm0.34$ \\
      \Kst{}{1410}{-} & \Kst{}{892}{0} & $-1.4$ & $0.2\pm0.5$ \\
      \Kst{}{1410}{-} & \Kst{}{1410}{0} & $1.4$ & $0.3\pm0.6$ \\
      \Kst{}{1410}{0} & \Kst{}{892}{0} & $-1.4$ & $0.2\pm0.7$ \\
    \end{tabular}
  \end{subtable}%
  \label{tab:supinterference}
\end{sidewaystable}
\begin{table}
  \caption{Matrices $\mathbf{U}$ relating the fit coordinates $\mathbf{b'}$ to the \lassc form factor
           coordinates $\mathbf{b}=\mathbf{Ub'}$ defined in Sect.~\ref{sec:formalism}.}
  \begin{center}
    \begin{tabular}{c c}
      \chgswave & \neutswave \\[6pt]
      $\begin{pmatrix*}[r]%
        -0.460 & 0.702 & -0.543 \\%
        0.776 & 0.197 & -0.631 \\%
        -0.433 & -0.711 & -0.554 \\%
      \end{pmatrix*}$ & $\begin{pmatrix*}[r]%
        -0.452 & -0.676\phantom{0} & -0.582 \\%
        0.776 & 0.0243 & -0.631 \\%
        -0.440 & 0.737\phantom{0} & -0.513 \\%
      \end{pmatrix*}$ \\
    \end{tabular}
  \end{center}
  \label{tab:lassrotations}
\end{table}
\begin{figure}[b]
  \begin{center}
    \begin{tabular}{c c}
      \includegraphics[width=0.48\textwidth]{{{figs/fav_combinatorial}}} & \includegraphics[width=0.48\textwidth]{{{figs/sup_combinatorial}}} \\
    \end{tabular}
  \end{center}
  \caption{Smooth functions, $c_{\KSKpi}(\sKSK, \sKSpi)$, used to describe the combinatorial
           background component in the \decay{\Dz}{\KSKmpip} (left) and \decay{\Dz}{\KSKppim}
           (right) amplitude model fits.}
  \label{fig:2dcomb}
\end{figure}
\begin{table}
  \caption{Additional fit parameters for \glass models. This table does not include parameters
           that are fixed to their nominal values. The first uncertainties are statistical and the second systematic.}
  \begin{center}
    \begin{tabular}{l l r@{$\,\pm\,$}l l}
      \multicolumn{2}{c}{Parameter} & \multicolumn{2}{c}{Value} \\
      \midrule
      \multirow{2}{*}{\Kst{}{892}{\pm}} & $m_{\PR}$ & $893.1$ & $0.1\pm0.9$ & \mevcc \\
       & $\Gamma_{\PR}$ & $46.9$ & $0.3\pm2.5$ & \mevcc \\
      \cmidrule{2-5}
      \multirow{1}{*}{\Kst{}{1410}{\pm}} & $\Gamma_{\PR}$ & $210$ & $20\pm60$ & \mevcc \\
      \cmidrule{2-5}
      \multirow{5}{*}{\chgswave} & $F$ & \multicolumn{2}{c}{1.785 (fixed)} \\
	   & $a$ & $4.7$ & $0.4\pm1.0$ & \invgevc \\
       & $\phi_F$ & $0.28$ & $0.05\pm0.19$ & \rad \\
       & $\phi_S$ & $2.8$ & $0.2\pm0.5$ & \rad \\
       & $r$ & $-5.3$ & $0.4\pm1.9$ & \invgevc \\
      \cmidrule{2-5}
      \multirow{2}{*}{\Kst{}{1410}{0}} & $m_{\PR}$ & $1426$ & $8\pm24$ & \mevcc \\
       & $\Gamma_{\PR}$ & $270$ & $20\pm40$ & \mevcc \\
      \cmidrule{2-5}
      \multirow{5}{*}{\neutswave} & $F$ & $0.15$ & $0.03\pm0.14$ &  \\
       & $a$ & $4.2$ & $0.3\pm2.8$ & \invgevc \\
       & $\phi_F$ & $-2.5$ & $0.2\pm1.0$ & \rad \\
       & $\phi_S$ & $-1.1$ & $0.6\pm1.3$ & \rad \\
       & $r$ & $-3.0$ & $0.4\pm1.7$ & \invgevc \\
      \cmidrule{2-5}
      \multirow{1}{*}{\ameson{0}{1450}{\pm}} & $m_{\PR}$ & $1430$ & $10\pm40$ & \mevcc \\
      \cmidrule{2-5}
      \multirow{1}{*}{\rmeson{1450}{\pm}} & $\Gamma_{\PR}$ & $410$ & $19\pm35$ & \mevcc \\
      \cmidrule{2-5}
      \multirow{1}{*}{\rmeson{1700}{\pm}} & $m_{\PR}$ & $1530$ & $10\pm40$ & \mevcc \\
    \end{tabular}
  \end{center}
  \label{tab:extraglassparameters}
\end{table}
\begin{table}
  \caption{Additional fit parameters for \lassc models. This table does not include parameters that are fixed to their nominal values.
           The first uncertainties are statistical and the second systematic.}
  \begin{center}
    \begin{tabular}{l l r@{$\,\pm\,$}l l}
      \multicolumn{2}{c}{Parameter} & \multicolumn{2}{c}{Value} \\
      \midrule
      \multirow{2}{*}{\Kst{}{892}{\pm}} & $m_{\PR}$ & $893.4$ & $0.1\pm1.1$ & \mevcc \\
       & $\Gamma_{\PR}$ & $47.4$ & $0.3\pm2.0$ & \mevcc \\
      \cmidrule{2-5}
      \multirow{1}{*}{\Kst{}{1410}{\pm}} & $m_{\PR}$ & $1437$ & $8\pm16$ & \mevcc \\
      \cmidrule{2-5}
      \multirow{3}{*}{\chgswave} & $b'_1$ & $60$ & $30\pm40$ &  \\
       & $b'_2$ & $4$ & $1\pm5$ &  \\
       & $b'_3$ & $3.0$ & $0.2\pm0.7$ &  \\
      \cmidrule{2-5}
      \multirow{1}{*}{\Kst{}{1410}{0}} & $m_{\PR}$ & $1404$ & $9\pm22$ & \mevcc \\
      \cmidrule{2-5}
      \multirow{3}{*}{\neutswave} & $b'_1$ & $130$ & $30\pm80$ &  \\
       & $b'_2$ & $-6$ & $1\pm14$ &  \\
       & $b'_3$ & $2.5$ & $0.1\pm1.4$ &  \\
      \cmidrule{2-5}
      \multirow{1}{*}{\kpswave} & $r$ & $1.2$ & $0.3\pm0.4$ & \invgevc \\
      \cmidrule{2-5}
      \multirow{1}{*}{\ameson{0}{980}{\pm}} & $m_{\PR}$ & $925$ & $5\pm8$ & \mevcc \\
      \cmidrule{2-5}
      \multirow{2}{*}{\ameson{0}{1450}{\pm}} & $m_{\PR}$ & $1458$ & $14\pm15$ & \mevcc \\
       & $\Gamma_{\PR}$ & $282$ & $12\pm13$ & \mevcc \\
      \cmidrule{2-5}
      \multirow{1}{*}{\rmeson{1450}{\pm}} & $m_{\PR}$ & $1208$ & $8\pm9$ & \mevcc \\
      \cmidrule{2-5}
      \multirow{1}{*}{\rmeson{1700}{\pm}} & $m_{\PR}$ & $1552$ & $13\pm26$ & \mevcc \\
    \end{tabular}
  \end{center}
  \label{tab:extralassparameters}
\end{table}

The changes in \ntll obtained in alternative models where one \rhomeson contribution
is removed are given in Table~\ref{tab:norhontll}.
\begin{table}
  \caption{Change in \ntll value when removing a \rhomeson resonance from
           one of the models.}
  \begin{center}
    \begin{tabular}{l l l c}
      \kpswave  & & Removed & \\
      parameterization & Decay mode & resonance & $\Delta(\ntll)$ \\
      \midrule
      \multirow{2}{*}{\lassc} & \decay{\Dz}{\KSKmpip} & \rmeson{1450}{-} & 338 \\
      & \decay{\Dz}{\KSKppim} & \rmeson{1700}{+} & 235 \\
      \cmidrule{2-4}
      \multirow{4}{*}{\glass} & \multirow{2}{*}{\decay{\Dz}{\KSKmpip}} & \rmeson{1450}{-} & 238 \\
      & & \rmeson{1700}{-} & 162 \\
      \cmidrule{3-4}
      & \multirow{2}{*}{\decay{\Dz}{\KSKmpip}} & \rmeson{1450}{+} & 175 \\
      & & \rmeson{1700}{+} & 233 \\
    \end{tabular}
  \end{center}
  \label{tab:norhontll}
\end{table}
\FloatBarrier
\section{Systematic uncertainty tables}
\label{sec:app_systematics}
This appendix includes tables summarizing the various contributions to the systematic
uncertainties assigned to the various results.
The table headings correspond to the uncertainties discussed in Sect.~\ref{sec:systematics}
with some abbreviations to allow the tables to be typeset compactly. Definitions of the
various abbreviations are given in Table~\ref{sec:systematics_abbreviations}.
The quantity `DFF' listed in the tables is the sum of fit fractions from the various
resonances, excluding interference terms.
Tables~\ref{tab:syst_glass_fav} (\glass)
and~\ref{tab:syst_lass_fav} (\lassc) show the results for the complex amplitudes and
fit fractions in the \decay{\Dz}{\KSKmpip} models, Tables~\ref{tab:syst_glass_sup} (\glass)
and~\ref{tab:syst_lass_sup} (\lassc) show the corresponding values for the
\decay{\Dz}{\KSKppim} models and Tables~\ref{tab:syst_glass} and \ref{tab:syst_lass}
summarize the uncertainties for the parameters that are not specific to a decay mode.

In each of these tables the parameter in question is listed on the left, followed by the central
value and the corresponding statistical (first) and systematic (second) uncertainty.
The subsequent columns list the contributions to this systematic uncertainty, and are
approximately ordered in decreasing order of significance from left to right.
\begin{table}
	\caption{Listing of abbreviations required to typeset the systematic uncertainty tables.}
	\begin{center}
		\begin{tabular}{l p{0.8\textwidth}}
			Abbreviation & Description \\
			\midrule
			$\max(|\cos|)$ & Variation of the cut that excludes the boundary regions of the Dalitz plot. \\
			Efficiency & Two efficiency modelling uncertainties added in quadrature: using an alternative parameterization, and accounting for the
							    limited size of the simulated event sample. \\
			Joint & Uncertainty obtained by simultaneously fitting disjoint sub-sets of the dataset, separated by the year of data-taking and type
                                of \KS daughter track, with distinct efficiency models. \\			
			Weights & Three uncertainties related to the re-weighting of simulated events used to generate the efficiency model \efffn, added in quadrature.
							 These account for: incorrect simulation of the underlying $\proton\proton$ interaction, uncertainty in the relative yield of long
							 and downstream \KS candidates, and uncertainty in the efficiency of selection requirements using information from the
							 RICH detectors. \\
			Comb. & Using an alternative combinatorial background model. \\
			\ntll & Using a more complex alternative model where the threshold in $\Delta(\ntll)$ for a resonance to be retained is reduced to $9$ units. \\
			\flatte & Variation of the \flatte lineshape parameters for the \ameson{0}{980}{\pm} resonance according to their nominal uncertainties. \\
			$f_{\text{m}}, f_{\text{c}}$ & Variation of the mistag and combinatorial background rates according to their uncertainties in the mass fit. \\
			$d_{\Dz}, d_{\PR}$ & Variation of the meson radius parameters. \\
			$T_{\rhomeson^{\pm}}$ & Switching to a Breit-Wigner dynamical function to describe the \rmeson{1450,1700}{\pm} resonances. \\
		\end{tabular}
	\end{center}
	\label{sec:systematics_abbreviations}
\end{table}
\begin{sidewaystable}
  \caption{Systematic uncertainties for complex amplitudes and fit fractions in the \decay{\Dz}{\KSKmpip} model using the \glass parameterization. The headings are defined in Table~\ref{sec:systematics_abbreviations}.}
  \vspace{-8mm}
  \footnotesize
  \begin{center}
    \begin{adjustbox}{width=\textheight,totalheight=0.7\textwidth,keepaspectratio}
      \begin{tabular}{l l r@{$\,\pm\,$}l l l l l l l l l l l l}
        Resonance & Var & \multicolumn{2}{c}{Baseline} & \rothead{$d_{\Dz}, d_{\PR}$} & \rothead{$\max(|\cos|)$} & \rothead{Comb.} & \rothead{$T_{\rhomeson^{\pm}}$} & \rothead{\ntll} & \rothead{Weights} & \rothead{Efficiency} & \rothead{\flatte} & \rothead{$f_{\text{m}}, f_{\text{c}}$} & \rothead{Joint} \\
        \midrule
        \multirow{1}{*}{\Kst{}{892}{+}} &FF [\%] & $57.0$ & $0.8\pm2.6$ & $1.76$ & $1.48$ & $0.56$ & $0.37$ & $0.23$ & $0.13$ & $0.66$ & $0.03$ & $0.14$ & $0.75$ &  \\
        \cmidrule{2-14}
        \multirow{3}{*}{\Kst{}{1410}{+}} &\aR & $4.3$ & $0.3\pm0.7$ & $0.39$ & $0.32$ & $0.16$ & $0.18$ & $0.30$ & $0.14$ & $0.22$ & $0.01$ & $0.04$ & $0.06$ &  \\
         &\pR $(^{\circ})$ & $-160$ & $6\pm24$ & $6.7$ & $11.3$ & $13.7$ & $14.4$ & $0.4$ & $0.2$ & $3.0$ & $1.4$ & $1.5$ & $0.3$ &  \\
         &FF [\%] & $5$ & $1\pm4$ & $4.08$ & $1.01$ & $0.84$ & $0.84$ & $0.65$ & $0.30$ & $0.36$ & $0.05$ & $0.08$ & $0.14$ &  \\
        \cmidrule{2-14}
        \multirow{3}{*}{\posswave} &\aR & $0.62$ & $0.05\pm0.18$ & $0.12$ & $0.08$ & $0.07$ & $0.05$ & $0.03$ & $0.02$ & $0.02$ & $0.03$ & $0.02$ & $0.01$ &  \\
         &\pR $(^{\circ})$ & $-67$ & $5\pm15$ & $12.2$ & $2.8$ & $6.7$ & $2.2$ & $1.4$ & $0.8$ & $2.4$ & $2.6$ & $1.6$ & $0.5$ &  \\
         &FF [\%] & $12$ & $2\pm9$ & $4.61$ & $4.14$ & $4.26$ & $4.17$ & $2.33$ & $1.26$ & $1.39$ & $1.39$ & $1.04$ & $0.43$ &  \\
        \cmidrule{2-14}
        \multirow{3}{*}{\Kbarst{}{892}{0}} &\aR & $0.213$ & $0.007\pm0.018$ & $0.01$ & $0.01$ & $0.00$ & $0.01$ & $0.00$ & $0.00$ & $0.00$ & $0.00$ & $0.00$ & $0.00$ &  \\
         &\pR $(^{\circ})$ & $-108$ & $2\pm4$ & $2.4$ & $1.0$ & $1.5$ & $0.6$ & $1.3$ & $0.2$ & $1.8$ & $0.2$ & $0.2$ & $0.2$ &  \\
         &FF [\%] & $2.5$ & $0.2\pm0.4$ & $0.19$ & $0.26$ & $0.07$ & $0.16$ & $0.07$ & $0.07$ & $0.05$ & $0.05$ & $0.06$ & $0.06$ &  \\
        \cmidrule{2-14}
        \multirow{3}{*}{\Kbarst{}{1410}{0}} &\aR & $6.0$ & $0.3\pm0.5$ & $0.16$ & $0.17$ & $0.00$ & $0.06$ & $0.44$ & $0.07$ & $0.08$ & $0.09$ & $0.07$ & $0.00$ &  \\
         &\pR $(^{\circ})$ & $-179$ & $4\pm17$ & $4.8$ & $9.1$ & $6.7$ & $9.1$ & $6.8$ & $2.2$ & $2.1$ & $0.9$ & $1.4$ & $0.7$ &  \\
         &FF [\%] & $9$ & $1\pm4$ & $3.15$ & $0.86$ & $0.30$ & $0.61$ & $1.66$ & $0.30$ & $0.46$ & $0.23$ & $0.19$ & $0.28$ &  \\
        \cmidrule{2-14}
        \multirow{3}{*}{\Kbarst{2}{1430}{0}} &\aR & $3.2$ & $0.3\pm1.0$ & $0.44$ & $0.57$ & $0.29$ & $0.42$ & $0.34$ & $0.16$ & $0.17$ & $0.02$ & $0.03$ & $0.00$ &  \\
         &\pR $(^{\circ})$ & $-172$ & $5\pm23$ & $17.2$ & $7.8$ & $1.4$ & $0.6$ & $10.5$ & $5.7$ & $3.8$ & $1.1$ & $1.2$ & $1.9$ &  \\
         &FF [\%] & $3.4$ & $0.6\pm2.7$ & $1.99$ & $1.14$ & $0.59$ & $0.84$ & $0.68$ & $0.36$ & $0.32$ & $0.05$ & $0.07$ & $0.13$ &  \\
        \cmidrule{2-14}
        \multirow{3}{*}{\neutswave} &\aR & $2.5$ & $0.2\pm1.3$ & $0.90$ & $0.18$ & $0.60$ & $0.56$ & $0.35$ & $0.13$ & $0.16$ & $0.08$ & $0.04$ & $0.06$ &  \\
         &\pR $(^{\circ})$ & $50$ & $10\pm80$ & $71.0$ & $13.6$ & $5.5$ & $2.6$ & $5.4$ & $21.9$ & $10.7$ & $13.9$ & $8.8$ & $0.8$ &  \\
         &FF [\%] & $11$ & $2\pm10$ & $5.28$ & $4.42$ & $4.67$ & $4.49$ & $0.13$ & $1.27$ & $0.76$ & $0.98$ & $0.44$ & $0.29$ &  \\
        \cmidrule{2-14}
        \multirow{3}{*}{\ameson{2}{1320}{-}} &\aR & $0.19$ & $0.03\pm0.09$ & $0.01$ & $0.06$ & $0.04$ & $0.04$ & $0.00$ & $0.01$ & $0.03$ & $0.00$ & $0.00$ & $0.00$ &  \\
         &\pR $(^{\circ})$ & $-129$ & $8\pm17$ & $5.7$ & $8.5$ & $5.8$ & $8.7$ & $5.8$ & $2.5$ & $3.4$ & $1.3$ & $1.6$ & $1.3$ &  \\
         &FF [\%] & $0.20$ & $0.06\pm0.21$ & $0.06$ & $0.14$ & $0.08$ & $0.10$ & $0.00$ & $0.01$ & $0.06$ & $0.01$ & $0.00$ & $0.00$ &  \\
        \cmidrule{2-14}
        \multirow{3}{*}{\ameson{0}{1450}{-}} &\aR & $0.52$ & $0.04\pm0.15$ & $0.01$ & $0.08$ & $0.07$ & $0.07$ & $0.03$ & $0.05$ & $0.04$ & $0.02$ & $0.01$ & $0.01$ &  \\
         &\pR $(^{\circ})$ & $-82$ & $7\pm31$ & $3.7$ & $12.1$ & $19.9$ & $18.0$ & $3.2$ & $3.6$ & $8.5$ & $0.9$ & $0.9$ & $1.6$ &  \\
         &FF [\%] & $1.2$ & $0.2\pm0.6$ & $0.00$ & $0.33$ & $0.24$ & $0.27$ & $0.15$ & $0.24$ & $0.14$ & $0.07$ & $0.06$ & $0.03$ &  \\
        \cmidrule{2-14}
        \multirow{3}{*}{\rmeson{1450}{-}} &\aR & $1.6$ & $0.2\pm0.5$ & $0.06$ & $0.15$ & $0.03$ & $0.35$ & $0.28$ & $0.07$ & $0.11$ & $0.04$ & $0.01$ & $0.04$ &  \\
         &\pR $(^{\circ})$ & $-177$ & $7\pm32$ & $2.5$ & $15.7$ & $15.9$ & $19.5$ & $8.3$ & $4.1$ & $4.7$ & $2.2$ & $1.9$ & $1.5$ &  \\
         &FF [\%] & $1.3$ & $0.3\pm0.7$ & $0.21$ & $0.28$ & $0.04$ & $0.27$ & $0.32$ & $0.13$ & $0.22$ & $0.08$ & $0.03$ & $0.24$ &  \\
        \cmidrule{2-14}
        \multirow{3}{*}{\rmeson{1700}{-}} &\aR & $0.38$ & $0.08\pm0.15$ & $0.11$ & $0.03$ & $0.06$ & $0.02$ & $0.06$ & $0.01$ & $0.03$ & $0.01$ & $0.00$ & $0.02$ &  \\
         &\pR $(^{\circ})$ & $-70$ & $10\pm60$ & $50.1$ & $28.3$ & $17.0$ & $17.6$ & $6.1$ & $7.3$ & $5.6$ & $3.2$ & $1.0$ & $0.8$ &  \\
         &FF [\%] & $0.12$ & $0.05\pm0.14$ & $0.12$ & $0.01$ & $0.04$ & $0.00$ & $0.02$ & $0.01$ & $0.02$ & $0.00$ & $0.00$ & $0.04$ &  \\
        \midrule
        \chisqbin & & \multicolumn{2}{c}{1.12} & $1.12$ & $1.08$ & $1.11$ & $1.10$ & $1.11$ & $-$ & $-$ & $1.11$ & $1.11$ & $1.20$ \\
        DFF [\%] & & \multicolumn{2}{c}{103.0} & $103.7$ & $110.6$ & $111.7$ & $111.8$ & $103.7$ & $-$ & $-$ & $101.0$ & $102.1$ & $102.3$ \\
      \end{tabular}
    \end{adjustbox}
  \end{center}
  \label{tab:syst_glass_fav}
\end{sidewaystable}
\begin{sidewaystable}
 \caption{Systematic uncertainties for complex amplitudes and fit fractions in the \decay{\Dz}{\KSKmpip} model using the \lassc parameterization. The headings are defined in Table~\ref{sec:systematics_abbreviations}.}
 \vspace{-8mm}
 \footnotesize
 \begin{center}
 \begin{adjustbox}{width=\textheight,totalheight=0.7\textwidth,keepaspectratio}
  \begin{tabular}{l l r@{$\,\pm\,$}l l l l l l l l l l l l}
    Resonance & Var & \multicolumn{2}{c}{Baseline} & \rothead{$d_{\Dz}, d_{\PR}$} & \rothead{\ntll} & \rothead{Efficiency} & \rothead{$\max(|\cos|)$} & \rothead{Weights} & \rothead{Comb.} & \rothead{\flatte} & \rothead{$T_{\rhomeson^{\pm}}$} & \rothead{$f_{\text{m}}, f_{\text{c}}$} & \rothead{Joint} \\
    \midrule
    \multirow{1}{*}{\Kst{}{892}{+}} &FF [\%] & $56.9$ & $0.6\pm1.1$ & $0.03$ & $0.10$ & $0.40$ & $0.10$ & $0.19$ & $0.07$ & $0.06$ & $0.16$ & $0.17$ & $1.00$ &  \\
    \cmidrule{2-14}
    \multirow{3}{*}{\Kst{}{1410}{+}} &\aR & $5.83$ & $0.29\pm0.29$ & $0.04$ & $0.02$ & $0.26$ & $0.05$ & $0.06$ & $0.02$ & $0.06$ & $0.02$ & $0.05$ & $0.06$ &  \\
     &\pR $(^{\circ})$ & $-143$ & $3\pm6$ & $1.9$ & $0.2$ & $3.7$ & $2.8$ & $1.6$ & $1.6$ & $0.7$ & $1.1$ & $0.1$ & $0.1$ &  \\
     &FF [\%] & $9.6$ & $1.1\pm2.9$ & $2.79$ & $0.06$ & $0.80$ & $0.22$ & $0.15$ & $0.14$ & $0.23$ & $0.02$ & $0.18$ & $0.21$ &  \\
    \cmidrule{2-14}
    \multirow{3}{*}{\posswave} &\aR & $1.13$ & $0.09\pm0.21$ & $0.11$ & $0.11$ & $0.03$ & $0.09$ & $0.06$ & $0.06$ & $0.04$ & $0.04$ & $0.03$ & $0.02$ &  \\
     &\pR $(^{\circ})$ & $-59$ & $4\pm13$ & $10.6$ & $6.2$ & $3.2$ & $0.9$ & $0.6$ & $0.1$ & $1.3$ & $0.5$ & $0.8$ & $0.3$ &  \\
     &FF [\%] & $11.7$ & $1.0\pm2.3$ & $0.68$ & $0.53$ & $0.35$ & $1.35$ & $0.87$ & $0.90$ & $0.33$ & $0.71$ & $0.30$ & $0.56$ &  \\
    \cmidrule{2-14}
    \multirow{3}{*}{\Kbarst{}{892}{0}} &\aR & $0.210$ & $0.006\pm0.010$ & $0.01$ & $0.00$ & $0.00$ & $0.01$ & $0.00$ & $0.00$ & $0.00$ & $0.00$ & $0.00$ & $0.00$ &  \\
     &\pR $(^{\circ})$ & $-101.5$ & $2.0\pm2.8$ & $0.2$ & $1.8$ & $2.0$ & $0.0$ & $0.2$ & $0.4$ & $0.3$ & $0.4$ & $0.2$ & $0.1$ &  \\
     &FF [\%] & $2.47$ & $0.15\pm0.23$ & $0.16$ & $0.00$ & $0.06$ & $0.10$ & $0.03$ & $0.09$ & $0.03$ & $0.03$ & $0.04$ & $0.03$ &  \\
    \cmidrule{2-14}
    \multirow{3}{*}{\Kbarst{}{1410}{0}} &\aR & $3.9$ & $0.2\pm0.4$ & $0.13$ & $0.07$ & $0.24$ & $0.24$ & $0.17$ & $0.05$ & $0.07$ & $0.05$ & $0.05$ & $0.09$ &  \\
     &\pR $(^{\circ})$ & $-174$ & $4\pm7$ & $0.4$ & $3.5$ & $2.3$ & $3.3$ & $4.0$ & $1.9$ & $0.4$ & $0.4$ & $0.3$ & $0.2$ &  \\
     &FF [\%] & $3.8$ & $0.5\pm2.0$ & $1.75$ & $0.38$ & $0.70$ & $0.38$ & $0.08$ & $0.00$ & $0.15$ & $0.01$ & $0.10$ & $0.45$ &  \\
    \cmidrule{2-14}
    \multirow{3}{*}{\neutswave} &\aR & $1.28$ & $0.12\pm0.23$ & $0.10$ & $0.07$ & $0.13$ & $0.07$ & $0.10$ & $0.06$ & $0.03$ & $0.03$ & $0.03$ & $0.01$ &  \\
     &\pR $(^{\circ})$ & $75$ & $3\pm8$ & $3.7$ & $3.7$ & $3.1$ & $2.5$ & $3.0$ & $1.6$ & $1.2$ & $1.4$ & $0.9$ & $0.0$ &  \\
     &FF [\%] & $18$ & $2\pm4$ & $2.36$ & $0.57$ & $1.97$ & $1.10$ & $1.08$ & $0.54$ & $0.30$ & $0.72$ & $0.22$ & $1.58$ &  \\
    \cmidrule{2-14}
    \multirow{3}{*}{\ameson{0}{980}{-}} &\aR & $1.07$ & $0.09\pm0.14$ & $0.12$ & $0.02$ & $0.07$ & $0.02$ & $0.02$ & $0.01$ & $0.01$ & $0.02$ & $0.01$ & $0.02$ &  \\
     &\pR $(^{\circ})$ & $82$ & $5\pm7$ & $2.5$ & $2.1$ & $4.6$ & $1.2$ & $0.9$ & $0.9$ & $3.0$ & $0.6$ & $0.6$ & $0.5$ &  \\
     &FF [\%] & $4.0$ & $0.7\pm1.1$ & $0.91$ & $0.19$ & $0.50$ & $0.06$ & $0.20$ & $0.03$ & $0.17$ & $0.18$ & $0.14$ & $0.05$ &  \\
    \cmidrule{2-14}
    \multirow{3}{*}{\ameson{2}{1320}{-}} &\aR & $0.17$ & $0.03\pm0.05$ & $0.02$ & $0.01$ & $0.04$ & $0.00$ & $0.01$ & $0.01$ & $0.00$ & $0.01$ & $0.00$ & $0.01$ &  \\
     &\pR $(^{\circ})$ & $-128$ & $10\pm8$ & $3.9$ & $3.9$ & $2.9$ & $2.8$ & $1.9$ & $2.5$ & $1.0$ & $0.5$ & $1.1$ & $1.4$ &  \\
     &FF [\%] & $0.15$ & $0.06\pm0.13$ & $0.09$ & $0.02$ & $0.08$ & $0.01$ & $0.02$ & $0.01$ & $0.01$ & $0.01$ & $0.01$ & $0.02$ &  \\
    \cmidrule{2-14}
    \multirow{3}{*}{\ameson{0}{1450}{-}} &\aR & $0.43$ & $0.05\pm0.10$ & $0.07$ & $0.06$ & $0.03$ & $0.02$ & $0.02$ & $0.00$ & $0.00$ & $0.01$ & $0.01$ & $0.01$ &  \\
     &\pR $(^{\circ})$ & $-49$ & $11\pm19$ & $3.0$ & $12.5$ & $8.2$ & $5.7$ & $7.1$ & $1.2$ & $4.0$ & $2.2$ & $4.1$ & $0.5$ &  \\
     &FF [\%] & $0.74$ & $0.15\pm0.34$ & $0.25$ & $0.18$ & $0.10$ & $0.07$ & $0.08$ & $0.00$ & $0.00$ & $0.04$ & $0.01$ & $0.01$ &  \\
    \cmidrule{2-14}
    \multirow{3}{*}{\rmeson{1450}{-}} &\aR & $1.3$ & $0.1\pm0.4$ & $0.27$ & $0.05$ & $0.06$ & $0.01$ & $0.05$ & $0.01$ & $0.03$ & $0.24$ & $0.01$ & $0.00$ &  \\
     &\pR $(^{\circ})$ & $-144$ & $7\pm9$ & $1.5$ & $6.0$ & $3.2$ & $2.8$ & $2.2$ & $3.3$ & $1.0$ & $3.7$ & $1.0$ & $0.8$ &  \\
     &FF [\%] & $1.4$ & $0.2\pm0.7$ & $0.63$ & $0.21$ & $0.13$ & $0.06$ & $0.12$ & $0.05$ & $0.07$ & $0.10$ & $0.04$ & $0.07$ &  \\
    \midrule
    \chisqbin & & \multicolumn{2}{c}{1.10} & $1.11$ & $1.10$ & $-$ & $1.08$ & $-$ & $1.10$ & $1.10$ & $1.10$ & $1.10$ & $1.17$ \\
    DFF [\%] & & \multicolumn{2}{c}{109.1} & $114.1$ & $108.5$ & $-$ & $107.0$ & $-$ & $107.7$ & $108.7$ & $110.8$ & $109.4$ & $110.0$ \\
  \end{tabular}
  \end{adjustbox}
  \end{center}
  \label{tab:syst_lass_fav}
\end{sidewaystable}
\begin{sidewaystable}
  \caption{Systematic uncertainties for complex amplitudes and fit fractions in the \decay{\Dz}{\KSKppim} model using the \glass parameterization. The headings are defined in Table~\ref{sec:systematics_abbreviations}.}
  \vspace{-8mm}
  \footnotesize
  \begin{center}
  \begin{adjustbox}{width=\textheight,totalheight=0.7\textwidth,keepaspectratio}
  \begin{tabular}{l l r@{$\,\pm\,$}l l l l l l l l l l l l}
    Resonance & Var & \multicolumn{2}{c}{Baseline} & \rothead{$d_{\Dz}, d_{\PR}$} & \rothead{\ntll} & \rothead{Comb.} & \rothead{$\max(|\cos|)$} & \rothead{$T_{\rhomeson^{\pm}}$} & \rothead{Efficiency} & \rothead{Weights} & \rothead{\flatte} & \rothead{$f_{\text{m}}, f_{\text{c}}$} & \rothead{Joint} \\
    \midrule
    \multirow{1}{*}{\Kst{}{892}{-}} &FF [\%] & $29.5$ & $0.6\pm1.6$ & $1.30$ & $0.15$ & $0.32$ & $0.49$ & $0.27$ & $0.25$ & $0.41$ & $0.26$ & $0.25$ & $0.47$ &  \\
    \cmidrule{2-14}
    \multirow{3}{*}{\Kst{}{1410}{-}} &\aR & $4.7$ & $0.5\pm1.1$ & $0.30$ & $0.81$ & $0.29$ & $0.28$ & $0.45$ & $0.22$ & $0.09$ & $0.06$ & $0.04$ & $0.05$ &  \\
     &\pR $(^{\circ})$ & $-106$ & $6\pm25$ & $6.3$ & $18.1$ & $11.0$ & $8.0$ & $7.4$ & $5.3$ & $0.9$ & $1.9$ & $1.5$ & $0.6$ &  \\
     &FF [\%] & $3.1$ & $0.6\pm1.6$ & $1.13$ & $0.99$ & $0.13$ & $0.22$ & $0.36$ & $0.21$ & $0.11$ & $0.06$ & $0.07$ & $0.11$ &  \\
    \cmidrule{2-14}
    \multirow{3}{*}{\negswave} &\aR & $0.58$ & $0.05\pm0.11$ & $0.07$ & $0.07$ & $0.01$ & $0.01$ & $0.05$ & $0.02$ & $0.02$ & $0.00$ & $0.01$ & $0.00$ &  \\
     &\pR $(^{\circ})$ & $-164$ & $6\pm31$ & $19.0$ & $2.0$ & $13.2$ & $13.0$ & $12.3$ & $7.7$ & $1.8$ & $3.2$ & $2.5$ & $0.2$ &  \\
     &FF [\%] & $5.4$ & $0.9\pm1.7$ & $1.16$ & $0.94$ & $0.33$ & $0.59$ & $0.06$ & $0.33$ & $0.23$ & $0.11$ & $0.14$ & $0.12$ &  \\
    \cmidrule{2-14}
    \multirow{3}{*}{\Kst{}{892}{0}} &\aR & $0.410$ & $0.010\pm0.021$ & $0.01$ & $0.01$ & $0.01$ & $0.00$ & $0.00$ & $0.00$ & $0.01$ & $0.00$ & $0.00$ & $0.01$ &  \\
     &\pR $(^{\circ})$ & $176$ & $2\pm9$ & $4.4$ & $2.4$ & $4.4$ & $4.7$ & $3.6$ & $2.4$ & $0.2$ & $0.4$ & $0.3$ & $0.2$ &  \\
     &FF [\%] & $4.82$ & $0.23\pm0.35$ & $0.02$ & $0.27$ & $0.15$ & $0.08$ & $0.08$ & $0.08$ & $0.06$ & $0.04$ & $0.02$ & $0.03$ &  \\
    \cmidrule{2-14}
    \multirow{3}{*}{\Kst{}{1410}{0}} &\aR & $6.2$ & $0.5\pm1.4$ & $1.13$ & $0.09$ & $0.38$ & $0.22$ & $0.07$ & $0.17$ & $0.54$ & $0.26$ & $0.13$ & $0.03$ &  \\
     &\pR $(^{\circ})$ & $175$ & $4\pm14$ & $0.9$ & $13.4$ & $1.9$ & $2.9$ & $1.4$ & $1.1$ & $1.2$ & $1.8$ & $0.2$ & $0.7$ &  \\
     &FF [\%] & $5.2$ & $0.7\pm1.6$ & $0.25$ & $0.37$ & $0.83$ & $0.62$ & $0.32$ & $0.24$ & $0.90$ & $0.44$ & $0.22$ & $0.09$ &  \\
    \cmidrule{2-14}
    \multirow{3}{*}{\Kst{2}{1430}{0}} &\aR & $6.3$ & $0.5\pm1.7$ & $1.17$ & $0.82$ & $0.72$ & $0.35$ & $0.32$ & $0.13$ & $0.23$ & $0.27$ & $0.15$ & $0.09$ &  \\
     &\pR $(^{\circ})$ & $-139$ & $5\pm21$ & $16.2$ & $3.8$ & $6.5$ & $7.4$ & $4.8$ & $2.7$ & $4.4$ & $2.5$ & $2.2$ & $0.3$ &  \\
     &FF [\%] & $7$ & $1\pm4$ & $2.84$ & $1.66$ & $1.60$ & $0.71$ & $0.69$ & $0.30$ & $0.37$ & $0.55$ & $0.25$ & $0.41$ &  \\
    \cmidrule{2-14}
    \multirow{3}{*}{\neutswave} &\aR & $3.7$ & $0.3\pm1.8$ & $1.46$ & $0.65$ & $0.55$ & $0.03$ & $0.45$ & $0.12$ & $0.11$ & $0.09$ & $0.03$ & $0.01$ &  \\
     &\pR $(^{\circ})$ & $100$ & $10\pm70$ & $57.6$ & $13.1$ & $9.0$ & $13.6$ & $3.2$ & $10.6$ & $20.4$ & $11.9$ & $7.0$ & $0.4$ &  \\
     &FF [\%] & $12$ & $1\pm8$ & $6.60$ & $0.83$ & $2.75$ & $2.68$ & $2.20$ & $0.54$ & $0.92$ & $1.00$ & $0.22$ & $0.15$ &  \\
    \cmidrule{2-14}
    \multirow{3}{*}{\ameson{0}{980}{+}} &\aR & $1.8$ & $0.1\pm0.6$ & $0.25$ & $0.44$ & $0.11$ & $0.14$ & $0.06$ & $0.10$ & $0.04$ & $0.20$ & $0.03$ & $0.01$ &  \\
     &\pR $(^{\circ})$ & $64$ & $5\pm24$ & $14.6$ & $14.9$ & $5.9$ & $1.7$ & $2.4$ & $3.2$ & $6.1$ & $6.7$ & $1.7$ & $0.3$ &  \\
     &FF [\%] & $11$ & $1\pm6$ & $2.43$ & $4.86$ & $1.42$ & $1.75$ & $0.81$ & $1.15$ & $0.68$ & $1.40$ & $0.36$ & $0.27$ &  \\
    \cmidrule{2-14}
    \multirow{3}{*}{\ameson{0}{1450}{+}} &\aR & $0.44$ & $0.05\pm0.13$ & $0.10$ & $0.01$ & $0.01$ & $0.06$ & $0.03$ & $0.02$ & $0.03$ & $0.01$ & $0.01$ & $0.00$ &  \\
     &\pR $(^{\circ})$ & $-140$ & $9\pm35$ & $26.9$ & $3.9$ & $13.7$ & $9.2$ & $13.3$ & $6.2$ & $3.7$ & $1.0$ & $1.1$ & $2.6$ &  \\
     &FF [\%] & $0.45$ & $0.09\pm0.34$ & $0.27$ & $0.01$ & $0.00$ & $0.15$ & $0.09$ & $0.03$ & $0.08$ & $0.03$ & $0.02$ & $0.00$ &  \\
    \cmidrule{2-14}
    \multirow{3}{*}{\rmeson{1450}{+}} &\aR & $2.3$ & $0.4\pm0.8$ & $0.16$ & $0.26$ & $0.29$ & $0.14$ & $0.52$ & $0.30$ & $0.02$ & $0.23$ & $0.15$ & $0.08$ &  \\
     &\pR $(^{\circ})$ & $-60$ & $6\pm18$ & $10.0$ & $8.3$ & $4.3$ & $10.1$ & $1.7$ & $3.2$ & $2.8$ & $4.0$ & $1.4$ & $1.4$ &  \\
     &FF [\%] & $1.5$ & $0.5\pm0.9$ & $0.08$ & $0.16$ & $0.38$ & $0.23$ & $0.30$ & $0.35$ & $0.03$ & $0.25$ & $0.16$ & $0.51$ &  \\
    \cmidrule{2-14}
    \multirow{3}{*}{\rmeson{1700}{+}} &\aR & $1.04$ & $0.12\pm0.32$ & $0.23$ & $0.09$ & $0.07$ & $0.15$ & $0.09$ & $0.05$ & $0.06$ & $0.02$ & $0.01$ & $0.01$ &  \\
     &\pR $(^{\circ})$ & $4$ & $11\pm20$ & $4.6$ & $13.9$ & $7.3$ & $2.4$ & $8.4$ & $6.9$ & $2.2$ & $1.6$ & $1.8$ & $2.5$ &  \\
     &FF [\%] & $0.5$ & $0.1\pm0.5$ & $0.39$ & $0.12$ & $0.07$ & $0.13$ & $0.02$ & $0.06$ & $0.05$ & $0.02$ & $0.01$ & $0.11$ &  \\
    \midrule
    \chisqbin & & \multicolumn{2}{c}{1.07} & $1.09$ & $1.07$ & $1.06$ & $1.04$ & $1.06$ & $-$ & $-$ & $1.07$ & $1.07$ & $1.12$ \\
    DFF [\%] & & \multicolumn{2}{c}{80.7} & $78.5$ & $71.3$ & $84.9$ & $83.4$ & $82.9$ & $-$ & $-$ & $79.9$ & $80.5$ & $81.5$ \\
  \end{tabular}
  \end{adjustbox}
  \end{center}
  \label{tab:syst_glass_sup}
\end{sidewaystable}
\begin{sidewaystable}
  \caption{Systematic uncertainties for complex amplitudes and fit fractions in the \decay{\Dz}{\KSKppim} model using the \lassc parameterization. The headings are defined in Table~\ref{sec:systematics_abbreviations}.}
  \vspace{-8mm}
  \footnotesize
  \begin{center}
  \begin{adjustbox}{width=\textheight,totalheight=0.7\textwidth,keepaspectratio}
  \begin{tabular}{l l r@{$\,\pm\,$}l l l l l l l l l l l l}
    Resonance & Var & \multicolumn{2}{c}{Baseline} & \rothead{$d_{\Dz}, d_{\PR}$} & \rothead{\ntll} & \rothead{$\max(|\cos|)$} & \rothead{Weights} & \rothead{Efficiency} & \rothead{\flatte} & \rothead{Joint} & \rothead{Comb.} & \rothead{$T_{\rhomeson^{\pm}}$} & \rothead{$f_{\text{m}}, f_{\text{c}}$} \\
    \midrule
    \multirow{1}{*}{\Kst{}{892}{-}} &FF [\%] & $28.8$ & $0.4\pm1.3$ & $0.07$ & $0.08$ & $0.32$ & $0.21$ & $0.15$ & $0.08$ & $1.21$ & $0.01$ & $0.01$ & $0.19$ &  \\
    \cmidrule{2-14}
    \multirow{3}{*}{\Kst{}{1410}{-}} &\aR & $9.1$ & $0.6\pm1.5$ & $1.21$ & $0.06$ & $0.58$ & $0.41$ & $0.39$ & $0.08$ & $0.19$ & $0.09$ & $0.01$ & $0.03$ &  \\
     &\pR $(^{\circ})$ & $-79$ & $3\pm7$ & $5.2$ & $4.1$ & $2.7$ & $0.8$ & $0.6$ & $0.9$ & $0.2$ & $0.2$ & $0.0$ & $0.4$ &  \\
     &FF [\%] & $11.9$ & $1.5\pm2.2$ & $0.15$ & $0.00$ & $1.42$ & $1.11$ & $1.04$ & $0.17$ & $0.71$ & $0.15$ & $0.10$ & $0.08$ &  \\
    \cmidrule{2-14}
    \multirow{3}{*}{\negswave} &\aR & $1.16$ & $0.11\pm0.32$ & $0.20$ & $0.21$ & $0.10$ & $0.04$ & $0.04$ & $0.04$ & $0.01$ & $0.04$ & $0.02$ & $0.03$ &  \\
     &\pR $(^{\circ})$ & $-101$ & $6\pm21$ & $19.3$ & $6.2$ & $5.7$ & $2.7$ & $1.6$ & $2.3$ & $0.4$ & $0.0$ & $0.8$ & $0.6$ &  \\
     &FF [\%] & $6.3$ & $0.9\pm2.1$ & $1.30$ & $1.31$ & $0.81$ & $0.24$ & $0.42$ & $0.20$ & $0.19$ & $0.26$ & $0.20$ & $0.11$ &  \\
    \cmidrule{2-14}
    \multirow{3}{*}{\Kst{}{892}{0}} &\aR & $0.427$ & $0.010\pm0.013$ & $0.01$ & $0.01$ & $0.00$ & $0.00$ & $0.00$ & $0.00$ & $0.00$ & $0.00$ & $0.00$ & $0.00$ &  \\
     &\pR $(^{\circ})$ & $-175.0$ & $1.7\pm1.4$ & $0.2$ & $1.2$ & $0.3$ & $0.2$ & $0.2$ & $0.3$ & $0.1$ & $0.5$ & $0.3$ & $0.2$ &  \\
     &FF [\%] & $5.17$ & $0.21\pm0.32$ & $0.16$ & $0.20$ & $0.04$ & $0.08$ & $0.08$ & $0.01$ & $0.10$ & $0.10$ & $0.00$ & $0.02$ &  \\
    \cmidrule{2-14}
    \multirow{3}{*}{\Kst{}{1410}{0}} &\aR & $4.2$ & $0.5\pm0.9$ & $0.33$ & $0.20$ & $0.71$ & $0.16$ & $0.13$ & $0.15$ & $0.03$ & $0.08$ & $0.02$ & $0.06$ &  \\
     &\pR $(^{\circ})$ & $165$ & $5\pm10$ & $1.3$ & $7.1$ & $3.6$ & $4.3$ & $2.5$ & $0.7$ & $0.2$ & $1.8$ & $1.1$ & $0.6$ &  \\
     &FF [\%] & $2.2$ & $0.6\pm2.1$ & $1.82$ & $0.36$ & $0.77$ & $0.34$ & $0.19$ & $0.16$ & $0.12$ & $0.03$ & $0.07$ & $0.07$ &  \\
    \cmidrule{2-14}
    \multirow{3}{*}{\neutswave} &\aR & $1.7$ & $0.2\pm0.4$ & $0.18$ & $0.29$ & $0.17$ & $0.06$ & $0.09$ & $0.04$ & $0.00$ & $0.08$ & $0.00$ & $0.03$ &  \\
     &\pR $(^{\circ})$ & $144$ & $3\pm6$ & $3.6$ & $0.9$ & $2.2$ & $2.3$ & $3.3$ & $0.4$ & $0.9$ & $0.5$ & $0.7$ & $0.3$ &  \\
     &FF [\%] & $17$ & $2\pm6$ & $3.76$ & $4.31$ & $2.30$ & $0.70$ & $0.56$ & $0.60$ & $1.37$ & $0.54$ & $0.00$ & $0.32$ &  \\
    \cmidrule{2-14}
    \multirow{3}{*}{\ameson{0}{980}{+}} &\aR & $3.8$ & $0.2\pm0.7$ & $0.30$ & $0.64$ & $0.03$ & $0.11$ & $0.17$ & $0.06$ & $0.02$ & $0.10$ & $0.01$ & $0.03$ &  \\
     &\pR $(^{\circ})$ & $126$ & $3\pm6$ & $4.3$ & $0.4$ & $0.8$ & $1.3$ & $1.9$ & $2.7$ & $0.2$ & $0.6$ & $0.5$ & $0.5$ &  \\
     &FF [\%] & $26$ & $2\pm10$ & $3.61$ & $8.83$ & $0.01$ & $1.67$ & $1.44$ & $0.27$ & $1.12$ & $0.92$ & $0.15$ & $0.15$ &  \\
    \cmidrule{2-14}
    \multirow{3}{*}{\ameson{0}{1450}{+}} &\aR & $0.86$ & $0.10\pm0.12$ & $0.06$ & $0.00$ & $0.00$ & $0.04$ & $0.07$ & $0.02$ & $0.03$ & $0.04$ & $0.01$ & $0.02$ &  \\
     &\pR $(^{\circ})$ & $-110$ & $8\pm7$ & $1.7$ & $2.5$ & $3.5$ & $1.2$ & $3.9$ & $1.8$ & $1.1$ & $1.1$ & $0.5$ & $1.0$ &  \\
     &FF [\%] & $1.5$ & $0.3\pm0.4$ & $0.22$ & $0.07$ & $0.02$ & $0.15$ & $0.25$ & $0.06$ & $0.05$ & $0.13$ & $0.03$ & $0.05$ &  \\
    \cmidrule{2-14}
    \multirow{3}{*}{\rmeson{1700}{+}} &\aR & $1.25$ & $0.15\pm0.33$ & $0.22$ & $0.12$ & $0.07$ & $0.12$ & $0.12$ & $0.02$ & $0.02$ & $0.04$ & $0.11$ & $0.02$ &  \\
     &\pR $(^{\circ})$ & $39$ & $9\pm15$ & $9.5$ & $8.4$ & $5.1$ & $4.0$ & $2.0$ & $1.3$ & $0.1$ & $0.4$ & $2.1$ & $1.1$ &  \\
     &FF [\%] & $0.53$ & $0.11\pm0.23$ & $0.15$ & $0.08$ & $0.06$ & $0.07$ & $0.10$ & $0.01$ & $0.07$ & $0.03$ & $0.02$ & $0.01$ &  \\
    \midrule
    \chisqbin & & \multicolumn{2}{c}{1.09} & $1.10$ & $1.07$ & $1.08$ & $-$ & $-$ & $1.09$ & $1.14$ & $1.09$ & $1.09$ & $1.09$ \\
    DFF [\%] & & \multicolumn{2}{c}{99.0} & $99.1$ & $104.8$ & $95.6$ & $-$ & $-$ & $98.9$ & $100.7$ & $99.3$ & $99.3$ & $99.4$ \\
  \end{tabular}
  \end{adjustbox}
  \end{center}
  \label{tab:syst_lass_sup}
\end{sidewaystable}
\begin{sidewaystable}
  \caption{Systematic uncertainties for shared parameters, coherence and relative
           branching ratio observables in the \glass models. The systematic
           uncertainty on the \Kst{}{892}{\pm} width
           due to neglecting resolution effects in the nominal models is $0.5\mevcc$.}
  \footnotesize
  \resizebox{\textwidth}{!}{
  \begin{tabular}{l l r@{$\,\pm\,$}l l l l l l l l l l l l}
    Resonance & Var & \multicolumn{2}{c}{Baseline} & \rothead{$d_{\Dz}, d_{\PR}$} & \rothead{Comb.} & \rothead{$\max(|\cos|)$} & \rothead{$T_{\rhomeson^{\pm}}$} & \rothead{Joint} & \rothead{\ntll} & \rothead{Efficiency} & \rothead{Weights} & \rothead{\flatte} & \rothead{$f_{\text{m}}, f_{\text{c}}$} \\
    \midrule
    \multirow{3}{*}{Coherence} &$R_{\KSKpis}$ & $0.573$ & $0.007\pm0.019$ & $0.000$ & $0.004$ & $0.001$ & $0.003$ & $0.017$ & $0.004$ & $0.002$ & $0.004$ & $0.002$ & $0.004$ &  \\
     &$\delta_{\KSKpis} - \delta_{\Kstar\kaon}$ & $0.2$ & $0.6\pm1.1$ & $0.03$ & $0.32$ & $0.45$ & $0.39$ & $0.69$ & $0.35$ & $0.23$ & $0.21$ & $0.09$ & $0.22$ & $(^{\circ})$ \\
     &$R_{\Kstar\kaon}$ & $0.831$ & $0.004\pm0.011$ & $0.002$ & $0.004$ & $0.002$ & $0.002$ & $0.007$ & $0.002$ & $0.004$ & $0.003$ & $0.001$ & $0.002$ &  \\
    \cmidrule{2-14}
    \multirow{2}{*}{\BR} &$\BR_{\Kstar\kaon}$ & $0.372$ & $0.001\pm0.009$ & $0.001$ & $0.001$ & $0.000$ & $0.001$ & $0.008$ & $0.000$ & $0.002$ & $0.001$ & $0.001$ & $0.001$ &  \\
     &$\BR_{\KSKpis}$ & $0.655$ & $0.001\pm0.006$ & $0.001$ & $0.002$ & $0.003$ & $0.001$ & $0.002$ & $0.000$ & $0.003$ & $0.003$ & $0.000$ & $0.000$ &  \\
    \cmidrule{2-14}
    \multirow{1}{*}{\CP-even fraction} &$F_+$ & $0.777$ & $0.003\pm0.009$ & $0.000$ & $0.002$ & $0.000$ & $0.001$ & $0.008$ & $0.002$ & $0.001$ & $0.002$ & $0.001$ & $0.002$ &  \\
    \cmidrule{2-14}
    \multirow{2}{*}{\Kst{}{892}{\pm}} &$m_{\PR}$ & $893.1$ & $0.1\pm0.9$ & $0.3$ & $0.1$ & $0.1$ & $0.1$ & $0.9$ & $0.0$ & $0.1$ & $0.0$ & $0.0$ & $0.0$ & \mevcc \\
     &$\Gamma_{\PR}$ & $46.9$ & $0.3\pm2.5$ & $0.5$ & $0.4$ & $0.5$ & $0.4$ & $2.2$ & $0.1$ & $0.2$ & $0.2$ & $0.1$ & $0.1$ & \mevcc \\
    \cmidrule{2-14}
    \multirow{1}{*}{\Kst{}{1410}{\pm}} &$\Gamma_{\PR}$ & $210$ & $20\pm60$ & $8.8$ & $41.0$ & $27.0$ & $36.2$ & $11.0$ & $4.1$ & $12.5$ & $2.9$ & $7.1$ & $5.9$ & \mevcc \\
    \cmidrule{2-14}
    \multirow{4}{*}{\chgswave} &$a$ & $4.7$ & $0.4\pm1.0$ & $0.209$ & $0.168$ & $0.606$ & $0.311$ & $0.092$ & $0.470$ & $0.095$ & $0.360$ & $0.114$ & $0.150$ & \invgevc \\
     &$\phi_F$ & $0.28$ & $0.05\pm0.19$ & $0.1$ & $0.0$ & $0.1$ & $0.0$ & $0.0$ & $0.0$ & $0.0$ & $0.0$ & $0.0$ & $0.0$ & \rad \\
     &$\phi_S$ & $-3.5$ & $0.2\pm0.5$ & $0.2$ & $0.1$ & $0.0$ & $0.4$ & $0.1$ & $0.1$ & $0.1$ & $0.1$ & $0.1$ & $0.1$ & \rad \\
     &$r$ & $-5.3$ & $0.4\pm1.9$ & $0.68$ & $0.84$ & $0.87$ & $1.08$ & $0.15$ & $0.49$ & $0.37$ & $0.29$ & $0.19$ & $0.24$ & \invgevc \\
    \cmidrule{2-14}
    \multirow{2}{*}{\Kst{}{1410}{0}} &$m_{\PR}$ & $1426$ & $8\pm24$ & $2.3$ & $11.8$ & $5.8$ & $13.0$ & $10.6$ & --- & $7.4$ & $5.6$ & $2.9$ & $2.0$ & \mevcc \\
     &$\Gamma_{\PR}$ & $270$ & $20\pm40$ & $26.2$ & $5.4$ & $17.6$ & $1.4$ & $23.6$ & $8.2$ & $4.6$ & $3.0$ & $4.9$ & $0.9$ & \mevcc \\
    \cmidrule{2-14}
    \multirow{5}{*}{\neutswave} &$F$ & $0.15$ & $0.03\pm0.14$ & $0.06$ & $0.02$ & $0.06$ & $0.03$ & $0.02$ & $0.08$ & $0.05$ & $0.03$ & $0.00$ & $0.00$ &  \\
     &$a$ & $4.2$ & $0.3\pm2.8$ & $1.074$ & $1.563$ & $1.092$ & $1.548$ & $0.649$ & $0.260$ & $0.607$ & $0.143$ & $0.090$ & $0.030$ & \invgevc \\
     &$\phi_F$ & $-2.5$ & $0.2\pm1.0$ & $0.8$ & $0.2$ & $0.2$ & $0.2$ & $0.0$ & $0.1$ & $0.3$ & $0.2$ & $0.1$ & $0.1$ & \rad \\
     &$\phi_S$ & $-1.1$ & $0.6\pm1.3$ & $0.7$ & $0.3$ & $0.8$ & $0.4$ & $0.2$ & $0.3$ & $0.5$ & $0.3$ & $0.0$ & $0.0$ & \rad \\
     &$r$ & $-3.0$ & $0.4\pm1.7$ & $1.51$ & $0.34$ & $0.33$ & $0.30$ & $0.02$ & $0.45$ & $0.17$ & $0.11$ & $0.05$ & $0.04$ & \invgevc \\
    \cmidrule{2-14}
    \multirow{1}{*}{\ameson{0}{1450}{\pm}} &$m_{\PR}$ & $1430$ & $10\pm40$ & $4.1$ & $26.9$ & $16.6$ & $23.8$ & $12.6$ & $1.8$ & $7.1$ & $3.1$ & $0.8$ & $0.9$ & \mevcc \\
    \cmidrule{2-14}
    \multirow{1}{*}{\rmeson{1450}{\pm}} &$\Gamma_{\PR}$ & $410$ & $19\pm35$ & $11.1$ & $1.6$ & $4.1$ & $2.2$ & $31.3$ & --- & $8.1$ & $2.7$ & $2.6$ & $1.5$ & \mevcc \\
    \cmidrule{2-14}
    \multirow{1}{*}{\rmeson{1700}{\pm}} &$m_{\PR}$ & $1530$ & $10\pm40$ & $6.4$ & $2.6$ & $4.0$ & $0.8$ & $36.3$ & $14.7$ & $3.0$ & $4.3$ & $2.4$ & $2.6$ & \mevcc \\
  \end{tabular}}
  \label{tab:syst_glass}
\end{sidewaystable}
\begin{sidewaystable}
  \caption{Systematic uncertainties for shared parameters, coherence and
           relative branching ratio observables in the \lassc models. The
           systematic uncertainty on the \Kst{}{892}{\pm} width due to
           neglecting resolution effects in the nominal models is $0.6\mevcc$.}
  \footnotesize
  \resizebox{\textwidth}{!}{
  \begin{tabular}{l l r@{$\,\pm\,$}l l l l l l l l l l l l}
    Resonance & Var & \multicolumn{2}{c}{Baseline} & \rothead{Joint} & \rothead{\ntll} & \rothead{$d_{\Dz}, d_{\PR}$} & \rothead{Weights} & \rothead{Efficiency} & \rothead{$\max(|\cos|)$} & \rothead{Comb.} & \rothead{\flatte} & \rothead{$f_{\text{m}}, f_{\text{c}}$} & \rothead{$T_{\rhomeson^{\pm}}$} \\
    \midrule
    \multirow{3}{*}{Coherence} &$R_{\KSKpis}$ & $0.571$ & $0.005\pm0.019$ & $0.015$ & $0.011$ & $0.005$ & $0.003$ & $0.002$ & $0.001$ & $0.001$ & $0.002$ & $0.003$ & $0.000$ &  \\
     &$\delta_{\KSKpis} - \delta_{\Kstar\kaon}$ & $-0.0$ & $0.5\pm0.8$ & $0.00$ & $0.01$ & $0.58$ & $0.30$ & $0.24$ & $0.09$ & $0.16$ & $0.12$ & $0.19$ & $0.00$ & $(^{\circ})$ \\
     &$R_{\Kstar\kaon}$ & $0.835$ & $0.003\pm0.011$ & $0.009$ & $0.001$ & $0.003$ & $0.003$ & $0.003$ & $0.001$ & $0.001$ & $0.001$ & $0.002$ & $0.000$ &  \\
    \cmidrule{2-14}
    \multirow{2}{*}{\BR} &$\BR_{\Kstar\kaon}$ & $0.368$ & $0.001\pm0.011$ & $0.010$ & $0.001$ & $0.001$ & $0.001$ & $0.002$ & $0.000$ & $0.000$ & $0.001$ & $0.001$ & $0.000$ &  \\
     &$\BR_{\KSKpis}$ & $0.656$ & $0.001\pm0.006$ & $0.003$ & $0.001$ & $0.000$ & $0.003$ & $0.003$ & $0.001$ & $0.000$ & $0.000$ & $0.000$ & $0.000$ &  \\
    \cmidrule{2-14}
    \multirow{1}{*}{\CP-even fraction} &$F_+$ & $0.776$ & $0.003\pm0.009$ & $0.007$ & $0.005$ & $0.002$ & $0.001$ & $0.001$ & $0.001$ & $0.000$ & $0.001$ & $0.002$ & $0.000$ &  \\
    \cmidrule{2-14}
    \multirow{2}{*}{\Kst{}{892}{\pm}} &$m_{\PR}$ & $893.4$ & $0.1\pm1.1$ & $1.0$ & $0.1$ & $0.2$ & $0.0$ & $0.1$ & $0.0$ & $0.0$ & $0.0$ & $0.0$ & $0.0$ & \mevcc \\
     &$\Gamma_{\PR}$ & $47.4$ & $0.3\pm2.0$ & $1.9$ & $0.1$ & $0.1$ & $0.2$ & $0.2$ & $0.3$ & $0.1$ & $0.1$ & $0.1$ & $0.0$ & \mevcc \\
    \cmidrule{2-14}
    \multirow{1}{*}{\Kst{}{1410}{\pm}} &$m_{\PR}$ & $1437$ & $8\pm16$ & $8.8$ & $5.8$ & $8.9$ & $1.9$ & $5.6$ & $5.4$ & $3.3$ & $1.3$ & $1.2$ & $2.0$ & \mevcc \\
    \cmidrule{2-14}
    \multirow{3}{*}{\chgswave} &$b'_1$ & $60$ & $30\pm40$ & $14$ & $16$ & $20$ & $10$ & $22$ & $8$ & $6$ & $6$ & $6$ & $2$ &  \\
     &$b'_2$ & $4$ & $1\pm5$ & $4.8$ & $1.2$ & $0.6$ & $0.4$ & $0.6$ & $0.1$ & $0.4$ & $0.3$ & $0.2$ & $0.3$ &  \\
     &$b'_3$ & $3.0$ & $0.2\pm0.7$ & $0.4$ & $0.4$ & $0.3$ & $0.1$ & $0.1$ & $0.1$ & $0.1$ & $0.1$ & $0.1$ & $0.0$ &  \\
    \cmidrule{2-14}
    \multirow{1}{*}{\Kst{}{1410}{0}} &$m_{\PR}$ & $1404$ & $9\pm22$ & $7.1$ & --- & $14.9$ & $10.7$ & $6.6$ & $4.5$ & $3.5$ & $0.7$ & $0.5$ & $3.4$ & \mevcc \\
    \cmidrule{2-14}
    \multirow{3}{*}{\neutswave} &$b'_1$ & $130$ & $30\pm80$ & $49$ & $28$ & $2$ & $39$ & $13$ & $32$ & $12$ & $6$ & $5$ & $1$ &  \\
     &$b'_2$ & $-6$ & $1\pm14$ & $0.7$ & $14.2$ & $1.4$ & $1.2$ & $1.3$ & $0.2$ & $0.2$ & $0.1$ & $0.3$ & $0.3$ &  \\
     &$b'_3$ & $2.5$ & $0.1\pm1.4$ & $0.1$ & $1.3$ & $0.1$ & $0.3$ & $0.2$ & $0.1$ & $0.1$ & $0.1$ & $0.1$ & $0.0$ &  \\
    \cmidrule{2-14}
    \multirow{1}{*}{\kpswave} &$r$ & $1.2$ & $0.3\pm0.4$ & $0.04$ & $0.23$ & $0.28$ & $0.11$ & $0.10$ & $0.06$ & $0.03$ & $0.04$ & $0.04$ & $0.04$ & \invgevc \\
    \cmidrule{2-14}
    \multirow{1}{*}{\ameson{0}{980}{\pm}} &$m_{\PR}$ & $925$ & $5\pm8$ & $3.7$ & $1.6$ & $3.8$ & $1.6$ & $3.4$ & $3.3$ & $1.9$ & $2.0$ & $0.5$ & $0.0$ & \mevcc \\
    \cmidrule{2-14}
    \multirow{2}{*}{\ameson{0}{1450}{\pm}} &$m_{\PR}$ & $1458$ & $14\pm15$ & $4.4$ & --- & $4.2$ & $6.0$ & $8.2$ & $7.1$ & $2.2$ & $4.4$ & $4.1$ & $1.4$ & \mevcc \\
     &$\Gamma_{\PR}$ & $282$ & $12\pm13$ & $12.6$ & $1.0$ & $2.2$ & $1.2$ & $2.1$ & $1.4$ & $0.6$ & $1.9$ & $1.5$ & $0.7$ & \mevcc \\
    \cmidrule{2-14}
    \multirow{1}{*}{\rmeson{1450}{\pm}} &$m_{\PR}$ & $1208$ & $8\pm9$ & $2.7$ & $5.2$ & $3.5$ & $0.7$ & $4.7$ & $2.7$ & $1.9$ & $1.2$ & $0.8$ & --- & \mevcc \\
    \cmidrule{2-14}
    \multirow{1}{*}{\rmeson{1700}{\pm}} &$m_{\PR}$ & $1552$ & $13\pm26$ & $19.0$ & $5.1$ & $14.9$ & $7.0$ & $3.5$ & $3.2$ & $0.1$ & $2.5$ & $2.0$ & $1.1$ & \mevcc \\
  \end{tabular}}
  \label{tab:syst_lass}
\end{sidewaystable}
\section{{\boldmath\CP} violation fit results}
\label{sec:cpvappendix}
This appendix contains Table~\ref{tab:fullcpvresults}, which summarizes the full fit results of the
\CP violation searches described in Sect.~\ref{sec:cpv}.
\begin{sidewaystable}
  \caption{Full \CP violation fit results as described in Sect.~\ref{sec:cpv}.
           The only uncertainties included are statistical.}
  \begin{subtable}{\textwidth}
    \centering
    \caption{Model parameters for the \decay{\Dz}{\KSKmpip} mode.}
    \resizebox{\textwidth}{!}{
    \begin{tabular}{l r@{$\,\pm\,$}l r@{$\,\pm\,$}l r@{$\,\pm\,$}l r@{$\,\pm\,$}l r@{$\,\pm\,$}l r@{$\,\pm\,$}l r@{$\,\pm\,$}l r@{$\,\pm\,$}l}
      & \multicolumn{4}{c}{\aR} & \multicolumn{4}{c}{\daR} & \multicolumn{4}{c}{\pR} & \multicolumn{4}{c}{\dpR}\\
      Resonance & \multicolumn{2}{c}{\glass } & \multicolumn{2}{c}{\lassc } & \multicolumn{2}{c}{\glass } & \multicolumn{2}{c}{\lassc } & \multicolumn{2}{c}{\glass } & \multicolumn{2}{c}{\lassc } & \multicolumn{2}{c}{\glass } & \multicolumn{2}{c}{\lassc } \\
      \midrule
      \Kst{}{892}{+} & \multicolumn{2}{c}{1.0 (fixed)} & \multicolumn{2}{c}{1.0 (fixed)} & \multicolumn{2}{c}{0.0 (fixed)} & \multicolumn{2}{c}{0.0 (fixed)} & \multicolumn{2}{c}{0.0 (fixed)} & \multicolumn{2}{c}{0.0 (fixed)} & \multicolumn{2}{c}{0.0 (fixed)} & \multicolumn{2}{c}{0.0 (fixed)} \\
      \Kst{}{1410}{+} & $4.24$ & $0.30$ & $5.83$ & $0.30$ & $0.07$ & $0.05$ & $0.03$ & $0.05$ & $-160$ & $6$ & $-143.1$ & $3.4$ & $3.9$ & $2.9$ & $2.0$ & $2.1$ \\
      \posswave & $0.63$ & $0.05$ & $1.12$ & $0.09$ & $0.018$ & $0.034$ & $-0.053$ & $0.030$ & $-66$ & $5$ & $-59$ & $4$ & $2.0$ & $1.7$ & $2.0$ & $1.7$ \\
      \Kbarst{}{892}{0} & $0.213$ & $0.007$ & $0.210$ & $0.006$ & $-0.046$ & $0.031$ & $-0.051$ & $0.029$ & $-108.2$ & $2.2$ & $-101.5$ & $2.0$ & $1.2$ & $1.6$ & $1.5$ & $1.6$ \\
      \Kbarst{}{1410}{0} & $6.04$ & $0.27$ & $3.93$ & $0.19$ & $0.006$ & $0.029$ & $0.02$ & $0.04$ & $-179$ & $4$ & $-174$ & $4$ & $1.9$ & $1.3$ & $-3.3$ & $2.1$ \\
      \Kbarst{2}{1430}{0} & $3.31$ & $0.29$ & \multicolumn{2}{c}{---} & $-0.05$ & $0.04$ & \multicolumn{2}{c}{---} & $-171$ & $5$ & \multicolumn{2}{c}{---} & $1.8$ & $2.4$ & \multicolumn{2}{c}{---} \\
      \neutswave & $2.48$ & $0.23$ & $1.28$ & $0.11$ & $0.051$ & $0.031$ & $0.032$ & $0.031$ & $47$ & $13$ & $74.9$ & $2.9$ & $0.4$ & $1.4$ & $1.0$ & $1.3$ \\
      \ameson{0}{980}{-} & \multicolumn{2}{c}{---} & $1.07$ & $0.09$ & \multicolumn{2}{c}{---} & $-0.01$ & $0.06$ & \multicolumn{2}{c}{---} & $82$ & $5$ & \multicolumn{2}{c}{---} & $7.0$ & $2.8$ \\
      \ameson{2}{1320}{-} & $0.195$ & $0.030$ & $0.169$ & $0.030$ & $-0.25$ & $0.14$ & $-0.24$ & $0.13$ & $-128$ & $9$ & $-128$ & $10$ & $2$ & $8$ & $-1$ & $9$ \\
      \ameson{0}{1450}{-} & $0.51$ & $0.04$ & $0.44$ & $0.05$ & $-0.01$ & $0.07$ & $-0.13$ & $0.08$ & $-84$ & $7$ & $-52$ & $10$ & $0.2$ & $2.7$ & $-4$ & $4$ \\
      \rmeson{1450}{-} & $1.58$ & $0.18$ & $1.30$ & $0.11$ & $0.06$ & $0.07$ & $-0.05$ & $0.06$ & $-175$ & $7$ & $-144$ & $7$ & $-13$ & $4$ & $-4.7$ & $2.9$ \\
      \rmeson{1700}{-} & $0.39$ & $0.08$ & \multicolumn{2}{c}{---} & $-0.08$ & $0.15$ & \multicolumn{2}{c}{---} & $-65$ & $12$ & \multicolumn{2}{c}{---} & $7$ & $8$ & \multicolumn{2}{c}{---} \\
    \end{tabular}}
  \end{subtable}
  \\[12pt]
  \begin{subtable}{\textwidth}
    \centering
    \caption{Model parameters for the \decay{\Dz}{\KSKppim} mode.}
    \resizebox{\textwidth}{!}{
    \begin{tabular}{l r@{$\,\pm\,$}l r@{$\,\pm\,$}l r@{$\,\pm\,$}l r@{$\,\pm\,$}l r@{$\,\pm\,$}l r@{$\,\pm\,$}l r@{$\,\pm\,$}l r@{$\,\pm\,$}l}
      & \multicolumn{4}{c}{\aR} & \multicolumn{4}{c}{\daR} & \multicolumn{4}{c}{\pR} & \multicolumn{4}{c}{\dpR}\\
      Resonance & \multicolumn{2}{c}{\glass } & \multicolumn{2}{c}{\lassc } & \multicolumn{2}{c}{\glass } & \multicolumn{2}{c}{\lassc } & \multicolumn{2}{c}{\glass } & \multicolumn{2}{c}{\lassc } & \multicolumn{2}{c}{\glass } & \multicolumn{2}{c}{\lassc } \\
      \midrule
      \Kst{}{892}{-} & \multicolumn{2}{c}{1.0 (fixed)} & \multicolumn{2}{c}{1.0 (fixed)} & \multicolumn{2}{c}{0.0 (fixed)} & \multicolumn{2}{c}{0.0 (fixed)} & \multicolumn{2}{c}{0.0 (fixed)} & \multicolumn{2}{c}{0.0 (fixed)} & \multicolumn{2}{c}{0.0 (fixed)} & \multicolumn{2}{c}{0.0 (fixed)} \\
      \Kst{}{1410}{-} & $4.8$ & $0.5$ & $9.1$ & $0.6$ & $0.05$ & $0.09$ & $-0.03$ & $0.06$ & $-105$ & $6$ & $-78.5$ & $3.0$ & $-5.8$ & $3.3$ & $-3.0$ & $2.2$ \\
      \negswave & $0.59$ & $0.05$ & $1.14$ & $0.11$ & $0.10$ & $0.06$ & $-0.14$ & $0.07$ & $-163$ & $6$ & $-102$ & $6$ & $-7.7$ & $3.4$ & $-8$ & $4$ \\
      \Kst{}{892}{0} & $0.409$ & $0.010$ & $0.427$ & $0.010$ & $-0.010$ & $0.024$ & $-0.012$ & $0.022$ & $176.2$ & $2.2$ & $-174.6$ & $1.7$ & $-1.4$ & $1.8$ & $0.8$ & $1.6$ \\
      \Kst{}{1410}{0} & $6.2$ & $0.5$ & $4.2$ & $0.5$ & $0.10$ & $0.05$ & $0.19$ & $0.09$ & $175$ & $4$ & $165$ & $5$ & $-0.6$ & $3.3$ & $-9$ & $4$ \\
      \Kst{2}{1430}{0} & $6.2$ & $0.5$ & \multicolumn{2}{c}{---} & $-0.09$ & $0.05$ & \multicolumn{2}{c}{---} & $-139$ & $5$ & \multicolumn{2}{c}{---} & $6$ & $4$ & \multicolumn{2}{c}{---} \\
      \neutswave & $3.66$ & $0.29$ & $1.72$ & $0.16$ & $-0.075$ & $0.034$ & $-0.12$ & $0.04$ & $96$ & $12$ & $145.4$ & $3.0$ & $-1.9$ & $2.4$ & $1.9$ & $2.3$ \\
      \ameson{0}{980}{+} & $1.74$ & $0.10$ & $3.79$ & $0.20$ & $0.06$ & $0.04$ & $0.052$ & $0.024$ & $65$ & $5$ & $127.4$ & $3.5$ & $-3$ & $5$ & $-0.9$ & $2.3$ \\
      \ameson{0}{1450}{+} & $0.44$ & $0.05$ & $0.87$ & $0.10$ & $-0.11$ & $0.09$ & $-0.07$ & $0.06$ & $-139$ & $9$ & $-112$ & $7$ & $10$ & $6$ & $5$ & $4$ \\
      \rmeson{1450}{+} & $2.4$ & $0.4$ & \multicolumn{2}{c}{---} & $-0.06$ & $0.10$ & \multicolumn{2}{c}{---} & $-60$ & $6$ & \multicolumn{2}{c}{---} & $5$ & $4$ & \multicolumn{2}{c}{---} \\
      \rmeson{1700}{+} & $1.01$ & $0.12$ & $1.23$ & $0.15$ & $-0.03$ & $0.09$ & $-0.12$ & $0.09$ & $2$ & $11$ & $40$ & $9$ & $4$ & $6$ & $2$ & $5$ \\
    \end{tabular}}
  \end{subtable}
  \label{tab:fullcpvresults}
\end{sidewaystable}



\clearpage
\section*{Supplemental material}
\label{sec:Supplementary}
This is divided into two parts: lookup tables
for the complex amplitude and covariance information, each
for the four quoted amplitude models.
These are available at Ref.~\cite{SupplementalMaterial}.
No correlation information is included for systematic uncertainties.
\subsection*{Isobar model lookup tables}
The lookup table filenames are listed in Table~\ref{tab:lookupnames}.
\begin{table}[b]
  \caption{Lookup table filenames.}
  \begin{center}
    \begin{tabular}{l c c}
      & \multicolumn{2}{c}{\kpswave parameterization} \\
      \Dz decay mode & \glass & \lassc \\
      \midrule
      \decay{\Dz}{\KSKmpip} & \texttt{glass\_fav\_lookup.txt} & \texttt{lass\_fav\_lookup.txt} \\
      \decay{\Dz}{\KSKppim} & \texttt{glass\_sup\_lookup.txt} & \texttt{lass\_sup\_lookup.txt} \\
    \end{tabular}
  \end{center}
  \label{tab:lookupnames}
\end{table}
As an example, the first five lines of the file
\texttt{glass\_fav\_lookup.txt} are:
\begin{verbatim}
# S-wave: GLASS, mode: D0->KSK-pi+ (FAV)
# mD0 = 1.86486; mKS = 0.497614; mK = 0.493677; mPi = 0.13957018 GeV/c^2
# m^2(Kpi) GeV^2/c^4, m^2(KSpi) GeV^2/c^4, |amp|^2 arb. units, arg(amp) rad
0.300625,0.300625,0.000000e+00,0.000000
0.300625,0.301875,0.000000e+00,0.000000
\end{verbatim}
The first three lines are comments, describing which \Dz decay mode
and isobar model this file corresponds to, giving the precise
nominal masses used in the fit and, finally, defining the data
fields in the remainder of the file.
As this shows, the models are evaluated on a grid with a spacing of
$0.00125\gevgevcccc$.
\subsection*{Covariance information}
A reduced covariance matrix is presented for each isobar model,
tabulating the correlations between the complex
amplitudes $\aR e^{i\pR}$.
These are listed in files
named analogously to those in Table~\ref{tab:lookupnames},
\eg \texttt{glass\_fav\_covariance.txt}.
An example first four lines:
\begin{verbatim}
# S-wave: GLASS, mode: D0->KSK-pi+ (FAV)
# x , y , cov(x,y)
K(0)*(1430)+_Amp,K(0)*(1430)+_Amp,2.195e-03
K(0)*(1430)+_Amp,K(0)*(1430)+_Phase,1.147e-01
\end{verbatim}
\ie a similar format to the lookup tables. Note that the \kpswave
contributions are tabulated as \texttt{K(0)*(1430)+} and
\texttt{K(0)*(1430)bar0}.

\addcontentsline{toc}{section}{References}
\setboolean{inbibliography}{true}
\setboolean{uprightparticles}{false}
\bibliographystyle{LHCb}
\bibliography{kskpi,main,LHCb-PAPER,LHCb-CONF,LHCb-DP,LHCb-TDR}

\newpage

\centerline{\large\bf LHCb collaboration}
\begin{flushleft}
\small
R.~Aaij$^{38}$, 
B.~Adeva$^{37}$, 
M.~Adinolfi$^{46}$, 
A.~Affolder$^{52}$, 
Z.~Ajaltouni$^{5}$, 
S.~Akar$^{6}$, 
J.~Albrecht$^{9}$, 
F.~Alessio$^{38}$, 
M.~Alexander$^{51}$, 
S.~Ali$^{41}$
% ...

{\footnotesize \it
% ...
$ ^{5}$Clermont Universit\'{e}, Universit\'{e} Blaise Pascal, CNRS/IN2P3, LPC, Clermont-Ferrand, France\\
$ ^{6}$CPPM, Aix-Marseille Universit\'{e}, CNRS/IN2P3, Marseille, France\\
% ...
$ ^{9}$Fakult\"{a}t Physik, Technische Universit\"{a}t Dortmund, Dortmund, Germany\\
% ...
$ ^{37}$Universidad de Santiago de Compostela, Santiago de Compostela, Spain\\
$ ^{38}$European Organization for Nuclear Research (CERN), Geneva, Switzerland\\
% ...
$ ^{41}$Nikhef National Institute for Subatomic Physics, Amsterdam, The Netherlands\\
% ...
$ ^{46}$H.H. Wills Physics Laboratory, University of Bristol, Bristol, United Kingdom\\
% ...
$ ^{51}$School of Physics and Astronomy, University of Glasgow, Glasgow, United Kingdom\\
$ ^{52}$Oliver Lodge Laboratory, University of Liverpool, Liverpool, United Kingdom\\
% ...
}
\end{flushleft}

\end{document}